\documentclass[a4paper]{article}
\usepackage{epsfig}
\usepackage{epstopdf}                    
\usepackage{color}     
\usepackage{booktabs}
\usepackage{color}
\usepackage{lscape}
\usepackage{pdflscape}
\usepackage{amssymb}
\usepackage{tikz}
\usepackage{float}
\date{\today}
 \normalsize

\def\be {\begin{equation}}
\def\ee {\end{equation}}
\def\bea {\begin{eqnarray}}
\def\eea {\end{eqnarray}}
\def\bc {\begin{center}}
\def\ec {\end{center}}
\def\bfg {\begin{figure}}
\def\efg {\end{figure}}
\def\bi {\begin{itemize}}
\def\ei {\end{itemize}}
\def\nn {\nonumber}

\def\rt  {\tilde{\rho}}

\def\Ht  {\tilde{H}}

\def\pt  {\tilde{p}}
\def\Lat  {\tilde{\Lambda}}
\def\xit  {\tilde{\xi}}
\def\a  {\alpha}

\def\D  {\Delta}

\def\l  {\lambda}
\def\L  {\Lambda}
\def\m  {\mu}
\def\n  {\nu}
\def\o  {\omega}

\def\r  {\rho}
\def\th {\theta}

\def\tt {\tilde{t}}

\newcommand{\bdm}{\begin{displaymath}}
\newcommand{\edm}{\end{displaymath}}
\begin{document}
\renewcommand{\thefootnote}{\fnsymbol{footnote}}

\vspace{.3cm}

\title{\Large\bf Degenerate Bogdanov-Takens bifurcations in a bulk viscous cosmology}

\author
{   Asmaa Abdel Azim $^{1}$\thanks{asmaa\_m\_26@sci.asu.edu.eg},
  Adel Awad$^{1,2}$\thanks{awad.adel@aucegypt.edu} and  E.  I. Lashin$^{1,3,4}$\thanks{slashin@zewailcity.edu.eg, elashin@ictp.it} \\
\small$^1$ Department of Physics, Faculty of Science, Ain Shams
University, Cairo 11566,
Egypt.\\
\small$^2$ Department of Physics, School of Sciences and Engineering, American University in Cairo,\\
\small P.O. Box 74, AUC Avenue New Cairo, Cairo, Egypt.\\
\small$^3$ Centre for Fundamental Physics, Zewail City of Science and Technology,\\
\small Sheikh Zayed, 6 October City, 12578, Giza, Egypt.\\
\small$^4$ The Abdus Salam ICTP, P.O. Box 586, 34100 Trieste, Italy}

\maketitle

\begin{center}
\small{\bf Abstract}\\[3mm]
\end{center}
Using the dynamical system theory we show that the
Friedmann-Robertson-Walker (FRW) cosmological model with bulk
viscous fluid in the presence of cosmological constant is
equivalent to a degenerate two dimensional Bogdanov-Takens normal
form. The equation of state parameter, $\omega$, the bulk viscosity
coefficient, $\xi$, and the cosmological constant, $\Lambda$, define
the necessary parameters for unfolding the degenerate
Bogdanov-Takens system. The fixed points of the system are discussed
together with the variation of their stability properties upon
changing the relevant parameters $\o, \Lambda$ and $\xi$. The
variation of the stability properties are visualized by the
appropriate bifurcation diagrams. Phase portrait for finite domain
and global phase portrait are displayed and the issue of the
structural stability is discussed. Typical issues such as late
acceleration or inflation that can be induced by viscosity and could
have relevance to observational cosmology are also discussed.
\begin{minipage}[h]{14.0cm}
\end{minipage}
\vskip 0.3cm \hrule \vskip 0.5cm
\maketitle
\clearpage
\section{Introduction}
Dynamical systems techniques are important tools to classify, describe and analyze many systems and phenomena in physics \cite{Strogatz}. One of the physical systems that can be described through dynamical systems is our universe. Dynamical system tools applied to cosmology are valuable for
enabling a qualitative understanding of the behavior of
cosmological models. Through a careful suitable choice of the
dynamical variables one can capture all possible solutions and initial conditions in what is called phase portrait. These portraits shows the global behavior of all possible solutions of a specific model and where it ends through finding fixed points, or equilibrium points, without the need for obtaining
explicit form of solutions. The phase portraits can reveal the general
properties of trajectories, or how a solution evolves. This provides us with a wealth of
information about solutions especially their nature and how to classify them according to various initial conditions. These dynamical system tools provide us with not
only a qualitative understanding but also a quantitative one through
using powerful analytical and numerical methods applied to the
models under consideration.

The dynamical system tools was first
applied to anisotropic cosmological models as in \cite{stewart,collin1,collin2} while for viscous cosmology in \cite{ belin1,belin2} and for recent applications of these techniques see \cite{Adel,Andron1,Andron2}. For a review one can consult \cite{review1} and references therein.

Bogdanov-Takens bifurcations have  been shown to occur in Bianchi IX
cosmological models in the frame work of Gauss-Bonnet gravity
\cite{bonnet}. A more recent study \cite{kohli1} has also
demonstrated the occurrence of such a bifurcation in
Friedmann-Roberston-Walker (FRW) cosmology in the presence of
cosmological constant without considering viscosity.  The latter
study is of limited scope due to neglecting viscosity which is a
real physical dissipative effect which is essential  for getting
certain desirable properties of  Bogdanov-Takens system such as the
finiteness of the number of fixed points.  Up to the best of our
knowledge, the works in \cite{bonnet,kohli1} are the only two
instances in cosmological  studies where the Bogdanov-Takens
bifurcations occurred.  In fact, investigating and classifying all
possible solutions and their stability properties in cosmological
models enhances our understanding of the models.  Needless to say,
the identification of what kind of bifurcation we have  for our
cosmological models is important not only for spotting where we are
in the vast landscape of dynamical systems but also for learning how
to tune our models to have certain desired properties.

In the realm of cosmology, bulk viscosity provides the only
dissipative mechanism consistent with isotropy and homogeneity. For simplicity, we consider a  bulk viscosity model as described in the context of the Eckart
formalism \cite{ekart} rather than using the full  causal theory of viscosity that was developed in \cite{is1,is2}.
Several authors have investigated the introduction of viscosity
into cosmology for several reasons and motivations\cite{WZ,BdHOS}. For examples; in
\cite{murphy} the viscosity was introduced to resolve the big-bang
singularity, while in \cite{colist} for finding a unified model for
the dark component of universe (dark energy and dark matter) that
could fit cosmological observational data like type Ia supernovae \cite{nova1,nova2}
and power spectrum \cite{anis1,anis2}. Others  as in \cite{jou,zak,maart} introduced
viscosity as a source for deriving inflation in the early cosmology
or for deriving late acceleration as in \cite{orest_anp}. The
possibility of using some sort of viscous fluid to get a unified
cosmic history starting by inflation and ending by late acceleration
dominated by dark energy have been investigated in \cite{odin}.
Furthermore,  in \cite{mathew_2017}, it was shown that  a bulk
viscous model with constant coefficient of viscosity can give a
viable coherent description of the different phases of the universe.

The bulk viscosity besides its clear physical origin as a dissipative effect, it might also entails the cosmological  dynamical system with  structural stability  in the sense that the qualitative behaviour of the dynamical system doesn't change under small perturbation. The structural stability is a desirable property to be processed by any realistic system and thus  worthy to be studied and tested through applying the proper criteria.

The paper is structured  as follows: in Section~2, Friedmann equations for bulk viscous cosmology are presented and then expressed in terms of
$\rho$ fluid density and $H$ Hubble parameter as our suggested dynamical variables. In Section~3, The basic theories and notations of dynamical
systems are presented and explained. The theories and techniques developed in Section~3 are applied in Sections~4, 5 and 6.  Section~4 is
devoted for investigating  the case of perfect fluid with linear equation of state $p = \o \r$ where $p$ is the pressure.
Section~5 is devoted to the case of perfect fluid as in Section~4 with the inclusion of a cosmological constant $\L$. In Section~6,
the bulk viscous fluid is introduced in the presence of cosmological constant and  investigated. Thus, this  last  case amounts to having three parameter
namely $\o$, $\L$ and $\xi$ where $\xi$ is the viscosity coefficient that might be constant or linearly dependent on $\rho$. Finally Section~7 is devoted for discussion and conclusion.

\section{Einstein Equations for Bulk  Viscous Cosmology}
A homogenous and isotropic cosmological model is described by Fredimann-Roberston-Walker (FRW) metric whose line element is given as,
\be
ds^2 =g_{\m\n}\, dx^\m\, dx^\n = -c^2 dt^2 + R_0^2\,a(t)^2\,\left[{dr^2 \over 1 - k r^2} + r^2\, d\th^2 + r^2\,\sin^2{\left(\th\right)}\,d\phi^2\right],
\label{ds2}
\ee
where $x^\mu$ is the four dimensional coordinate, $x^\mu \equiv \left(x^0 = c\,t,\, x^1 = r,\, x^2 = \th,\, x^3 = \phi\right)$, $a(t)$ is the
scale factor, $c$ is the speed of light and $k=\left\{0,\pm1\right\}$ which is the curvature index, while $R_0$ is a constant carrying the dimension of length. The metric
tensor $g_{\m\n}$ can be easily read from Eq.(\ref{ds2}) to be diagonal and given by,
\be
 g_{\m\n}=\mbox{Diag}\left[-1,\; {R_0^2\, a(t)^2\over 1 - k\, r^2},\; R_0^2\,a(t)^2\, r^2,\; R_0^2\,a(t)^2\, r^2\, \sin^2{\left(\th\right)} \right].
 \ee
The scale factor $a(t)$ can be determined by applying field equations of General Relativity (GR) which , in the presence of cosmological
constant $\Lambda$, assumes the following form:
\be
R_{\m\n} -{1\over 2}\, g_{\m\n} R - \Lambda g_{\m\n} = -{8 \pi G\over c^4}\, T_{\m\n},
\label{gre0}
\ee
where $R_{\m\n}$ and $R$ are the Ricci tensor and scalar respectively. $G$ is the universal Newton gravitational constant while $c$ as before denotes the speed of light.
As to the energy-momentum tensor $T_{\m\n}$ describing a bulk viscous fluid, it assumes the form
\be \label{emt}
T_{\m\n} = \left(\r + {p-6\,\xi\, H\over c^2}\right)U_\m\, U_\n + \left(p - 6\, \xi\, H\right)\, g_{\m\n},
\label{em}
\ee
where the viscous fluid has density $\rho$, pressure $p$, viscosity coefficient $\xi$ and velocity $U_\m$.
Also, notice that $H$ is the Hubble parameter defined as $H \equiv \displaystyle{a^{-1}\,{d a\over d t}}$.

The resulting Einstein field equations stemming from Eq.(\ref{gre0}), in the comoving frame, are;
\bea
 H^2 &=& {8\,\pi G \over 3}\, \rho + {c^2 \, \Lambda\over 3} - {k\,c^2\over R_0^2 a^2},\nn \\
\displaystyle{ {1\over a}\, {d^2\, a \over d\,t^2}} &=& - {4\,\pi G\over c^2}\,\left( {\rho\, c^2 \over 3} + p - 6\,\xi \, H\right) +
{c^2 \, \Lambda\over 3}.
\label{gre1}
\eea
The above equations, Eqs.(\ref{gre1}), can be written as a first order equations for $H$ and $\rho$ as,
\bea
\displaystyle{d H \over d t}  &=& - H^2 -{4\,\pi\, G\over c^2}\,\left({\rho\, c^2 \over 3} + p - 6\,\xi\, H\right) + {c^2\, \Lambda \over 3},\nn\\
\displaystyle{d \rho \over d t} &=& - 3\,H\, \left( \rho + {p- 6\,\xi\,H \over c^2}\right).
\label{gre2}
\eea
It is advantageous to rewrite the cosmological equations  in dimensionless form by introducing dimensionless variables as,
\bea
&&\Ht = {H \over H_{ch}},\;   \rt = {\rho \over \rho_{ch}},\;  \pt = {p \over \rho_{ch}\, c^2},\;
 \Lat = {c^2 \Lambda \over 8 \, \pi G\,\rho_{ch}},\; \xit = {8\,\pi G\,\xi \over \ c^2\, H_{ch}},\; \tt = t\, H_{ch},\; \rt_k = -{k\,c^2\over   8 \, \pi G\,\rho_{ch} R_0^2\, a^2}
\label{gre3}
 \eea
where $\rho_{ch}$ is a some chosen constant characteristic density and the characteristic Hubble parameter $H_{ch}$ is chosen such that
$H_{ch}^2 = 8\,\pi\, G \rho_{ch}$.
Thus, the dimensionless form of Eqs.(\ref{gre2}) would take the form,
 \bea
 {d \Ht \over d\tt} &=& - \Ht^2 - {1\over 6}\,\left[\rt\,+ 3\,\left(\pt - 6\,\xit\,\Ht\right)\right] + {\Lat\over 3},\nn\\
 {d \rt \over d\tt } &=& -3\,\Ht\,\left(\rt + \pt - 6\,\xit\,\Ht \right),
\label{gre4}
 \eea
 while the first equation in Eqs.(\ref{gre1}) would assume the form,
 \be
 \Ht^2 = {1\over 3}\,\left(\rt + \Lat\right) + \rt_k.
 \label{constgr}
 \ee

 Assuming a barotropic equation of state, $\pt = \omega \rt$, then cosmological equations Eqs.(\ref{gre4}) become,
 \bea
 {d \Ht \over d\tt} &=& - \Ht^2 - {1\over 6}\,\rt\,\left(1 + 3\,\omega\right) + 3\,\xit\,\Ht  + {\Lat\over 3},\nn\\
 {d \rt \over d\tt } &=& -3\,\Ht\,\rt \left(1 + \omega\right)  + 18\,\xit\,\Ht^2.
\label{gre5}
 \eea
 Notice that $\omega$ is an equation of state parameter with physically motivated range given by $\omega \in \left[-1,1\right]$. As examples for
 some typical values, we have $\omega = 0$ (dust), $\omega = -1$ (dark energy), $\omega = 1/3$ (radiation), and $\omega = 1$ (stiff fluid).

The equations as given in Eq.(\ref{gre5}) constitute the dynamical system representing the cosmological model with dynamical variables $\rt$ and $\Ht$ that
determine the state of the dynamical system. It is clear that these two dynamical variables are unbounded. Before we start analyzing these cosmological models using dynamical systems techniques let us have a very brief introduction to this subject to present the basic concepts and set our notations.
\section{Basic Theories and Notations for Dynamical System Approach}
The main task of studying dynamical systems is to understand all possible behaviors of a generic solution of a set of $n$ first order differential equations without necessarily solving them. This system of $n$ first order differential equations can be written as
\be
 \dot{x}=f( x),
\ee
where,
 \be
x \equiv \left[x_1(t),.. ,x_n(t)\right]^T,  \hspace{0.2 in}\dot{x}\equiv \left[{dx_1(t)\over dt},.. ,{dx_n(t)\over dt}\right]^T,  \hspace{0.2 in} f( x)\equiv \left[f_1(x_1,..,x_n),.. ,f_n(x_1,..,x_n)\right]^T, \nonumber\\ \ee  and subject to the initial conditions $ x(t=t_0)= x_0$. This system is called {\it Autonomous} if $f( x)$ has no explicit dependence on $t$. For such a system there is a basic existence and uniqueness theorem that guarantees the existence and uniqueness of a solution in some neighborhood of a point $x_0$ as long as $ f( x)$ is differentiable at $x_0$ in its $n$ arguments, see for example \cite{Strogatz}. For example in two dimensional dynamical systems (i.e., $n=2$) by drawing $x_1$ and $x_2$ in a plan one can visualize the evolution of the system starting from some initial point $ x_0=\left[x_1(0),x_2(0)\right]^T$ at $t=0$, and see how it changes with time. This continuous collection of points forms a trajectory or a flow line which describes the evolution of the system up to any latter time. These flow lines never intersect because of the above existence and uniqueness theorem that governs this system.

In the context of our study we are interested in cosmological equations of the form found in Eq.(\ref{gre5}) which can be described by two dimensional dynamical system. Thus for convenience  and notational simplicity we introduce the vector state $x$,  vector parameter  $\a$  and vector function $f$ defined as follows,
\bea
 x \equiv \left[x_1, x_2\right]^T = \left[\Ht, \rt\right]^T, &\a \equiv \left[\omega, \Lat, \xit\right]^T,& f\left(x,\a\right) = \left[f_1\left(x,\a\right), f_2\left(x,\a\right)\right]^T
\label{xpar}
 \eea
 The system of equations given in Eqs.(\ref{gre5}) can be written compactly as,
\bea
\dot{x} &=& f\left(x,\a\right),\;\; \mbox{where}\;\; \dot{x}\equiv {d x \over d\tt}\equiv \left[{dx_1 \over d\tt}, {dx_2 \over d\tt}\right]^T\nn\\
f_1\left(x,\a\right)&=& - x_1^2 - {1\over 6}\,x_2\,\left(1 + 3\,\omega\right) + 3\,\xit\,x_1  + {\Lat\over 3},\nn\\
f_2\left(x,\a\right) &=& -3\,x_1\,x_2 \left(1 + \omega\right)  + 18\,\xit\,x_1^2.
\label{eqco}
\eea
\subsection{Fixed Point Analysis and Classification}
A natural question one might ask is whether these flow lines can go indefinitely to an infinite values of $ x$, or they can end
at some special points or curves? Also, how long it takes to reach either the infinite value of $ x$ or the finite fixed points, do
we need the full analytic or numerical solution to answer these questions or there are quantitative methods one can follow to draw
these important information about the system.

To answer the above questions we need to study the "fixed points" of the system, or the points (or possibly curves) that satisfy
$f( x)=0$. If our system starts exactly at a fixed point it will remain there forever. In fact, they are the equilibrium points of the
dynamical system, which could be stable, unstable or saddle equilibrium points. In order to understand the behavior of the system
around these points, one have to study the behavior of small linear perturbation around the fixed point under consideration to
test the stability of such a point.

\begin{figure}[H]
\centerline{\epsfig{file=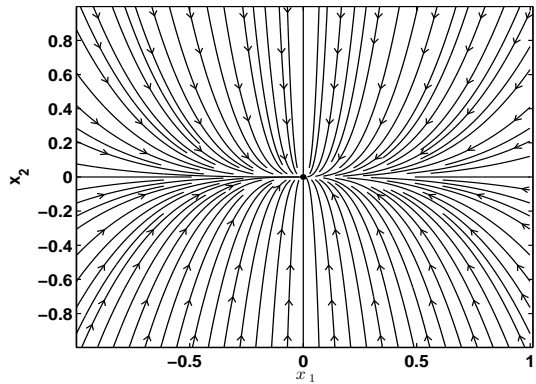,width=7cm,height=5cm}}
\caption{{\footnotesize
  Phase portraits for the system $(\dot{x}_1=-x_1,\;\;\;\;\dot{x}_2=-3\,x_2).$ The dotted circle at the origin represents a fixed point. }}
\label{fig1_simp}
\end{figure}

For any generic planer system, $\dot{x} = f\left(x,\a\right)$ not necessarily the one given in Eq.(\ref{eqco}), the existence of a fixed point is determined through $f\left(x_0,\a\right)=0$ and then the system can be expanded around the fixed point as,
\bea
\dot{x} &=& f\left(x_0,\a\right) + Df\left(x_0,\a\right)\left(x-x_0\right)+
O\left(x-x_0\right)^2,
\label{expan1}
\eea
 where $D f\left(x_0,\a\right)= \left[ \partial f_i\left(x_0,\a\right)\over \partial x_j\right]= J$ is the Jacobian matrix. For fixed points with non-vanishing $\mbox{det}\left(J \right)$, the stability of the planer system can be examined through the eigenvalues of the Jacobian matrix.
Here and  later, the eigenvalues of the Jacobian matrix are denoted by $\l_1$ and $\l_2$, they are conjugate to each other in case of being complex, while their corresponding eigenvectors by $\mathbf{e}_1$ and $\mathbf{e}_2$.
 The stability of the fixed point can be decided according to the following criteria:
\begin{itemize}
 \item Stable node (Sink), if $\l_1$ and $\l_2$ are real negative and attractive center (stable spiral) in case of being complex with negative real parts.
  \item Unstable node (Source), if $\l_1$ and $\l_2$  are real positive and  repulsive center (unstable spiral) in case of being complex with positive real parts.
  \item Saddle point, if   $\l_1$ and $\l_2$  are real and  have opposite sign.
  \item Center, if $\l_1$ and $\l_2$ are purely imaginary.
\end{itemize}
For the sake of illustration, we consider the following system,
\be \dot{x}_1=-x_1,\;\;\;\;\; \dot{x}_2=-3\,x_2.
\ee
This system has a fixed point at $(x_1,\,x_2)\equiv(0,0)$ and from the Jacobian matrix it has $\lambda_1=-1$ and $\lambda_2=-3$,
then it is a stable(sink)  node as can envisaged from Fig.(\ref{fig1_simp}).

We distinguish different types of dynamical systems through their phase portrait which could be topologically different only if the
number or/and the nature of their fixed points are different. If the number and the nature of their fixed points are the same but
in one system they are shifted or displaced compared to the other they are considered equivalent. More generally, if there is a
homeomorphic map (i.e., continuous deformations with continuous inverse) that takes one phase portrait to the other, they are considered
topologically equivalent.

Fixed points with the feature $Re(\lambda_i)\neq 0$ for all $\lambda_i$ are called hyperbolic fixed points. In hyperbolic cases we know
that the local behaviors of flow lines near fixed points are completely governed by the above linearized analysis. Furthermore, there is
an important theorem (due to Hartman and Grobman, see \cite{Strogatz,kuz}) which states that in the neighborhood of these fixed points the system is topologically equivalent to the linearized system, as a result, the nonlinear terms do not affect the system behavior near these points. Another important fact about systems with hyperbolic fixed points is that if we change
the values of the parameters in the system, (i.e., equation of state parameter $w$, cosmological constant $\Lambda$, etc..) the system
will not change its topology and its topology is still captured by the linearized system. If this happens to all the system fixed points we call
it structurally stable.

For cases where one of the two  eigenvalues or both equal to zero, degenerate fixed points (or non-hyberbolic), the stability can't be decided without knowing the nonlinear terms which means the failure of the linear stability theory. Classification of non-hyperbolic fixed points can be found in \cite{coley1}. In fact, theses  non-hyperbolic fixed points are known to form the germs of bifurcation. The term bifurcation will be explained later.

In this work we are going to see that the dynamical system defined above for cosmology contains fixed points with double zero eigenvalues (non-hyperbolic points). These cases have been classified in literature, here we follow Ref.{\cite{kuz} in classifying these planer dynamical systems whose fixed point lies at $\left( x , \a\right) = \left( 0 , 0\right)$ with double zero eigenvalues $\l_{1,2}\left(0\right)=0$. The Jacobian of this system can be brought into the form $J= \left(
\begin{array}{cc}
0 & 1\\
0 & 0
\end{array}
 \right)
$ by introducing new variables $\left(y_1, y_2\right)$ related linearly to $\left(x_1, x_2\right)$.  Then the entire system can be written and organized  as a power series in terms of $\left(y_1, y_2\right)$  as
\bea
\dot{y}_1 &=& y_2 + a_{00}\left(\a\right) + a_{10}\left(\a\right)\,y_1 + a_{01}\left(\a\right)\,y_2 + {1\over 2}\,a_{20}\left(\a\right)\,y_1^2 + a_{11}\left(\a\right)\,y_1\,y_2  + {1\over 2}\,a_{02}\left(\a\right)\,y_2^2 + O\left(y^3\right),\nn\\
\dot{y}_2 &=&  b_{00}\left(\a\right) + b_{10}\left(\a\right)\,y_1 + b_{01}\left(\a\right)\,y_2 + {1\over 2}\,b_{20}\left(\a\right)\,y_1^2 + b_{11}\left(\a\right)\,y_1\,y_2  + {1\over 2}\,b_{02}\left(\a\right)\,y_2^2 + O\left(y^3\right),
\label{expan2}
\eea
where the coefficients $ a_{ij}\left(\a\right)$ and  $ b_{ij}\left(\a\right)$  are smooth functions of $\a$ and satisfying
\be
a_{00}\left(0\right) = a_{10}\left(0\right) = a_{01}\left(0\right) = b_{00}\left(0\right) = b_{10}\left(0\right) = b_{01}\left(0\right) =0.
\ee
The nondegeneracy conditions for the system are the following,
\begin{description}
\item{($BT.0$)} $\; \mbox{the Jacobian matrix} \left[{\partial f_i \over \partial x_j}\right]\left(0,0\right) \neq 0$,
\item{($BT.1$)} $\; a_{20}\left(0\right) + b_{11}\left(0\right) \neq 0 $,
\item{($BT.2$)} $\; b_{20}\left(0\right) \neq 0 $,
\item{($BT.3$)} the map   $\left(x, \a\right) \rightarrow \left[ f\left(x, \a\right), \mbox{tr}\left(\left[{\partial f_i \over \partial x_j}\right]\right),
\mbox{det}\left(\left[{\partial f_i \over \partial x_j}\right]\right)\right] $ is regular at point $\left(x, \a\right) = \left(0, 0\right)$.
\end{description}

In our specific case, one can introduce the linear transformation $\left( y_1 = x_1,\; y_2 = -{1\over 6}\,x_2\right)$ and then Eq.(\ref{eqco}), for constant $\xit$, can be expressed in terms of $y's$  as,
\bea
\dot{y}_1 &=&  {\Lat \over 3} +  3\,\xit\,y_1 +\left(1 + 3\,\o\right) \,y_2  - y_1^2 ,\nn\\
\dot{y}_2 &=&   -3 \,\xit\,y_1^2 -3\,\left(1+\o\right)\,y_1\,y_2.
\label{expany}
\eea
One can notice the absence of $O\left( y^3\right)$ terms and the coefficients $ a_{ij}\left(\a\right)$ and  $ b_{ij}\left(\a\right)$ as  defined in Eq.(\ref{expan2}) assume the following forms,
\bea
&&a_{00}\left(\a\right) = {\Lat \over 3},\;\; a_{10}\left(\a\right) = 3\,\xit,\;\; a_{01}\left(\a\right) =  3\,\o\;\;  a_{20}\left(\a\right) = -2,\;\;
b_{20}\left(\a\right) = -\, 6\, \xit,\nn\\
&& b_{11}\left(\a\right) = -3\,\left(1 + \o\right),\;\; a_{11}\left(\a\right) = a_{02}\left(\a\right) =  b_{00}\left(\a\right) = b_{01}\left(\a\right) = b_{10}\left(\a\right)=0.
\label{coeffab}
\eea
In order to check the nondegeneracy conditions one needs the Jacobian matrix  $\left[{\partial f_i \over \partial x_j}\right]$ corresponding to the system in Eq.(\ref{eqco}) which is easily found to be,
\bea
\left[{\partial f_i \over \partial x_j}\right] &=&
\left(
\begin{array}{cc}
-2\,x_1 + 3\,\xit  & -{1\over 6}\,\left(1 + 3\,\o\right)\\
-3\,x_2\left(1 + \o\right) + 36\,\xit\, x_1 & -3\,x_1\,\left(1 + \o\right)
\end{array}
 \right).
\label{jacobfull}
 \eea
All nondegeneracy conditions are fulfilled except the condition ($BT.2$) where  $b_{20}\left(0\right) = 0$, thus the dynamical system described
in Eq.(\ref{eqco}) is a degenerate Bogdanov-Taken system.
\subsection{Bifurcations and normal forms}
As we have mentioned earlier, the dynamical systems which represents our cosmological models has a vanishing $\mbox{det}\left(J \right)$ at the point $\left(x, \a\right) = \left(0, 0\right)$. In addition, we could have  $\mbox{Re}\left(\lambda_i\right)=0$ for other possible fixed point as will be shown later. Therefore, the nature of theses equilibrium points depends on the behavior of the higher order terms in eqn.(\ref{eqco}) not the linear terms. The analysis of such cases is more interesting because of the existence of these degenerate fixed points, they are the seeds of a very nice phenomena called bifurcation. A bifurcation of a dynamical system happens when a change in a value of one of the system parameters produces a topologically nonequivalent phase portrait, i.e., changes the number or the nature of the system fixed points.

For illustrating the concept of bifurcation, let us consider the following two-dimensional system
\be
\label{PF}
\dot{x}_1=\mu\, x_1-x_1^3,\;\;\;\; \dot{x}_2=-x_2.
\ee
For $\mu<0$ this system has a fixed point at $x_0=\left(0,0\right)$, which is a stable node as one can check. The same fixed point survives the limit $\mu\rightarrow 0$, therefore, it is still there, but as $\mu$ becomes positive the system suddenly has two extra fixed points, $x_{\pm}=(\pm\sqrt{\mu},0)$ which are stable and the $ x_0$ one becomes unstable. This is known as pitchfork bifurcation in which fixed points exchange their nature as a parameter changes sign.
\begin{figure}[H]
\centerline{\epsfig{file=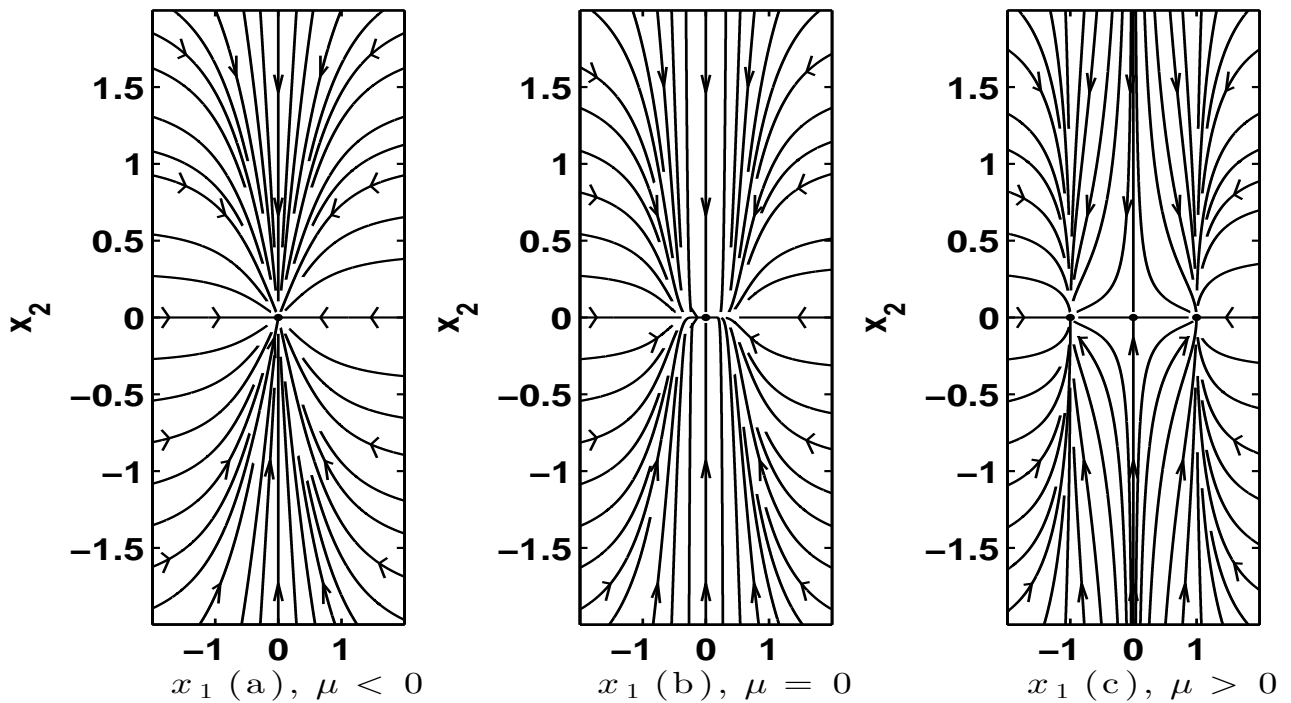,width=20cm,height=10cm}}
\caption{{\footnotesize
  Phase portraits for the system $(\dot{x}_1=\mu\,x_1 - x_1^3,\;\;\;\;\dot{x}_2=-x_2)$ revealing the pitchfork bifurcation behaviour. The dotted circles at the origin and $(\pm\sqrt{\mu},0)$ represents fixed points. }}
\label{fig_pitch}
\end{figure}

In certain sense our previous example of pitchfork bifurcation contains representative nonlinear terms (for all systems undergo this bifurcation), since if we go close enough to the fixed point and Taylor expand $ f( x)$ around it the leading nonlinear terms obtained are the terms in the example. These terms control the local behaviors of trajectories around the fixed points. They capture topologically different behaviors that might arise upon changing the values of the parameters, therefore, one might ask is it possible to classify all possible bifurcations and their nonlinear terms. In fact, most local bifurcations in two-dimensional systems with one system parameter (i.e., codimension-1) are classified into four known classes, for each class of bifurcation we write its nonlinear terms in a standard simple form which is called the normal form. The list of four classes of bifurcations are
\be
\left.
\begin{array}{llll}
\mbox{Saddle node:} & \dot{x}_1 =\mu\, \pm x_1^2, & \dot{x}_2=-x_2,\\
\mbox{Transcritical:} & \dot{x}_1=\mu\, x_1\pm x_1^2, &  \dot{x}_2=-x_2,\\
\mbox{Pitchfork:}      & \dot{x}_1 =\mu\, x_1 \pm x_1^3,&  \dot{x}_2=-x_2,\\
\mbox{Andronov-Hopf:}& \dot{x}_1=\mu\, x_1- x_2 + x_1 \,(x_1^2+x_2^2),& \dot{x}_2= x_1+\mu\, x_2 + x_2\, (x_1^2+x_2^2).
\end{array}
\right\}
\label{bifur_types}
\ee

As we increase the number of independent parameters and the number of dynamical variables we get more complicated classifications and new types of bifurcations. For example there is no Andronov-Hopf bifurcation in one-dimensional systems it starts to appear only in two-dimensions. Another example is Bogdanov-Taken bifurcation which appears only in two-dimensional systems with at least two system parameters. This latter bifurcation is a combination of saddle node, Andrnov-Hopf and Homoclinic bifurcations. In this work we are going to show that FRW cosmological equations with cosmological constant and bulk viscosity can be brought to a codimension-3 degenerate Bogdanov-Taken normal form. In the following subsection we are going to show the procedure of calculating normal forms for a generic dynamical system.
\subsection{Normal Forms and Simplifications}
Here we introduce the normal form technique which enables us to simplify the equations describing the dynamical system. In this subsection we follow closely the notation found in \cite{wigg}. In order to understand what we mean by a simplification, it is important to separate the equations describing the  dynamical system into linear and nonlinear parts as,
\bea
\dot{x} &=& J\,x + F\left(x\right),
\label{nonsim1}
\eea
where $J$, which determines the linear part of the system, is simplified into one of the Jordon canonical forms. As to the nonlinear part, it is organized as,
\bea
\dot{x} &=& J\,x + F_2\left(x\right) + F_3\left(x\right) + \cdots + F_{r-1}\left(x\right) + O\left(x^r\right),
\label{nonsim2}
\eea
where $F_i\left(x\right)$ means terms of order $x^i$. Starting with simplifying the second order term by introducing the nonlinear transformation,
\bea
x &=& y + h_2\left(y\right),
\label{ts}
\eea
where $h_2\left(y\right)$ is of order $y^2$, when applied to Eq.(\ref{nonsim1}) leads to,
\bea
\dot{y} &=& J\,y + J\,h_2\left(y\right) - Dh_2\left(y\right)\, \dot{y} + F\big(y + h_2\left(y\right)\big) \Rightarrow\nn\\
 \dot{y} &=&\bigg( id +  Dh_2\left(y\right)\bigg)^{-1}\,\bigg( J\,y + J\,h_2\left(y\right) + F\big(y + h_2\left(y\right)\big)\bigg) \Rightarrow\nn\\
\dot{y} &=&\bigg( id -  Dh_2\left(y\right) + O\left(y^2\right)\bigg)\,\bigg( J\,y + J\,h_2\left(y\right) + F\big(y + h_2\left(y\right)\big)\bigg) \Rightarrow\nn\\
\dot{y} &=& J\, y +  J\,h_2\left(y\right) + F\big(y + h_2\left(y\right)\big)  - Dh_2\left(y\right) J y + \cdots .
\label{ysim1}
\eea
Keeping terms up to second order amounts to,
\bea
\dot{y} &=& J\, y +  J\,h_2\left(y\right)  - Dh_2\left(y\right) J y  + F_2\left(y \right).
\label{ysim2}
\eea
To eliminate  the second order term, one need to impose
\be
 Dh_2\left(y\right) J y - J\,h_2\left(y\right)  = F_2\left(y \right).
\label{ysim3}
 \ee
To be more concrete we introduce $H_2$, the space of  homogenous two column polynomials of degree 2, and the map $L_J^{(2)}$ acting on $H_2$ defined as,
\bea
L_J^{(2)} : H_2 &\rightarrow & H_2,\nn \\
L_J^{(2)}\big(h_2\left(y\right)\big) &=& - Dh_2\left(y\right) J y
+ J\,h_2\left(y\right),\;\;\;h_2\left(y\right) \in H_2 .
\label{mapdef}
\eea
Using the map $L_J^{(2)}$ the space $H_2$ can be nonuniquely decomposed, direct sum composition, as
\bea
H_2 &=& L_J^{(2)}\left(H_2\right) \oplus G_2,
\label{dirsum2}
\eea
where $G_2$ represents the space complementary to $L_J^{(2)}\left(H_2\right)$.
Thus the simplification takes place by eliminating  $F_2$, if it is in the range of $L_J^{(2)}$, through choosing a suitable $h_2\left(y\right)$ leaving terms belonging to $G_2$.

Applying the technique of the normal form to the case of interest where $J$ and $H_2$ are respectively  given as,
\bea
J & =&
\left(
\begin{array}{ll}
0 & \a \\
0 & 0
\end{array}
\right), \;\;\; \a \neq 0,
\label{defj}
\eea
and
\bea
H_2 &=& \mbox{Span}\left\{
\left(
\begin{array}{c}
x_1^2\\
0
\end{array}
\right),
\left(
\begin{array}{c}
x_1\,x_2\\
0
\end{array}
\right),
\left(
\begin{array}{c}
x_2^2\\
0
\end{array}
\right),
\left(
\begin{array}{c}
0\\
x_1^2
\end{array}
\right),
\left(
\begin{array}{c}
0\\
x_1\, x_2
\end{array}
\right),
\left(
\begin{array}{c}
0\\
x_2^2
\end{array}
\right)
\right\},
\label{H2}
\eea
the parameter $\a$ is kept without normalization for the sake of clarity and simplicity. The resulting $L_J^{(2)}\left(H_2\right)$ according
to the map in Eq.(\ref{mapdef}) is found to be
\bea
L_J^{(2)}\left(H_2\right) &=&
\mbox{Span}\left\{
\left(
\begin{array}{c}
x_1\, x_2\\
0
\end{array}
\right),
\left(
\begin{array}{c}
x_2^2\\
0
\end{array}
\right),
\left(
\begin{array}{c}
x_1^2\\
-2\,x_1\,x_2
\end{array}
\right),
\left(
\begin{array}{c}
x_1\, x_2\\
- x_2^2
\end{array}
\right)
\right\}.
\label{Ljh2}
\eea
The construction of $G_2$ is a little bit more involved as we have to find the orthogonal complement of $L_J^{(2)}\left(H_2\right)$. The determining properties are;
\bea
\forall\;  V \in G_2\; \mbox{and}\; \forall\;  X \in H_2\;\; \langle V\left|\right.L_J^{(2)}\, X \rangle =  \langle V\, L_J^{(2)}\left|\right.\, X \rangle =0,
\eea
where the bracket $\langle \cdots\left|\right. \cdots\rangle$ indicates the  Euclidean inner product. The vanishing of $\langle V\, L_J^{(2)}\left|\right.\, X \rangle$ for any $X\, \in  H_2$ leads to the vanishing of $\langle V\, L_J^{(2)}\left|\right.$  which when written in a matrix form becomes
 $ L_J^{(2)\,T} \, V=0$, where $T$ indicates the transpose of the matrix. Thus $V$ are just right zero eigenvectors of $L_J^{(2)\,T}$. The easier way to
 get $V$ is to construct a $6\times 6$  matrix representation for $L_J^{(2)}$  where considering the vector space corresponding to $H_2$ as
\bea
 \left(
\begin{array}{c}
x_1^2\\0
\end{array}
\right)
&\equiv&
\left(
\begin{array}{cccccc}
1 &0 &0&0 &0 & 0
\end{array}
\right)^T, \;\;
\left(
\begin{array}{c}
x_1\,x_2\\0
\end{array}
\right)
\equiv
\left(
\begin{array}{cccccc}
0 &1 &0&0 &0 & 0
\end{array}
\right)^T, \nn \\
\left(
\begin{array}{c}
x_2^2\\0
\end{array}
\right)
&\equiv &
\left(
\begin{array}{cccccc}
0 &0 &1&0 &0 & 0
\end{array}
\right)^T, \;\;
\left(
\begin{array}{c}
0\\x_1^2
\end{array}
\right)
\equiv
\left(
\begin{array}{cccccc}
0 &0 &0& 1 &0 & 0
\end{array}
\right)^T, \nn \\
\left(
\begin{array}{c}
0\\x_1\, x_2
\end{array}
\right)
&\equiv &
\left(
\begin{array}{cccccc}
0 &0 &0& 0 &1 & 0
\end{array}
\right)^T, \;\;
\left(
\begin{array}{c}
0\\x_2^2
\end{array}
\right)
\equiv
\left(
\begin{array}{cccccc}
0 &0 &0& 0 &0 & 1
\end{array}
\right)^T.
\label{6vec}
\eea
The resulting matrix representation of $L_J^{(2)}$ is found to be,
\bea
L_J^{(2)} &=&
\left(
\begin{array}{cccccc}
0 & 0 & 0 & \a & 0 & 0\\
- 2 \a  & 0 & 0 & 0 & \a & 0\\
0  & - \a & 0 & 0 & 0 & \a\\
0  & 0 & 0 & 0 & 0 & 0\\
0  & 0 & 0 & -2 \a & 0 & 0\\
0  & 0 & 0 & 0 & - \a & 0
\end{array}
\right),
\label{LJ26by6}
\eea
and the resulting zero eigen-space for $L_J^{(2) T}$ and hence $G_2$ are found to be spanned by
\bea
  G_2 &=& \mbox{Span}\,\left\{
\left(
\begin{array}{c}
x_1^2\\ {1\over 2}\;x_1\,x_2
\end{array}
\right)
\equiv
\left(
\begin{array}{cccccc}
1 &0 &0&0 &{1\over 2} & 0
\end{array}
\right)^T, \;\;
\left(
\begin{array}{c}
0\\ x_1^2
\end{array}
\right)
\equiv
\left(
\begin{array}{cccccc}
0 &0 &0& 1 &0 & 0
\end{array}
\right)^T
\right\}.
\label{G2span}
\eea
It is clear that $L_J^{(2)}\left(H_2\right)$ and $G_2$, as given respectively in Eq.(\ref{Ljh2}) and Eq.(\ref{G2span}),  are orthogonal but this
is not necessary in direct sum composition introduced in Eq.(\ref{dirsum2}). One can combines $\left( x_1^2,\, -2\,x_1\,x_2\right)^T$ from $L_J^{(2)}\left(H_2\right)$ with elements in $G_2$, found in Eq.(\ref{G2span}), to find additional two realization for $G_2$. Last, the two-dimensional dynamical systems characterized by  $J$, in Eq.(\ref{defj}), in their simplest possible form containing quadratic terms are,
\bea
G_2 = \left\{
\left(
\begin{array}{c}
x_1^2\\ {1\over 2}\;x_1\,x_2
\end{array}
\right),\;
\left(
\begin{array}{c}
0\\ x_1^2
\end{array}
\right)
\right\}
& \Rightarrow &
\left.
\begin{array}{lll}
\dot{y}_1 & = & \a\, y_2 + a \, y_1^2\\
\dot{y}_2 & = & \displaystyle{{a\over 2}}\,y_1\,y_2 + b \, y_1^2
\end{array}
\right],\nn\\
G_2 = \left\{
\left(
\begin{array}{c}
x_1^2\\ 0
\end{array}
\right),\;
\left(
\begin{array}{c}
0\\ x_1^2
\end{array}
\right)
\right\}
& \Rightarrow &
\left.
\begin{array}{lll}
\dot{y}_1 & = & \a\, y_2 + a \, y_1^2\\
\dot{y}_2 & = &  b \, y_1^2
\end{array}
\right],\nn\\
G_2 = \left\{
\left(
\begin{array}{c}
0\\ x_1\,x_2
\end{array}
\right),\;
\left(
\begin{array}{c}
0\\ x_1^2
\end{array}
\right)
\right\}
& \Rightarrow &
\left.
\begin{array}{lll}
\dot{y}_1 & = & \a\, y_2 \\
\dot{y}_2 & = & a\, y_1\,y_2 + b \, y_1^2
\end{array}
\right],
\label{possG2}
\eea
where $a$ and $b$ are two independent constants.

The processes of simplification using normal forms can be continued to the terms of $O\left(y^3\right)$ and that is the maximum we need in our present
work. All procedures followed previously  for simplifying second order terms can be straight forwardly applied to third order terms. The dynamical system, after simplifying second order terms, is
\bea
\dot{y} & =& J\, y + F_2^r\left(y\right) + \tilde{F}_3\left(y\right) + \cdots ,
\eea
 where  $F_2^r\left(y\right)$ are the simplified $O\left(y^2\right)$ terms while $\tilde{F}_3\left(y\right)$ are the $O\left(y^3\right)$ terms in their unsimplified forms. The simplification of $\tilde{F}_3\left(y\right)$ terms  is achieved by making the following transformation,
 \be
 y \Rightarrow y + h_3\left(y\right),
\label{ty3}
 \ee
where for the notational simplicity we use the same name for $y$ for new and old variables describing the dynamical system. The resulting necessary condition to simplify $O\left(y^3\right)$ terms is,
\be
Dh_3\left(y\right) J y - J\,h_3\left(y\right)  = \tilde{F}_3\left(y \right).
\label{ysim4}
\ee
One can define analogous to $L_J^{(2)}$, Eq.(\ref{defj}), the corresponding $L_J^{(3)}$ which acts on the space of two columns homogeneous polynomials of
degree 3 denoted by $H_3$.
\bea
L_J^{(3)} : H_3 &\rightarrow & H_3,\nn \\
L_J^{(3)}\big(h_3\left(y\right)\big) &=& - Dh_3\left(y\right) J y
+ J\,h_3\left(y\right),\;\;\;h_3\left(y\right) \in H_3 .
\label{mapdef3}
\eea
The composition of $H_3$ as a direct sum of $L_J^{(3)}\left(H_3\right)$ and $G_3$ can be worked out for $J$, Eq.(\ref{defj}), to yield
\bea
L_J^{(3)}\left(H_3\right) &=&
\mbox{Span}\left\{
\left(
\begin{array}{c}
y_1^2\, y_2\\
0
\end{array}
\right),
\left(
\begin{array}{c}
y_1\,y_2^2\\
0
\end{array}
\right),
\left(
\begin{array}{c}
y_2^3\\
0
\end{array}
\right),
\left(
\begin{array}{c}
y_1^3\\
-3\,y_1^2\,y_2
\end{array}
\right),
\left(
\begin{array}{c}
y_1^2\, y_2\\
-2\,y_1\,y_2^2
\end{array}
\right),
\left(
\begin{array}{c}
y_1\, y_2^2\\
- y_2^3
\end{array}
\right)
\right\},
\label{Ljh3}
\eea
while $G_3$ which is orthogonal to $L_J^{(3)}\left(H_3\right)$ is found to be,
\bea
G_3 &=&
\mbox{Span}\left\{
\left(
\begin{array}{c}
3\,y_1^3\\
y_1^2\, y_2
\end{array}
\right),
\left(
\begin{array}{c}
0\\
y_1^3
\end{array}
\right)
\right\}.
\label{poss1G3}
\eea
As we know that $G_3$ is not necessarily to be orthogonal to $L_J^{(3)}\left(H_3\right)$ so we can combine $\left(y_1^3, - 3\, y_1^2 y_2\right)^T$ from $L_J^{(3)}\left(H_3\right)$
with $G_3$ to get other two alternatives for $G_3$ which are namely,
\bea
G_3 =
\mbox{Span}\left\{
\left(
\begin{array}{c}
y_1^3\\
0
\end{array}
\right),
\left(
\begin{array}{c}
0\\
y_1^3
\end{array}
\right)
\right\}
&\mbox{OR}&
G_3 =
\mbox{Span}\left\{
\left(
\begin{array}{c}
0\\
y_1^2\,y_2
\end{array}
\right),
\left(
\begin{array}{c}
0\\
y_1^3
\end{array}
\right)
\right\}.
\label{poss23G3}
\eea
\subsection{Behavior at Infinity and Poincar\'{e} Sphere}
As we mentioned earlier, it is quite useful to draw the phase portrait of the system which includes all possible solution curves in the $(x_1,x_2)$ plane and gives a clear visual representation of the solutions behavior for various initial conditions. As a matter of fact, this visual representation is limited to a finite domain in the $(x_1, x_2)$ planer phase space. Thus, one should seek an alternative visual representation that provides a global picture of the solution curves' behavior, specially at infinity. This global picture can be achieved by introducing the so-called Poincar\'{e} sphere \cite{perko,wigg} where one
projects from the center of the unit sphere $S^2=\left\{ \left(X,Y,Z\right) \in R^3\, | \, X^2 + Y^2 + Z^2 =1 \right\}$ onto the $(x_1,x_2)$-plane tangent to $S^2$ at either north or
south pole as shown in Fig.(\ref{figpoin}). Projecting the upper hemisphere of $S^2$ onto the $(x_1,x_2)$-plane, then one can derive the following relations
between $(x_1, x_2)$ and $(X, Y, Z)$,
\bea
X = \displaystyle{{x_1\over \sqrt{1 + x_1^2 + x_2^2}}}, & Y = \displaystyle{{x_2\over \sqrt{1 + x_1^2 + x_2^2}}}, & Z = \displaystyle{{1\over \sqrt{1 + x_1^2 + x_2^2}}},\nn\\
x_1 = \displaystyle{{X\over Z}}, & x_2 = \displaystyle{{Y \over Z}}.&
\label{XYZpoin}
\eea
\begin{figure}[hbtp]
\centerline{\epsfig{file=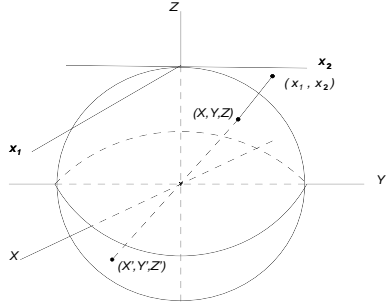,width=5cm,height=4cm}}
\caption{{\footnotesize
Central projection of the upper hemisphere of $S^2$ (Poincar\'{e}  sphere) onto the $(x_1, x_2)$ plane  }}
\label{figpoin}
\end{figure}
These clearly define a one-to-one correspondence between points $(X, Y, Z)$ on the upper hemisphere of $S^2$ with $Z>0$ and points $(x_1, x_2)$ in the plane. The origin $(0,0)$ in the $(x_1, x_2)$-plane corresponds to the north pole $(0, 0, 1)\,\in\, S^2$; The circle $x_1^2 + x_2^2 = a^2$ on the  $(x_1, x_2)$-plane corresponds to points on the circle $X^2 + Y^2 = \displaystyle{{a^2\over a^2 + 1}}$, $Z=\displaystyle{{1\over \sqrt{1 + a^2}}}$ on $S^2$; The circle at infinity of $(x_1, x_2)$-plane corresponds to the equator of $S^2$. The whole orbits induced by the dynamics described by Eqs.(\ref{eqco}) can be mapped onto the upper hemisphere  of the Poincar\'{e} sphere which is difficult to draw. In contrast,  the orthogonal projection of the upper hemisphere  of the Poincar\'{e} sphere on the unit disk in the $(X, Y)$ plane is much easier to draw and still captures all of the information about the behavior at infinity. Such a kind of flow on the unit disk , $X^2 + Y^2 < 1$, when drawn is called a global (or compact) phase portrait. It is possible to obtain the dynamical system in terms of $(X, Y)$ that corresponds to the dynamical system given in Eqs.(\ref{eqco}) and after simple algebra one can get,
\bea
\dot{X} &=& Z\,f_1\left({X\over Z}, {Y\over Z}, \a\right)- Z\,X\left[X\,f_1\left({X\over Z}, {Y\over Z}, \a \right) + Y\,f_2\left({X\over Z}, {Y\over Z}, \a \right)\right],\nn\\
\dot{Y} &=& Z\,f_2\left({X\over Z}, {Y\over Z}, \a\right)- Z\,Y\left[X\,f_1\left({X\over Z}, {Y\over Z}, \a\right) + Y\,f_2\left({X\over Z}, {Y\over Z}, \a \right)\right],\nn\\
Z &=& \sqrt{1 - X^2 - Y^2}.
\label{dynXY}
\eea
The determination of the fixed points, at infinity, is rather involved if one works in terms of the coordinates $(X, Y, Z)$. Fortunately, there is a simpler approach where one can introduce plane polar coordinates $(r, \th)$ where $x_1 = r\,\cos{\th}$ and $x_2 = r\,\sin{\th}$ and the dynamical system represented by Eqs.(\ref{eqco}) takes the following form,
\bea
\dot{r} &=& \cos{\th}\, f_1\left(r\,\cos{\th}, r\,\sin{\th}, \a\right) +
    \sin{\th}\, f_2\left(r\,\cos{\th}, r\,\sin{\th}, \a\right),\nn\\
 \dot{\th} &=& {1\over r}\,\left[\cos{\th}\, f_2\left(r\,\cos{\th}, r\,\sin{\th}, \a\right) -
    \sin{\th}\, f_1\left(r\,\cos{\th}, r\,\sin{\th}, \a\right)\right].
\label{dynrth}
\eea
Assuming $f_1$ and $f_2$ are multinomial in $x_1$ and $x_2$ and organized as,
\bea
f_1\left(x_1, x_2, \a \right) &=& f_1^{1}\left(x, y, \a\right) + \cdots + f_1^{\mbox{m}}\left(x, y, \a\right),\nn\\
f_2\left(x_1, x_2, \a\right) &=& f_2^{1}\left(x, y, \a\right) + \cdots + f_2^{\mbox{m}}\left(x, y, \a\right),
\label{exan}
\eea
where the integer superscripts, in $f's$,  indicate the power of the associated multinomial and $m$ is the maximum power in the expansion.
Then as $r\rightarrow \infty$ the evolution of $\th$ is dominated by terms of maximum power $f_{1,2}^{\mbox{m}}\left(x, y, \a\right)$\footnote{ Here we assume that the maximum power in $f_1$ and $f_2$ are the same for simplicity, but if they are different then the largest one would control the behaviour at infinity and the same analysis applies}, contained in the expansion of $f_{1,2}\left(x, y, \a\right)$, leading to
\bea
 \dot{\th} &\approx& {1\over r}\,\left[\cos{\th}\, f_2^{\mbox{m}}\left(r\,\cos{\th}, r\,\sin{\th}, \a\right) -
    \sin{\th}\, f_1^{\mbox{m}}\left(r\,\cos{\th}, r\,\sin{\th}, \a\right)\right],
\label{dynth1}
\eea
 Furthermore, one can factor $r$ from Eq.(\ref{dynth1}) since it doesn't affect the sign of $\dot{\th}$ to get,
\bea
 \dot{\th} &\sim & G^{\mbox{m}+1}\left(\th\right) =  \cos{\th}\, f_2^{\mbox{m}}\left(\cos{\th}, \sin{\th}, \a\right) -
    \sin{\th}\, f_1^{\mbox{m}}\left(\cos{\th}, \sin{\th}, \a\right).
\label{dynth2}
\eea
The function $G^{\mbox{m}+1}\left(\th\right)$ having only total powers of $\left(\mbox{m}+1\right)$ in $\sin{\th}$ and $\cos{\th}$ and thus $G^{\mbox{m}+1}\left(\th + \pi\right)=
\pm\; G^{\mbox{m}+1}\left(\th\right)$ for odd and even $\mbox{m}$ respectively.
The zeros of $G^{\mbox{m}+1}\left(\th\right)$ determine the fixed points at infinity and now it is evident if $\th_j$ is a zero of $G^{\mbox{m}+1}\left(\th\right)$ then so $\th_j + \pi$. For more details
about Poincar\'{e} Sphere and capturing the behavior at infinity one can consult \cite{perko,wigg}.
\section{Analysis of Universe Filled with Perfect Fluid}
It is tempting to apply the dynamical system theory to the system of Eqs.(\ref{eqco}) in its full generality, but it might be better to first study
special cases in order to get some insight into the dynamical system represented by these equations. The first simple case is to set cosmological constant and viscosity coefficient to zero $\a=\left( \o, \Lat =0, \xit =0\right)^T$. Thus, the resulting equations are , $(x_1 = \Ht, x_2 = \rt)$,
 \bea
 \dot{x}_1 &=& - x_1^2 - {1\over 6}\,x_2\,\left(1 + 3\,\omega\right),\nn\\
 \dot{x}_2 &=& -3\,x_1\,x_2 \left(1 + \omega\right).
\label{greo}
 \eea
 Unless $\omega \neq -1$ nor $\omega \neq -{1\over 3}$, the system has only one finite fixed point at $x=x_0=\left( 0, 0\right)$. Then the  Jacobian matrix evaluated at the fixed point turns out to be,
\bea
\left[ \partial f_i\left(x_0,\a_0\right)\over \partial x_j\right] &=&
\left(
\begin{array}{cc}
0 & -{1\over 6}\,\left(1 +3\,\omega\right)\\
0 & 0
\end{array}
 \right).
\label{jaco}
\eea
This clearly shows that the  Jacobian matrix has a double zero eigen values, $\l_1=\l_2=0$, while the corresponding generalized eigenvectors are  determined to be $\mathbf{e}_1=\left(1, 0\right)^T$ and $\mathbf{e}_2=\left(0, 1\right)^T$. Such a kind of system, where there are two zero eigenvalues, is termed as a Bogdanov-Taken system. The stability of such a system  can't be decided according to the linear stability theory.

Now let us turn to the $\o = -1$ case, where we find an infinite number of fixed points along the curve, $x_2 = 3\, x_1^2$, and the resulting Jacobian is,
 \bea
\left[ \partial f_i\left(x_0,\a_0\right)\over \partial x_j\right] &=&
\left(
\begin{array}{cc}
-2\,x_1 & {1\over 3}\\
0 & 0
\end{array}
 \right),
\label{jacom1}
\eea
where $x_0=\left( x_1, 3\,x_1^2\right)^T$ and $\a_0=\left( \o=-1, \Lat =0, \xit =0\right)^T$. The eigenvalues resulting from this Jacobian are $\l_1=-2\,x_1$ and $\l_2= 0$ while their corresponding eigenvectors are respectively $\mathbf{e}_1=\left(1, 0\right)^T$ and $\mathbf{e}_2=\left(1, 6\,x_1\right)^T$. The direction $\mathbf{e}_1$ is a stable when  $(x_1>0)$ and unstable for $(x_1<0)$.  The other direction $\mathbf{e}_2$ is along the tangent of the parabola curve $(x_2 = 3\, x_1^2)$ where all points along the parabola are fixed points.

The last remaining special case is that $\o = -{1\over 3}$, where we find an infinite number of fixed points, this time, along the $x_2$ axis and leading to the following Jacobian,
 \bea
\left[ \partial f_i\left(x_0,\a_0\right)\over \partial x_j\right] &=&
\left(
\begin{array}{cc}
0 & 0\\
-2\,x_2 & 0
\end{array}
 \right),
\label{jacom13}
\eea
where $x_0=\left( 0, x_2\right)^T$ and $\a_0=\left( \o=-{1\over 3}, \Lat =0, \xit =0\right)^T$. The  Jacobian matrix has a double zero eigenvalues, $\l_1=\l_2=0$, while the corresponding generalized eigenvectors are  determined to be $\mathbf{e}_1=\left(1, 0\right)^T$ and $\mathbf{e}_2=\left(0, 1\right)^T$. Once again, the occurrence of the double zero eigenvalues makes the stability analysis not possible according to the linear stability theory.

The fixed points  at infinity and as explained in Section 3.4 can be determined by the zeros of the  function $G^{\mbox{m}+1}\left(\th\right)$, defined in Eq.(\ref{dynth2}), which for  Eqs.(\ref{greo}) amounts to
\be
 G^{\mbox{m}+1}\left(\th\right) \stackrel{\mbox{m}=2}{{\scalebox{3}[1]{=}}} G^{3}\left(\th\right) =   - \cos^2{\th}\, \sin{\th}\, \left(2 + 3\,\omega\right).
\label{infomeg}
 \ee
 For $\omega = -{2\over 3}$, all points at the circle of infinity are fixed points otherwise there are finite number of fixed points corresponding to
 $\th =\left\{0, {\pi\over 2}, \pi, {3\,\pi\over 2}\right\}$. Considering the flow only along the circle at infinity and  provided that
$\left(2 + 3\,\omega\right) > 0$, the points $(\th =0)$ and $(\th =\pi)$
 can be shown to be respectively stable and unstable while the points $(\th = {\pi\over 2})$ and $(\th = {3\,\pi\over 2})$ are found to behave
 as saddle but of non-hyperbolic type since ${d G^{3}\left(\th\right)\over d \th}$ is vanishing  at $\th = {\pi\over 2}$ or $ {3\,\pi\over 2}$ .
Having $\left(2 + 3\,\omega\right) < 0$, all directions of flow are reversed on the circle at infinity leading to switching fixed point from stable
to unstable and vice versa. The saddle points keep their type unchanged.

As to the normal forms, the system in Eqs.(\ref{greo}) when compared to the form in Eq.(\ref{nonsim1}), $J$ has the form  of Eq.(\ref{defj}) with $\a = -{1\over 6}\left(1 + 3\,\o \right)$, $F(x)$ turns out to be
\bea
F(x) &=&
\left(
\begin{array}{c}
-x_1^2\\
-3\,x_1\,x_2\,\left(1 +\o\right)
\end{array}
\right)
= -2\,(1 +{3\over 5}\,\o)
\left(
\begin{array}{c}
x_1^2\\
{1\over 2}\,x_1\,x_2\,
\end{array}
\right)
 +
 (1 +{6\over 5}\,\o)
\left(
\begin{array}{c}
x_1^2\\
-2\,x_1\,x_2\,
\end{array}
\right).
\eea
It is evident that $F(x)$ contains two pieces the first one belongs to $G_2$, see Eq.(\ref{Ljh2}), while the second one to $L_J^{(2)}$ ,see Eq.(\ref{Ljh2}). Thus the piece belonging to $L_J^{(2)}$ can be shown to be canceled by the following transformation,
\bea
x_1 = y_1, && x_2 = y_2 + {6\left(5 + 6 \,\o\,\right)\over 5\left(1 + 3\, \o\right)} y_1^2.
\eea
The resulting equations in terms of $y_i$s variables turn out to be,
\bea
\dot{y}_1 & = & -{1\over 6}\left(1 + 3\,\o \right)\, y_2 -  \left(2 +{6\over 5}\,\o\right)\,y_1^2,\nn \\
\dot{y}_2 & = &  -\left(1 +{3\over 5}\,\o\right)\, y_1\,y_2 - {6\over 25} {\left(5 +6\,\o\right)\,\left(3\,\o - 5\right) \over
\left(1 + 3\,\o\right)}\, y_1^3.
\label{normomg1}
\eea
One can get another alternative normal form as
\bea
\dot{y}_1 & = & y_2 ,\nn \\
\dot{y}_2 & = &  -\left(5 +3\,\o\right)\, y_1\,y_2 - 3\,\left(1 + \o\right)\,y_1^3,
\label{normomg2}
\eea
which  can be achieved  by the following transformation,
\bea
x_1 = y_1, && x_2 = -{6\over \left(1+ 3 \,\o\,\right)}\,\left( y_2 + y_1^2 \right).
\eea
As is clear the reduction to normal forms produces terms of $O\left(y^3\right)$ which, in our case, belongs to $G_3$ (see. Eqs.(\ref{poss1G3}--\ref{poss23G3})) and thus can't be further simplified. The two normal forms, in Eqs.(\ref{normomg1}--\ref{normomg2}), are normal form for a degenerate Bogdanov-Takens bifurcation when condition (BT.2) is violated. The case corresponding to $\o = -{1\over 3}$ needs a careful treatment, since matrix $J$ equals to zero when  $x_0=\left( 0, 0\right)^T$ and $\a_0=\left( \o=-{1\over 3}, \Lat =0, \xit =0\right)^T$ and thus $G_2 = H_2$. Having $G_2 = H_2$, which means any quadratic term can't be simplified. Upon deciding to choose $x_0=\left( 0, x_2\right)^T$ and $\a_0=\left( \o=-{1\over 3}, \Lat =0, \xit =0\right)^T$ where $x_2 \neq 0$ we get
$J$ in the form found in Eq.(\ref{jacom13}) for which we can apply the same analysis carried out for the $J$ defined in Eq.(\ref{defj}).

As the fixed point analysis shows critical behavior occurs at $\o =
\{-1,-{2\over 3}, -{1\over 3}\}$ as revealed  by the presence of
infinite number of fixed points. In a more detailed terms, all
points along the curve $\rt = 3\,\Ht^2$ are fixed points for $\o =
-1$, while all points on the $\rt$ axis are fixed points  for $\o
=-{1\over 3}$ and finally all points at the circle at infinity are
fixed points for $\o =-{2\over 3}$. Other values for $\o$ has a one
fixed point at the origin besides four fixed points on the circle at
infinity. To sum up, the parameter space $\o$ can be divided into
four regions namely $] -\infty, -1[$ , $]-1 , -{2\over 3}[$ ,
$]-{2\over 3} , -{1\over 3}[$ and $]-{1\over 3} , \infty[$ where the
phase portraits are qualitatively the same within each region but
critical behaviors occurs  at $\o = \{-1,-{2\over 3}, -{1\over 3}\}$
revealed by changing  the number of fixed points to become infinite
at these values for $\o$. All these features are presented  in the
phase portraits (noncompact and compact) displayed in
Fig.(\ref{figomg1}) and Fig.(\ref{figomg2}) for seven representative
cases.

In more physical terms, the fixed points for a finite domain in this model have important features that can be summarized as,
\begin{itemize}
\item $w\neq -1$  and  $w\neq -1/3$ case: we have only one fixed point , $ x=(0,0)$, which is an empty Minkowski space.
\item $w=-1$ case: we have a whole curve of fixed points satisfying $\rt=3\,\Ht^2$, which is a collection of de Sitter points a part from the origin.
\item $w=-1/3$ case: we have a whole line of fixed points, which is the $\rt$-axis, or $ x=(0,\rt)$, which is a collection of Einstein Static universe a part from the origin.
\end{itemize}

Another important feature of this model is that the $\Ht$-axis is a solution by itself which is a Milne universe. More precisely, it consists of two solutions, one interpolate in the region $\Ht\geq 0$ and the other is its mirror image. This feature can be easily observed in the phase diagrams as depicted in Fig.(\ref{figomg1}) and Fig.(\ref{figomg2}). These solutions prevents any trajectory from crossing the $\Ht$-axis which disjoints the $\rt>0$ and $\rt<0$ regions.  In fact the presence of that particular solution, i.e. Milne universe, serves as a phantom divide separating zone $ ( \rt + \pt =0)$, which can never be crossed.

The above case of a perfect fluid contains a collection of
interesting cosmologies that includes different types of bounce
cosmologies including nonsingular ones. For example, in the cases presented
in Fig.\ref{figomg1}(A,a), if we started with an
expanding universe at some point in time i.e., $\Ht>0$ (where,
$\rt>0$) the Hubble rate will keep decreasing till it vanishes,
then becomes negative describing a collapsing universe. This case
has a maximum scale factor $a_{\mbox{\tiny max}}$ and a minimum density $\rt$, in
addition, the whole evolution occurs in a finite time since it does
not contains any fixed points. Furthermore, the values for only
$\rt>0$ is bounded from below but not bounded from above. But the
most interesting cases are presented in Fig.\ref{figomg1}(C,c),
Fig.\ref{figomg1}(D,d) and Fig.\ref{figomg2}(A,a) which describe centers with infinite periods which are also cosmological bounces. In this case the values of
$\Ht$ and $\rt>0$ are bounded from below and from above. One expects these solutions to be geodesically complete.\clearpage
\begin{figure}[btp]
\centerline{\epsfig{file=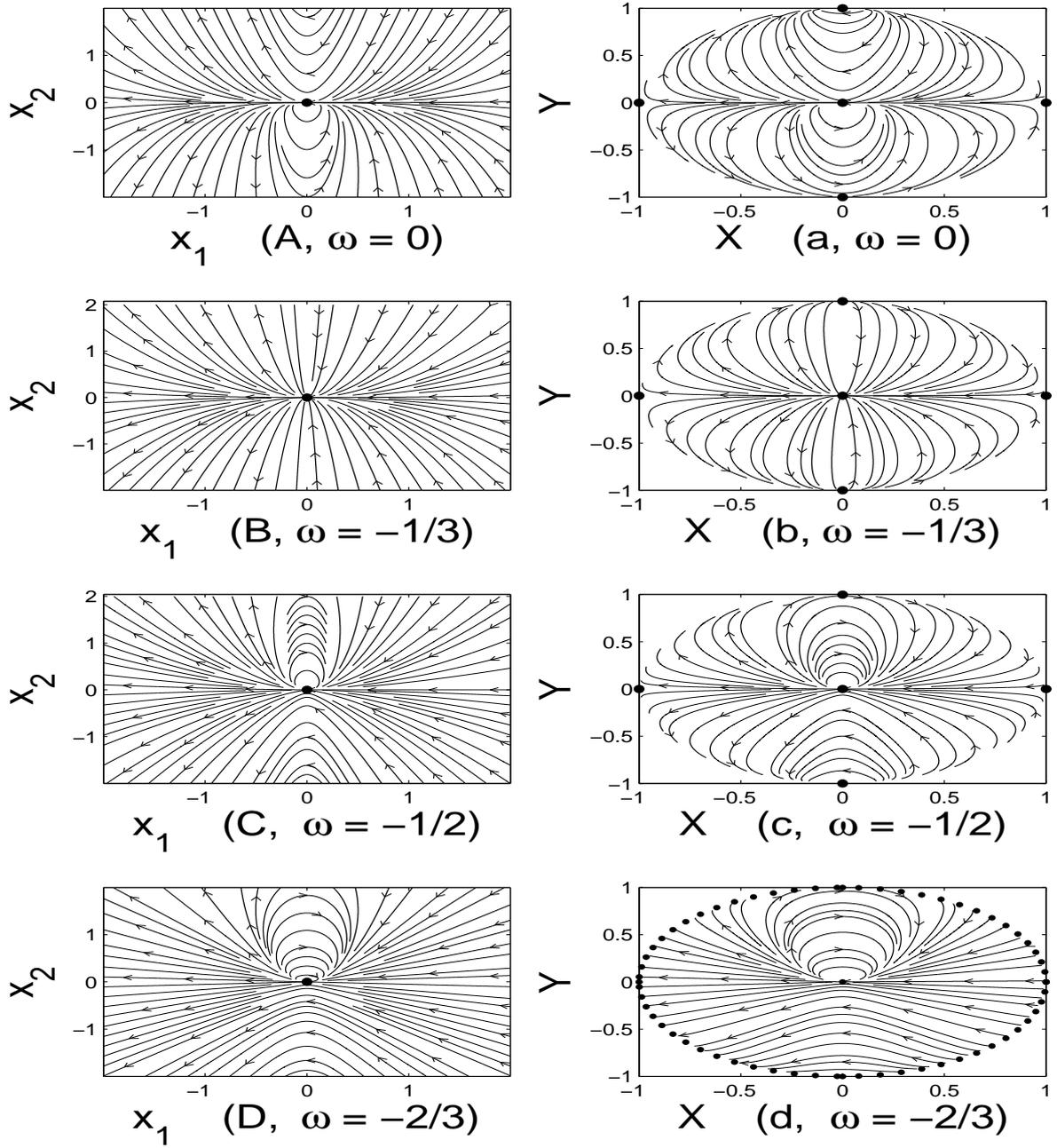,width=18cm,height=20cm}}
\caption{{\footnotesize
Uncompact (left panel) and compact (right panel)  phase portraits for the cases $\o = 0, -{1\over 3}, -{1\over 2}$ and $-{2\over 3}$. $x_1$ and $x_2$ respectively
denote the dimensionless $\Ht$ and $\rt$ as defined in Eq.(\ref{gre3}). $X$ and $Y$ are the coordinates on the Poincar\'{e}
sphere as defined in Eq.(\ref{XYZpoin}). The dotted circles represent fixed points. }}
\label{figomg1}
\end{figure}
\begin{figure}[btp]
\centerline{\epsfig{file=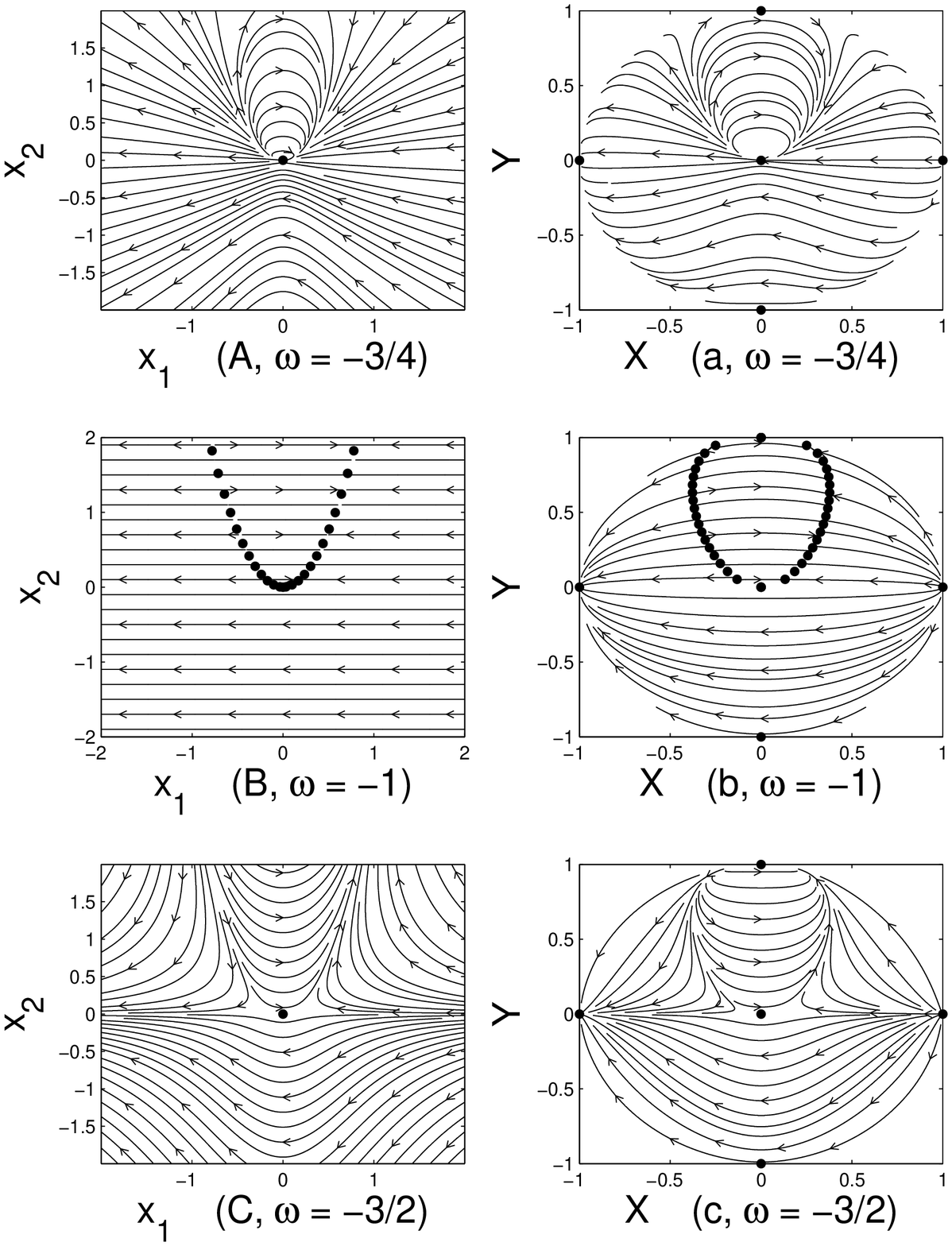,width=18cm,height=20cm}}
\caption{{\footnotesize Uncompact (left panel) and compact (right panel) phase portraits for the cases $\o = -{3\over 4}, -1$ and $-{3\over 2}$. $x_1$ and $x_2$ respectively
denote the dimensionless $\Ht$ and $\rt$ as defined in Eq.(\ref{gre3}). $X$ and $Y$ are the coordinates on the Poincar\'{e}
sphere as defined in Eq.(\ref{XYZpoin}). The dotted circles represent fixed points.
}}
\label{figomg2}
\end{figure}
\clearpage

\section{Analysis of Universe Filled with Perfect Fluid in the Presence of Cosmological Constant}
The second simple case is to ignore  viscosity in Eq.(\ref{eqco}) and thus the dynamical system reduces  to, $(x_1 = \Ht, x_2 = \rt)$,
\bea
\dot{x}_1 &=& - x_1^2 - {1\over 6}\,x_2\,\left(1 + 3\,\omega\right)  + {\Lat\over 3},\nn\\
\dot{x}_2  &=& -3\,x_1\,x_2\, \left(1 + \omega\right).
\label{eqomglam}
\eea
The fixed points are determined to be three fixed points. The first two fixed points together with their Jacobians are,
\bea
\left( x_1 = \pm\, \sqrt{{\Lat\over 3}},\; x_2 =0\right), && \left[{\partial f_i \over \partial x_j}\right]_{\left( x_1 = \pm\, \sqrt{{\Lat\over 3}},\; x_2 =0\right)} =
\left(
\begin{array}{cc}
\mp\, 2\,\sqrt{{\Lat\over 3}} & -{1\over 6}\,\left(1 + 3\,\o\right)\\
0 & \mp\, \sqrt{3\,\Lat}\,\left(1 + \o\right)
\end{array}
 \right).
\label{twofixI}
\eea
The reality of fixed points necessitates that $\Lat \ge 0$ and hence the real eigenvalues for the Jacobian in Eq.(\ref{twofixI}) together with their corresponding eigenvectors are,
\bea
\l_1 = -2\,\sqrt{\Lat \over 3},\; \l_2 = -\sqrt{3\,\Lat}\,\left(1 + \o\right), &&
\mathbf{e}_1 = \left( 1,\; 0\right)^T,\;  \mathbf{e}_2 = \left( {1\over 2\,\sqrt{3\,\Lat}},\; 1\right)^T,\, (+),\nn\\
\l_1 = +2\,\sqrt{\Lat \over 3},\; \l_2 = +\sqrt{3\,\Lat}\,\left(1 + \o\right), && \mathbf{e}_1 = \left( 1,\; 0\right)^T,\;
\mathbf{e}_2 = \left(- {1\over 2\,\sqrt{3\,\Lat}},\; 1\right)^T,\,(-),
\label{twolamI}
\eea
where the sign $(\pm)$ indicates to fixed points having $x_1 = \pm\, \sqrt{{\Lat\over 3}}$. The types of fixed points
are controlled by $\o$ as follows; the fixed point $\left( x_1 = +\,
\sqrt{{\Lat\over 3}},\; x_2 =0\right)$  is a stable  (sink) one for $\o > -1
$ and a saddle otherwise while the fixed point $\left( x_1 = -\,
\sqrt{{\Lat\over 3}},\; x_2 =0\right)$  is a unstable (source) one for $\o >
-1 $ and a saddle otherwise. Here a typical behaviour of saddle-node bifurcation is observed, where for $\Lat <0$ there is no fixed point  but at $\Lat = 0$  a single fixed point
appears  at the origin and then for $\Lat > 0$ two fixed point appear along the $\Ht (\mbox{or}\;x_1)$ axis. The stability of the two appearing fixed points depends on
the value of $\o$ as just discussed previously. This finding concerning the  saddle-node bifurcation can be conveniently depicted in the following diagram,
Fig.(\ref{figblam1}), consisting of two parts depending on the value of $\o$.
\begin{figure}[hbtp]
\centerline{\epsfig{file=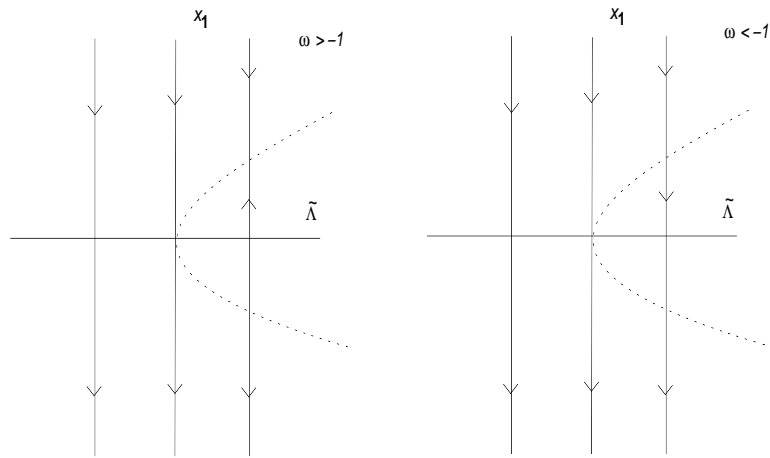,width=10cm,height=6cm}}
\caption{{\footnotesize
Saddle-node bifurcation diagram where the dashed curve, $x_1^2 = {\Lat\over 3}$, determining the fixed points along the $x_1$ axis. The arrows represent the
flow along the $x_1$ axis.}}
\label{figblam1}
\end{figure}

The third fixed point together with  its Jacobian matrix are,
\bea
 \left( x_1 = 0,\; x_2 ={2\,\Lat \over 1 + 3\,\o}\right), && \left[{\partial f_i \over \partial x_j}\right]_{\left( x_1 = 0,\; x_2 ={2\,\Lat \over 1 + 3\,\o}\right)} =
\left(
\begin{array}{cc}
0 & -{1\over 6}\,\left(1 + 3\,\o\right)\\
{-6\,\Lat\,\left(1 + \o\right) \over 1 + 3\,\o } & 0
\end{array}
 \right).
\label{onefixI}
\eea
The reality of this fixed point is ensured for all real values of $\Lat$ and $\o$ while the reality is not guaranteed  for the eigenvalues of the associated  Jacobian matrix. For
this case the eigenvalues together with their eigenvectors are,
\bea
\l_{1} = - \sqrt{\Lat\,\left(1 + \o\right)} ,\; \l_{2} = \sqrt{\Lat\,\left(1 + \o\right)}&&
\mathbf{e}_{1} = \left( {\left(1 + 3\,\o\right)\over 6\,\sqrt{\Lat\,\left(1 + \o\right)}},\; 1\right)^T,\; \mathbf{e}_{2} =
\left( -{\left(1 + 3\,\o\right)\over 6\,\sqrt{\Lat\,\left(1 + \o\right)}},\; 1\right)^T.\hspace{1cm}
\label{onefixlamI}
\eea
The fixed point is of a saddle type for $\Lat\,\left( 1 + \o\right) > 0$ while of a center type for $\Lat\,\left( 1 + \o\right) < 0$. This persistent fixed point along the $x_2$  axis
changes its type from saddle to center according the sign of  $\Lat\,\left( 1 + \o\right) $ and this is also a typical behavior of bifurcation called degenerate
Hopf bifurcation. The bifurcation behavior can be neatly and conveniently depicted in the following diagram consisting of three parts depending on the value of $\o$.
\begin{figure}[hbtp]
\centerline{\epsfig{file=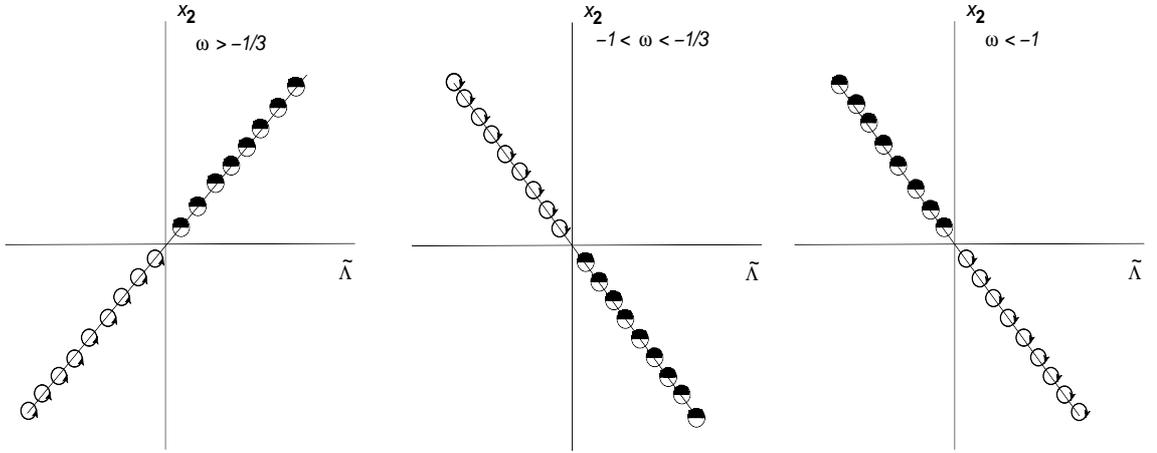,width=15cm,height=6cm}}
\caption{{\footnotesize
The degenerate Hopf bifurcation diagram  for all the three possible regions of $\o$. the solid curve, $x_2 = {2\,\Lat\over 1 + 3\,\o}$, determining the fixed points along the $x_2$ axis
as a function of $\Lat$ for a fixed value of $\o$ in the range specified. The half-filled circle and arrowed circle  represent a saddle and  a center respectively.}}
\label{figbhopf}
\end{figure}

It is worthy to stress that the flow depicted by bifurcation diagrams in Fig.~(\ref{figblam1}) is restricted to the flow along the $x_1$ axis while the proper
flow should be inferred from a kind of graphs as provided in Fig.~(\ref{figomglam1}) and Fig.~(\ref{figomglam2}) where the true flow is a two dimensional one.
Needless to mention that the flow depicted in Fig.~(\ref{figbhopf}) should be viewed in the proper context of two dimensional flow in the $x_1-x_2$
plane where a fixed point as a center along the $x_2$ can
have a meaning. In fact, this kind of reduction is intended for simplification and more clarification otherwise one should work in a plane describing the
parameter space for $\o$ and $\Lat$ divided into regions according to the behavior of the emerging  fixed points.
One should not take this kind of reduction too literally and keep in mind that the whole picture that these emerging fixed point whatever saddle, stable,
unstable and center are coexisting together as shown in various figures like Fig.~(\ref{figomglam1}) and
Fig.~(\ref{figomglam2}). This  kind of reduction proves to be more useful and convenient when viscosity is included where the parameter
space would be a three dimensional one leading  to a difficulty in visualization.
Another remark, in both bifurcations diagrams in Fig.~(\ref{figblam1}) and Fig.~(\ref{figbhopf}),
the nature of the fixed point when $\Lat = 0$, namely the origin except at $\o=-1$ where there an infinite number of fixed points,
should be inferred from the graphs in Fig.~(\ref{figomg1}) and Fig.~(\ref{figomg2}).

A careful treatment is required for the special case where $\o = -1$ which leads to,
\bea
\dot{x}_1 &=& - x_1^2 + {1\over 3}\,x_2  + {\Lat\over 3},\nn\\
\dot{x}_2  &=& 0.
\label{eqomg_m1_lam}
\eea
There is a family of fixed points determined by the relation $x_{1(0)}^2 = {1\over 3}\,\left(x_{2(0)} + \Lat\right)$. The reality of these fixed points
is ensured by requiring $\left(x_{2(0)} + \Lat\right) \ge 0$. The fixed points and their associated Jacobian matrices are,
\bea
 \left( x_1 = \pm\,\sqrt{{x_{2(0)} + \Lat\over 3}},\; x_2 = x_{2(0)}\right), && \left[{\partial f_i \over \partial x_j}\right]_{\mbox{fixed points}}=
\left(
\begin{array}{cc}
\mp\, 2\,\sqrt{{x_{2(0)} + \Lat\over 3}}  & {1\over 3}\\
0 & 0
\end{array}
 \right).
\label{twofixsp}
\eea
The eigenvalues for the Jacobian in Eq.(\ref{twofixsp}) together with their corresponding eigenvectors turn out to be,
\bea
 \l_1 = 0,\;\; \l_2 = -\, 2\,\sqrt{{x_{2(0)} + \Lat\over 3}}, && \mathbf{e}_1 = \left( \,{1\over 2}\,\left(3\,\left(x_{2(0)} +
 \Lat\right)\right)^{-1/2},\; 1\right)^T,\;\;  \mathbf{e}_2 = \left(1 ,\; 0\right)^T,(+)\nn\\
 \l_1 = 0,\;\; \l_2 = +\, 2\,\sqrt{{x_{2(0)} + \Lat\over 3}}, && \mathbf{e}_1 = \left( -\,{1\over 2}\,\left(3\,\left(x_{2(0)} +
 \Lat\right)\right)^{-1/2},\; 1\right)^T,\;\;  \mathbf{e}_2 = \left(1 ,\; 0\right)^T, (-),
\label{twolamsp}
\eea
where the sign $(\pm)$ denotes fixed points having $x_1 =  \pm\,\sqrt{{x_{2(0)} + \Lat\over 3}}$. The direction $\mathbf{e}_2$ is a stable when  $(x_1>0)$ and
unstable for $(x_1<0)$.  The other direction $\mathbf{e}_1$ is along the tangent of the parabola curve
$x_{1(0)}^2 = {1\over 3}\,\left(x_{2(0)} + \Lat\right)$, where all points along the parabola are fixed points.
As expected, we see here the presence of cosmological constant doesn't prohibit the occurrence of infinitely fixed points for $\o = -1$ since it is equivalent to introducing
cosmological constant. The behavior would be the same as for $\o = -1$  in the absence of cosmological constant and the sole effect is shifting vertically the flat
curve solution upward or downward depending on the sign of $\Lat$.

The other special case for $\o =-{1\over 3}$ also requires a careful treatment and here is the equations governing this case as obtained from Eqs.(\ref{eqomglam}) after substituting  $\o =-{1\over 3}$,
 \bea
\dot{x}_1 &=& - x_1^2   + {\Lat\over 3},\nn\\
\dot{x}_2  &=& -2\,x_1\,x_2.
\label{eqomg_mthird_lam}
\eea
There are only two fixed points that are given as $\left( x_1 = \pm\,\sqrt{{\Lat\over 3}},\; x_2 =0\right)$ as opposed to case, in the absence of cosmological constant, where there an infinite number
of fixed points along the $\rt$ axis.  Thus, the issue of the presence of an  infinite number of fixed points is cured
for that case of $\o = -{1\over 3}$ after including cosmological constant.

In order to get real fixed points one should impose $\Lat \ge 0$. The fixed points and their associated Jacobian matrices are,
\bea
\left( x_1 = \pm\,\sqrt{{\Lat\over 3}},\; x_2 =0\right), && \left[{\partial f_i \over \partial x_j}\right]_{\left( x_1 = \pm\,\sqrt{{\Lat\over 3}},\; x_2 =0\right)}=
\left(
\begin{array}{cc}
\mp\, 2\,\sqrt{{\Lat\over 3}}  & 0\\
0 & \mp\, 2\,\sqrt{{\Lat\over 3}}
\end{array}
 \right).
\label{twofix_lam_omthird}
\eea
As is clear the system has degenerate eigenvalues $\mp\, 2 \sqrt{{\Lat\over 3}}$ and their corresponding eigenvectors are
$\mathbf{e}_1 = \left(1 ,\; 0\right)^T$ and $\mathbf{e}_2 = \left(0 ,\; 1\right)^T$. The fixed point $\left( x_1 = +\,\sqrt{{\Lat\over 3}},\; x_2 =0\right)$ is of
a stable (sink) type while the other   $\left( x_1 = -\,\sqrt{{\Lat\over 3}},\; x_2 =0\right)$ is unstable (source) one. Furthermore, the system here at $\o= -{1\over 3}$ is not of Bogdanov-Taken type
since the Jacobian is proportional to the identity.

In Figs.(\ref{figomglam1}) and (\ref{figomglam2}), all possible behavior are illustrated in the presence of cosmological constant. Fig.(\ref{figomglam1}) (A,a,B,b) represents the cases
for $\o = -{1\over 3}$ with respectively positive and  negative cosmological constant. As evident from the figure, in the finite domain, there are only two fixed
points along the $x_1$ axis for positive $\Lat$ while none for the negative one. The fixed points at infinity are the same as in the case without including cosmological
constant. Regarding to Figs.(\ref{figomglam1})(C,c,D,d) where $\Lat$ is positive and assuming ${1\over 2}$ and  $1$ but the combination $\Lat\,\left(1 + \o \right)$ flips sign as positive
for $\o = 0$ and negative for $\o = -{3\over 2}$. In these cases, there are three fixed points, namely, two along the $x_1$ axis $\left( x_1 = \pm\,\sqrt{{\Lat\over 3}},\; x_2 =0\right)$
and the third one along $x_2$ axis $\left( x_1 = 0,\; x_2 ={2\,\Lat \over 1 + 3\,\o}\right)$. For $\Lat\,\left(1 + \o \right)>0$.  The stability of the two fixed points along the $x_1$
axis are, the right one is stable (sink) while the left one is unstable (source). In contrast, for $\Lat\,\left(1 + \o \right) < 0$, the two fixed points along the $x_1$ axis are of saddle type.
Now, the third fixed point along $x_2$ axis, it is a saddle for $\Lat\,\left(1 + \o \right)>0$ and a center otherwise. The rest of figures in Fig.(\ref{figomglam2})(A,a,B,b,C,c) confirms
the analytical analysis revealing that when $\Lat < 0$ and $\o \neq -{1\over 3}$, there is no fixed points along the $x_1$ axis but only one point along the $x_2$ axis
being a saddle for  $\Lat\,\left(1 + \o \right)>0$ and a center for $\Lat\,\left(1 + \o \right) < 0$.
\clearpage
\begin{figure}[tbp]
\centerline{\epsfig{file=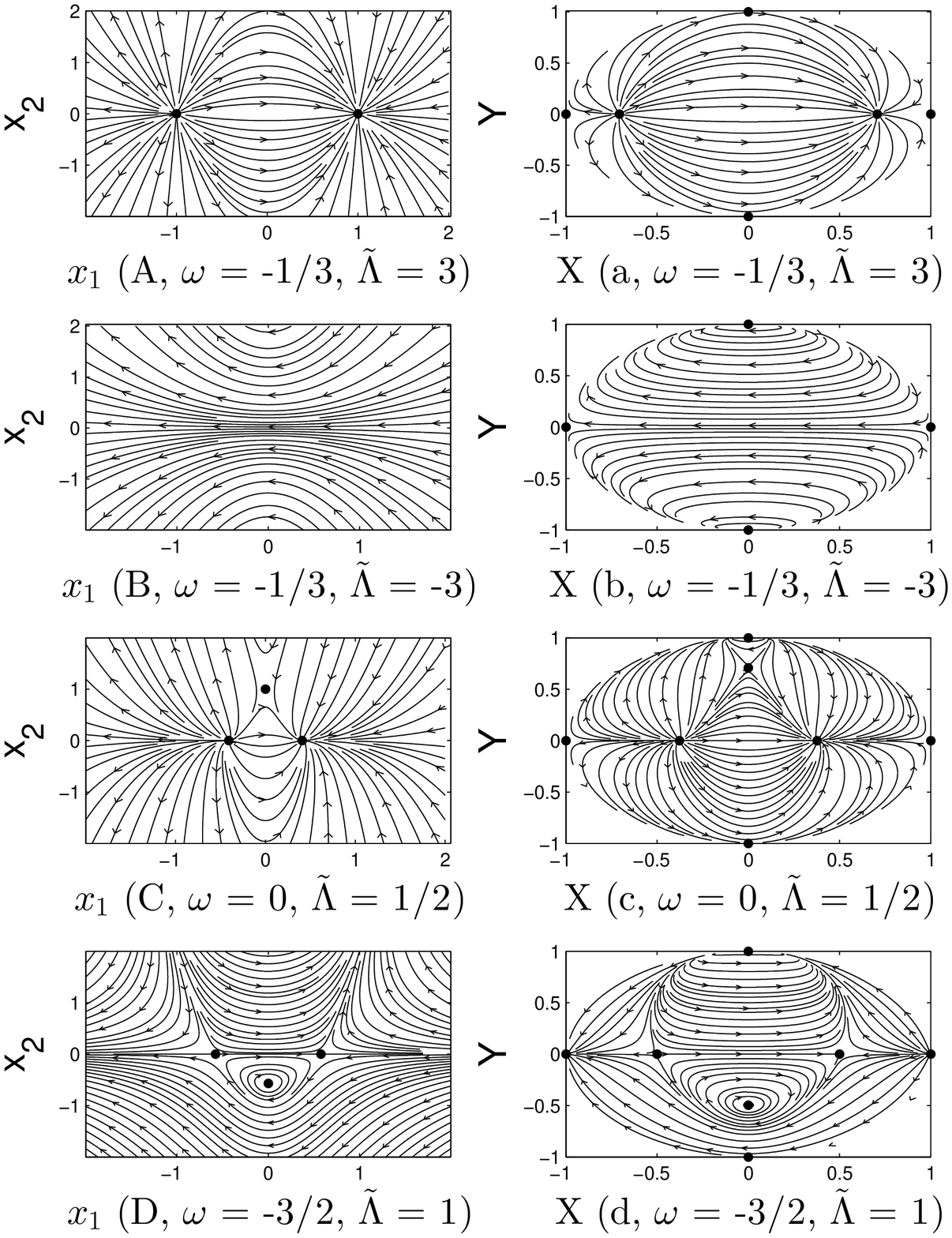,width=18cm,height=20cm}}
\caption{{\footnotesize
Uncompact (left panel) and compact (right panel) phase portraits  when cosmological constant is included. Representative cases are  $\left(\o = -{1\over 3}, \Lat = \pm 3\right)$,  $\left(\o = 0, \Lat = {1\over 2}\right)$ and $\left(\o = -{3\over 2}, \Lat = 1\right)$. $x_1$ and $x_2$ respectively
denote the dimensionless $\Ht$ and $\rt$ as defined in Eq.(\ref{gre3}). $X$ and $Y$ are the coordinates on the Poincar\'{e}
sphere as defined in Eq.(\ref{XYZpoin}). The dotted circles represent fixed points. }}
\label{figomglam1}
\end{figure}
\begin{figure}[tbp]
\centerline{\epsfig{file=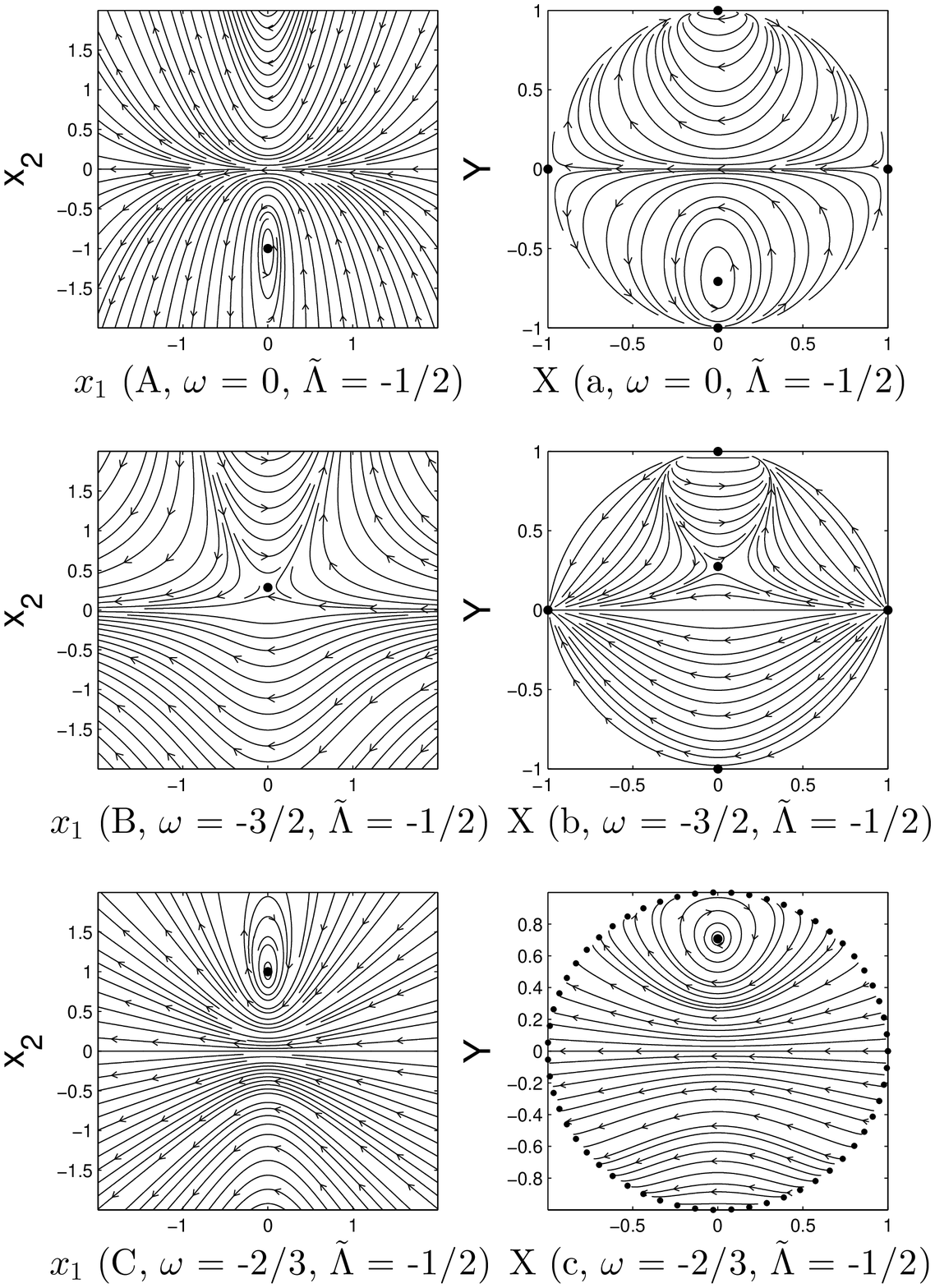,width=18cm,height=20cm}}
\caption{{\footnotesize Uncompact (left panel) and compact (right panel) phase portraits  when cosmological constant is included. Representative cases are
 $\left(\o = 0, \Lat = -{1\over 2}\right)$, $\left(\o = -{3\over 2}, \Lat = -{1\over 2}\right)$ and  $\left(\o = -{2\over 3}, \Lat = -{1\over 2}\right)$. $x_1$ and $x_2$ respectively
denote the dimensionless $\Ht$ and $\rt$ as defined in Eq.(\ref{gre3}). $X$ and $Y$ are the coordinates on the Poincar\'{e}
sphere as defined in Eq.(\ref{XYZpoin}). The dotted circles represent fixed points.
}}
\label{figomglam2}
\end{figure}
\clearpage
Remarks concerning the fixed points at infinity and the normal forms
are in order. First, we find the same fixed points as the case
without including the cosmological constant and the fixed points are
determined by the same function found in Eq.(\ref{infomeg}). This is
can be easily understood since the introduction of the cosmological
constant adds only zero order terms and thus doesn't affect the
behavior at infinity compared to the other present higher order
ones. All figures in Figs.(\ref{figomglam1}) and Figs.(\ref{figomglam2}) for the compact phase
portraits confirm this finding concerning the fixed points at infinity. A particular emphasis on
the case, $\o = -{2\over 3}$ is needed where the circle at infinity in its totality are fixed points as clear from
Fig.~\ref{figomglam2}(c).

Second, as to the normal form one can use the following
transformation, \bea x_1 = y_1, && x_2 = -{6\over \left(1+ 3
\,\o\,\right)}\,\left( y_2 + y_1^2 - {\Lat\over 3} \right),
\label{tnorm_lam_om} \eea then the system in Eq.(\ref{eqomglam})
will reduces to, \bea
\dot{y}_1 & = & y_2 ,\nn \\
\dot{y}_2 & = & \Lat\,\left(1 + \o \right)\,y_1  -\left(5 +3\,\o\right)\, y_1\,y_2 - 3\,\left(1 + \o\right)\,y_1^3.
\label{norm_lam_omg2}
\eea
The normal form corresponding to the case where $\o = -{1\over 3}$ needs a careful treatment since the transformation in Eq.(\ref{tnorm_lam_om}) is singular.
Introducing the variables $z_1 = x_1 - \sqrt{\Lat\over 3}$ and $z_2 = x_2$ then  Eq.(\ref{eqomg_mthird_lam})  would transform into,
\bea
\dot{z}_1 &=& -2\,\sqrt{\Lat \over 3}\,z_1 -  z_1^2 ,\nn\\
\dot{z}_2  &=& -2\,\sqrt{\Lat \over 3}\,z_2 -2\,z_1\,z_2.
\label{eqomg_mthird_lamz}
\eea
In this new form  described  by Eq.(\ref{eqomg_mthird_lamz}), the Jacobian, J, is clearly proportional to the identity and thus $L_J^{(2)}\left(H_2\right) = H_2$ which enables
us to remove any quadratic terms. Removal of quadratic terms is not for free but at the expense of introducing higher order terms. As an example one can try the following
transformation that has a validity not at the whole region of the coordinates but at small neighborhood around the origin whose size is depending on $\Lat$,
\be
\begin{array}{l}
\begin{array}{lll}
 y_1 = z_1-{1\over 2}\,\sqrt{3\over \Lat}\,z_1^2, &&  y_2 = z_2 - \sqrt{3\over \Lat}\, z_1\,z_2, \;\;  \mbox{(Transformation)},
 \end{array}
\\
\begin{array}{ll}
\left.
 \begin{array}{lll}
z_1 = \sqrt{\Lat \over 3} - F &=& y_1 + {1\over 2}\,\sqrt{3\over \Lat}\,y_1^2 + \cdots,\\
z_2 = \sqrt{{\Lat\over 3}}\,\left(y_2/F\right) &=&  y_2 + \sqrt{3\over \Lat}\,y_1\,y_2 + \cdots,
\end{array}
\right]
& { \rm ( Inverse Transformation),}
\end{array}
\end{array}
\label{tnorm_lam_ommthirdex}
\ee
where $F = \sqrt{{\Lat\over 3} - 2\,\sqrt{{\Lat\over 3}}\;y_1}$ . The above
transformation when applied to Eq.(\ref{eqomg_mthird_lamz}) results
in the following, \bea
\dot{y}_1 &=& -2\,\sqrt{\Lat \over 3}\,y_1 - \sqrt{3 \over \Lat}\, y_1^3 + \cdots ,\nn\\
\dot{y}_2  &=& -2\,\sqrt{\Lat \over 3}\,y_2 -3\,\sqrt{3 \over \Lat}\,y_1^2\,y_2 + \cdots.
\label{eqomg_mthird_lamy}
\eea
The dots in Eq.(\ref{tnorm_lam_ommthirdex}) and Eq.(\ref{eqomg_mthird_lamy}) indicates the neglected higher order terms.
It is important to stress that there are two extreme cases for the Jacobian where it is zero or proportional to the identity. In both cases the simplification
introduced through normal forms losses its appealing and the reason behind is detailed as follows; For $J=0$ we have $L_J^{(2)}\left(H_2\right) = 0$  implying that any
 $F_2$  (second order terms) can't be transformed away, while for $J$ proportional to the identity we have $L_J^{(2)}\left(H_2\right) = H_2$
which means that we can remove any second order terms but at the expense of introducing other higher order terms as obtained in Eq.(\ref{eqomg_mthird_lamy}).

The case of a perfect fluid with cosmological constant contains new
interesting features in addition to bounce cosmologies, which is the
appearance of a pair of fixed points along the $\Ht$-axis. This pair
admits new type of cosmological models in which the universe is
interpolating between two fixed points one in the negative $\Ht$
region and another in the positive $\Ht$ region. As presented in Fig.\ref{figomglam1}(A,a,C,c), the universe could start with a
fixed point along the negative $\Ht$-axis and end up with another
fixed point along the positive $\Ht$-axis passing through a bounce,
i.e., $\Ht=0$ point. Another new feature here is the existence of
oscillating cosmological solutions as shown in
Fig.\ref{figomglam2}(C,c). In this interesting case for positive $\rt$, all solutions are either bounces
or oscillating cosmologies with finite evolution time and a minimum
density $\rt$ in the case of bounce or minimum and maximum values
for both $\Ht$ and $\rt$ in the case of oscillating cosmologies.

The physical attributes of the fixed points at a finite domain can be summed up as,
\begin{itemize}
\item $w\neq -1$  and  $w\neq -1/3$ case: \\
a) We have a fixed point, $x=(0,{2 \Lat \over 1+3w})$, which is non-expanding universe $\displaystyle{ \Lat \, \left(1 + \o\right)\over 1+3w}={k\,c^2 \over 8\,\pi \, G\, \r_{\mbox{ch}}\,R_0^2\,a^2}$, which is Einstein Static universe if $\Lat {1+w \over 1+3w}>0$, with ${R} \times S^3$ topology, or a static universe with ${R} \times H^3$\footnote{$H^3$ is a hyperbolic three dimensional space}, if $\Lat {1+w \over 1+3w}<0$.\\
b) We have two fixed points, $ x=\left(\pm \sqrt{{\Lat\over 3}},0\right)$, which are de Sitter universes. There is a region which is filled with trajectories interpolating between these two point (one is a stable node and the other is unstable node). These trajectories are nonsingular and geodesically complete since they start from $t=-\infty$ and end at $t=+\infty$.
\item $w=-1$ case: We have a whole curve of fixed points, $\rt=3\,\Ht^2-\Lat$, which is a collection of de Sitter points.
\item $w=-1/3$ case: We have two de Sitter fixed points, $ x=\left(\pm \sqrt{{\Lat\over 3}},0\right)$, which allow for nonsingular solutions to interpolate between them.
\end{itemize}

Finally, it is worthy to mention  that the $\Ht$-axis is a collection of three solutions, for $\Lat >0$,  which describe cosmological evolution  governed by $\displaystyle {d^2\,a\over dt^2}={c^2\,a\,\Lambda \over 3}$.  One solution starts from Milne universe at $\Ht=\infty$ to a de Sitter universes at the fixed point, $ x=\left(+ \sqrt{{\Lat\over 3}},0\right)$. A second solution interpolates between the two de Sitter universes at $x=\left(\pm \sqrt{{\Lat\over 3}},0\right)$. A third solution is the mirror image of the first one flowing to the de Sitter universe at $ x=\left(+ \sqrt{{\Lat\over 3}},0\right)$.  The case corresponding to $\Lat <0$ is an oscillatory universe.These solutions prevents any trajectory from crossing the $\Ht$-axis except at the fixed point which takes an infinite amount of time to reach it.


\section{Analysis of Universe Filled with Bulk Viscous Fluid in the Presence of Cosmological Constant}
The cosmological equations in their full generality, in the presence of $\o$, $\Lat$ and $\xit$, are
\bea
\dot{x} _1&=& - x_1^2 - {1\over 6}\,x_2\,\left(1 + 3\,\omega\right) + 3\,\xit\,x_1  + {\Lat\over 3},\nn\\
\dot{x} _2&=& -3\,x_1\,x_2 \left(1 + \omega\right)  + 18\,\xit\,x_1^2,
\label{eqgen}
\eea
where the coefficient of bulk viscosity $\xit$ maybe dependent on $x_2$ and $\left(x_1 = \Ht,\;\;x_2 = \rt\right)$.

Here we are interested in two cases, one for which  $\xit$ is constant while the other where $\xit\, \propto x_2$.
The case of constant $\xit$ turns out to be rich and therefore it is discussed in its full generality.
The other case of variable viscosity is equally rich and deserves a sperate study which would be the subject of a future work. Although we would like to report on the case of variable viscosity coefficient, $\xit\, \propto x_2$, in a future work, we still want to show some of the interesting features of this case which are different from the previous cases. However to have a clearer picture on the impact of variable viscosity coefficient on models it is enough to consider only the spatially flat case.

\subsection{ Analysis of  bulk viscosity in models with spatial curvature}
In case of constant bulk  viscosity $(\xit)$, the general cosmological equations, as given in Eqs.(\ref{eqgen}),  can be transformed into one of the standard normal form  given as,
\bea
\dot{y}_1 & = & y_2 ,\nn \\
\dot{y}_2 & = & \a_1 + \a_2\,y_1 + \a_3\,y_2 + b\, y_1^3 + d\, y_1\,y_2 +
e\, y_1^2\,y_2.
\label{norm_lam_omg_xi}
\eea
This form corresponds to the normal form for a degenerate Bogdanov-Taken  bifurcation as classified in \cite{kuz,dumo}.
The form can be achieved by the following transformation, given here with its inverse,
\be
\begin{array}{l}
\begin{array}{lll}
y_1 = x_1 - {2\,\xit \over 3\,\left(1 +\o\right)}, && y_2 = -{1\over 6}\,\left(1 +3\,\o\right)\,x_2 + 3\,\xit\,x_1 - x_1^2 +{\Lat\over 3},\;\; {\rm  (Transformation)} \nn
\end{array}
\\
\begin{array}{ll}
\left.
\begin{array}{lll}
 x_1& =& y_1 +  {2\,\xit \over 3\,\left(1 +\o\right)},\nn\\
x_2& =& {2\over 3\left(1 +3 \o\right)\left(1 + \o\right)^2}\left[2\, \xit^2\left(7 + 9\,\o\right) + 3\,\Lat\,\left(1 + \o\right)^2\right.\nn\\
&& \left. +3\,\xit\,\left(1 + \o\right)\,
\left(5 +9 \o\right)\,y_1 - 9\,\left(1 +\o\right)^2\,y_2 - 9\,\left(1 +\o\right)^2\,y_1^2\right].
\end{array}
 \right\}
& {\rm  (Inverse Transformation)}
\end{array}
\end{array}
\label{trans_norm_oxls}
\ee
After performing the previous transformation, the parameters $\a_1, \a_2, \a_3, b, d$ and $e$ are found to be \bea
\a_1 ={\left[6\,\Lat\,\xit\,\left(1 + \o\right)^2 +
16\,\xit^3\right]\over 9\,\left(1 + \o\right)^2},&
\a_2 =   \displaystyle{{ \Lat\,\left(1 + \o\right)^2 + 4\,\xit^2\over \left(1 + \o\right)}} ,& \a_3 = { \left(-1 + 3\,\o\right)\,\xit \over 3\,\left(1 + \o\right)},\nn\\
b = - 3\,\left(1 + \o\right), &
d = -\left(5 + 3\,\o\right),&  e = 0.
\label{norforparas}
\eea
According to the classification and the study carried out in \cite{kuz,dumo}, the parameters $b$ and $d$  together with their combination $d^2 + 8 b$
shouldn't be vanishing. Actually, $b$ is vanishing for $\o = -1$ and $d$ for $\o = -{5\over 3}$ while $d^2 + 8 b$ for $\o =-{1\over 3}$. We also notice that the transformation as given in Eq.(\ref{trans_norm_oxls}) is problematic when $\o=-1$ or $\o =  -{1\over 3}$ but the resulting system of equations is still of a degenerate Bogdanov-Taken type.  Although the form in Eq.(\ref{norm_lam_omg_xi}) is relevant to recognize the classification of the system described in Eq.(\ref{eqgen}) as a degenerate Bogdanov-Taken
 bifurcation involving three parameter and there is no consensus in literature for organizing this kind of complicated bifurcation.
Therefore, we believe  that  it is more simpler and transparent to study the bifurcation of the system in its original form in Eq.(\ref{eqgen}) involving the parameters $\o$, $\Lat$ and $\xit$ .
It is worthy to mention that this is the first time to recognize that the system of cosmological equations, as given by Eq.(\ref{eqgen}), can be casted  into the form  of a degenerate Bogdanov-Taken bifurcation according to the best of our knowledge. Hence, it is highly recommended to study the cosmological equations
from that perspective which, as shown later, would enable us to easily extract results and identify regions in parameter space that are relevant in describing the actual physical universe.

The starting point for this bifurcation study is to find and classify fixed points and then investigate their behavior
under changing parameters $(\o, \Lat, \xit)$. The first possibility is where we have a fixed point along the
$x_2$ axis which is given together with its Jacobian as,
\bea
\left( x_1 = 0,\; x_2 ={2\,\Lat\over \left(1 + 3\,\o\right)}\right), &&
\left[{\partial f_i \over \partial x_j}\right]_{\left( x_1 = 0,\; x_2 ={2\,\Lat\over \left(1 + 3\,\o\right)}\right)} =
\left(
\begin{array}{cc}
3\,\xit & -{1\over 6}\,\left(1 + 3\,\o\right)\\
-{6\,\Lat\,\left(1 + \o\right)\over \left(1 + 3\,\o\right)} & 0
\end{array}
 \right).
\label{one_fix_gen}
\eea
The eigenvalues for the Jacobian in Eq.(\ref{one_fix_gen}) together with the corresponding eigenvectors are,
\bea
\l_{1} = {3\,\xit \over 2} + \, {\sqrt{\D_1}\over 2},\;\l_{2} = {3\,\xit \over 2} - {\sqrt{\D_1}\over 2},&&
\mathbf{e}_{1} = \left(1,\; { 9\,\xit - 3\,\sqrt{\D_1}\over \left(1 + 3\,\o\right)}\right)^T,\; \mathbf{e}_{2} =
\left(1,\; { 9\,\xit + 3\,\sqrt{\D_1}\over \left(1 + 3\,\o\right)}\right)^T,
\label{one_fix_lam_gen}
\eea
where,
\be
\D_1 = 9\,\xit^2 + 4\,\left(1 + \o\right)\,\Lat.
\label{Delta1def}
\ee
This fixed point is always present provided that $\o \neq -{1\over 3}$.
For nonvanishing $\xit > 0$ and where $\D_1 < 0$ the fixed point
is a repelling center. When  $\D_1 = 0$, the fixed point turns out to be  unstable (source) and
continues to be unstable (source) whenever $\D_1 < 9\,\xit^2$. When  $\D_1 \ge  9\,\xit^2$, the eigenvalue $\l_2$ vanish at $\D_1 = 9\,\xit^2$
and then start to be negative leading to a saddle fixed point. To simplify matter for depicting the behavior of the fixed point as
the parameters change, we fix $\o$ at a specific values and then the condition $\D_1=0$ turns out to define a parabola in the
$(\Lat ,\;\xit)$ plane given by $\Lat = {-9\,\xit^2 \over 4 (1 + \o)}$. This parabola together with $\xit$ axis  divide the $(\Lat ,\;\xit)$  plane into four distinct regions\footnote{Here the boundary is counted as a region if it has a distinct behaviour for the fixed point}
 and each region has a characteristic behavior for the fixed point. The nature of the fixed point is changing with the value of the parameter according to Hophf bifurcation. This typical kind of bifurcation is shown in a bifurcation diagram in Fig.(\ref{figbhopf_xi}).
\begin{figure}[hbtp]
\centerline{\epsfig{file=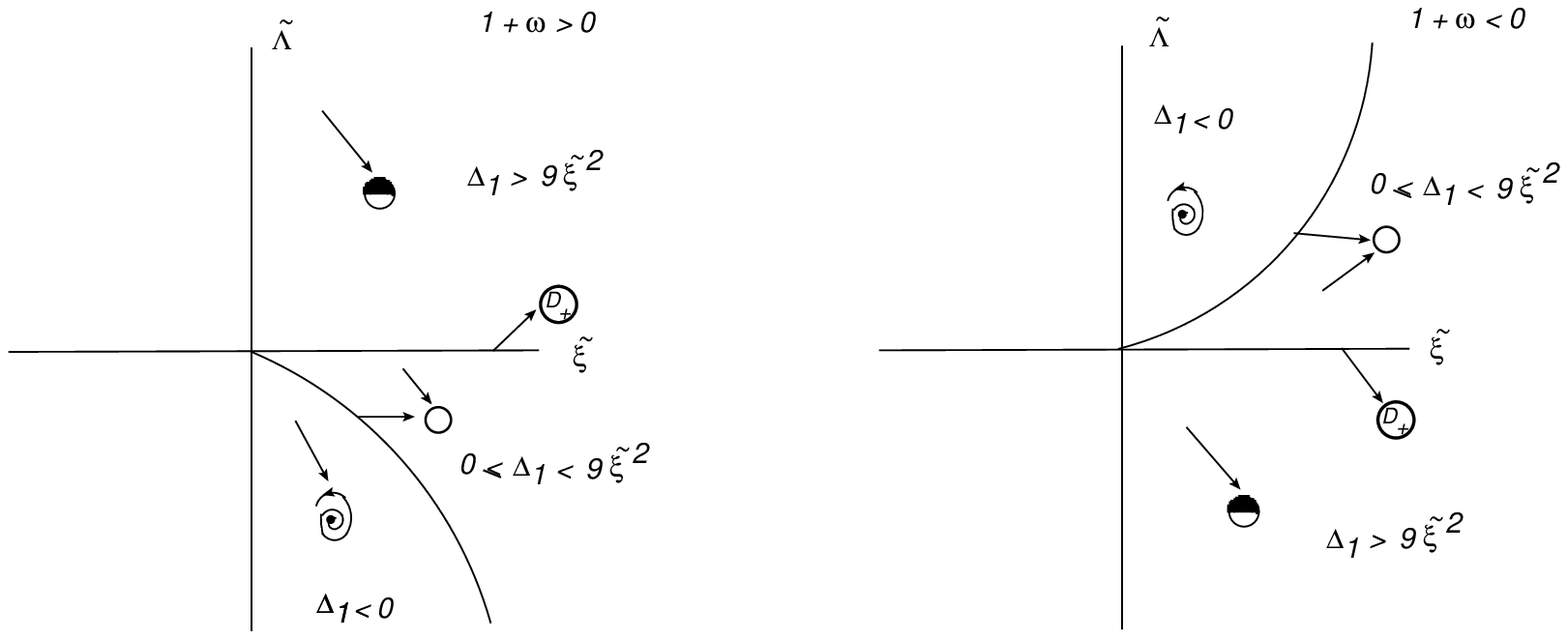,width=15cm,height=6cm}}
\caption{{\footnotesize
The Hopf bifurcation diagram  for all the four  possible regions in the $(\Lat ,\;\xit)$  plane as divided by the solid curve, $\Lat = -{9\,\xit^2\over 4\,\left( 1 + \o\right)}$ and the $\xit$ axis.
The outward spiral indicates a repulsive center, the hollow circle indicates an unstable (source) fixed point, the circled $D_+$ denotes  degenerate (non-hyperbolic) fixed point having one zero eigenvalue and one positive
eigenvalue and  half-filled circle  represent a saddle.}}
\label{figbhopf_xi}
\end{figure}
 The phase space diagrams, compact and noncompact ones,
 are also displayed for some representative  cases as in Figures Figs.~(\ref{fig1_cos_omega_lam_xi_ex_bihopf}--\ref{fig2_tot_cos_omega_lam_xi_bihopf}).
 For fixed value of $\xit=0.1$, we  choose the other parameters $(\Lat, \o)$ in such a way to have only, whenever possible,  a fixed point along the
 $x_2$ axis with a clear appearance as done in Figures Fig.~(\ref{fig1_cos_omega_lam_xi_ex_bihopf}) and Fig.~(\ref{fig2_cos_omega_lam_xi_ex_bihopf}).
 As to the fixed points not appearing along $x_2$, but along the flat curve solution, we anticipate the results which will be explained later in this
 section. The figures in Fig.(\ref{fig1_tot_cos_omega_lam_xi_bihopf}) and  Fig.(\ref{fig2_tot_cos_omega_lam_xi_bihopf}) are devoted to the case of stiff
 matter, $(\o =1)$ and thus $(1 + \o) > 0$, with varying  $\Lat$ to produce all possible scenarios for the fixed point along the $x_2$ axis.
 The figures in Fig.(\ref{fig_omeg_eq_min2_cos_omega_lam_xi_bihopf}) are devoted to the case of phantom matter, $(\o =-2)$ and thus $(1 + \o) < 0$,
 with varying  $\Lat$ to produce all possible scenarios for the fixed point along the $x_2$ axis.
It is worth mentioning that in these set of figures  we include the case for $(\o = -1)$ where a fixed point doesn't  occur except at the $x_2$ axis where $x_2 = - \Lat$ and has a saddle character.

To have more quantitative results, we present in Table~(\ref{Tab_hopf}), for fixed value of $\xit = 0.1$,
 the numerical values for $(\o, \Lat)$ together with their corresponding values $(\D_1$, $\D_2)$ as respectively defined in Eq.~(\ref{Delta1def}) and Eq.~(\ref{Delta2def}),
 the coordinates of fixed points are
 $\left\{\left( x_{1\pm}, x_{2\pm}\right),  \left( x_{1}, x_{2}\right)\right\}$  as respectively defined in Eq.~(\ref{one_fix_gen}) and Eq.~(\ref{two_fix_flat}),
 eigenvalues of the Jacobian at the fixed points
 $\left\{\left( \l_{\pm 1},  \l_{\pm 2} \right), \left( \l_{1},  \l_{2} \right)\right\} $ as respectively defined in
 Eq.~(\ref{one_fix_lam_gen}) and Eq.~(\ref{gen_two_fix_lam}), and fixed point characters.
{\sf
\begin{table}[h]
\centering
\scalebox{0.6}{
\begin{tabular}{lllllll}
\toprule
$\o$ & $\Lat$  & $\D_1$ & $\D_2$ & $\left( x_{1+}, x_{2+}\right)$ & $\left( x_{1-}, x_{2-}\right)$ & $\left( x_{1}, x_{2}\right)$ \\
     &         &         &        &  $\left( \l_{+1},  \l_{+2} \right)$ &  $\left( \l_{-1},  \l_{-2} \right)$ &  $\left( \l_{1},  \l_{2} \right)$ \\
\toprule
0 & -0.5   & -1.91 & -1.41 & None & None & $\left( 0, - 1\right)$, Repulsive center \\
  &        &       &       &      &       &  $\left( 0.15 - 0.691\,i, 0.15 + 0.691\,i\right)$ \\
\midrule
2/3 & -0.0135  & 0 & - 0.0225 & None & None & $\left( 0, - 0.009\right)$, Unstable (Source) \\
  &        &       &       &      &       &  $\left( 0.15 , 0.15 \right)$ \\
\midrule
2/3 & -0.0125  & 0.0067 & -0.0142 & None & None  & $\left( 0, -0.0083\right)$, Unstable (Source) \\
  &        &       &       &      &       &  $\left( 0.1908 , 0.1092 \right)$ \\
\midrule
0 & 0  & 0.09 & 0.09 & $\left( 0.2000, 0.1200\right)$, Stable (Sink) & $\left( 0, 0\right)$, Degenerate & $\left( 0, 0\right)$, Degenerate \\
  &        &       &       & $\left( -0.3, -0.4 \right)$     &   $\left( 0.3, 0 \right)$    &  $\left( 0.3 , 0 \right)$ \\
\midrule
-1 & 1  & 0.09 & 0.09 &  None & None  & $\left( 0, -1\right)$, Degenerate \\
  &        &       &       &      &       &  $\left( 0.3 , 0 \right)$ \\
\midrule
0 & 0.5  & 2.09 & 1.59 &  $\left( 0.5203, 0.3122\right)$, Stable (Sink) & $\left( -0.3203, -0.1922\right)$, Unstable (Source)  & $\left( 0, 1\right)$, Saddle \\
  &        &       &       & $\left( -1.261, -1.0406 \right)$     &   $\left( 1.2610, 0.6406 \right)$    &  $\left( 0.8728 , -0.5728\right)$ \\
\toprule
\toprule
1 & -0.02325   & -0.096 & -0.189 & None & None & $\left( 0, -0.0116\right)$, Repulsive center \\
   &        &       &       &      &       &  $\left( 0.15 + 0.1549\,i,  0.15 - 0.1549\,i\right)$ \\
\midrule
1 & -0.01125  & 0 & -0.045 & None & None & $\left( 0, - 0.0056\right)$, Unstable (Source)\\
 &        &       &       &      &       &  $\left( 0.15 , 0.15 \right)$ \\
\midrule
1 & -0.00925  & 0.016 & -0.021 & None & None  & $\left( 0, -0.0046\right)$, Unstable (Source) \\
  &        &       &       &      &       &  $\left( 0.2132 , 0.0868 \right)$ \\
\midrule
1 & 0 &  0.09 & 0.09 & $\left( 0.1 , 0.03 \right)$, Stable (Sink) & $\left( 0, 0\right)$, Degenerate & $\left( 0, 0\right)$, Degenerate \\
  &        &       &       & $\left( -0.3, -0.2 \right)$     &   $\left( 0.3, 0 \right)$    &  $\left( 0.3 , 0 \right)$ \\
\midrule
-1 & 0  & 0.09 & 0.09 &  None & None  & $\left( 0, 0\right)$, Degenerate \\
  &        &       &       &      &       &  $\left( 0.3 , 0 \right)$ \\
\midrule
1 & 0.01  & 0.17 & 0.21 &  $\left( 0.1264, 0.0379\right)$, Stable & $\left( -0.0264, -0.0079\right)$, Unstable  & $\left( 0, 0.005\right)$, Saddle \\
  &        &       &       & $\left( -0.4583, -0.2528 \right)$     &   $\left( 0.4583, 0.0528 \right)$    &  $\left( 0.3562 , -0.0562\right)$ \\
\bottomrule
\bottomrule
-2 & 2.5   & -9.91 & 7.59 & $\left( -1.0183, 0.61099\right)$, Saddle & $\left( 0.81833, -0.49099\right)$, Saddle & $\left( 0, -1\right)$, Repulsive center \\
  &        &       &       & $\left( -2.7550, 2.0367 \right)$     &   $\left( 2.7550, -1.6367 \right)$    &  $\left( 0.15 + 1.5740\,i , 0.15 - 1.5740\,i\right)$ \\
\midrule
-2 &  0.0225   & 0 & 0.1575 & $\left( -0.2323, 0.1394\right)$, Saddle & $\left( 0.0323, -0.0194\right)$, Saddle & $\left( 0, -0.009\right)$, Unstable (Source)\\
&        &       &       & $\left( -0.3969, 0.4646 \right)$     &   $\left( 0.3969, -0.0646 \right)$    &  $\left( 0.15 , 0.15 \right)$ \\
\midrule
-2 & 0.011   & 0.0450 & 0.12350 & $\left( -0.2109, 0.1301\right)$, Saddle & $\left( 0.0169, -0.0101\right)$, Saddle & $\left( 0, -0.0044\right)$, Unstable (Source) \\
&        &       &       & $\left( -0.3507, 0.4338 \right)$     &   $\left( 0.3507, -0.0338 \right)$    &  $\left( 0.2572 , 0.0428 \right)$ \\
\midrule
-2 & 0   & 0.09 & 0.09 & $\left( -0.2, 0.12\right)$, Saddle & $\left( 0, 0\right)$, Degenerate & $\left( 0, 0\right)$, Degenerate \\
  &        &       &       & $\left( -0.3, 0.4\right)$     &   $\left( 0.3, 0 \right)$    &  $\left( 0.3 ,  0 \right)$ \\
\midrule
-2 & -0.1   & 0.490 & -0.2100 & None  & None  & $\left( 0, -0.2100\right)$, Saddle \\
 &        &       &       &      &      &  $\left( 0.5 , -0.2 \right)$ \\
\bottomrule
\end{tabular}
}
\caption{\footnotesize Results for the representative cases of having a lone  fixed point along the $x_2$ axis but with
also including the possible ones  along the flat curve solution if they arise. The first set are for $\o =0, 2/3$ and $-1$ with
suitably chosen value of $\Lat$ to have a clear appearance of the fixed point along the $x_2$ axis.
The last two sets are respectively for $\o =1$ and $\o =-2$  and exhibiting all possible scenarios for the fixed point
along $x_2$ axis.  The quantities $(\D_1$, $\D_2)$ are respectively defined in Eq.~(\ref{Delta1def}) and Eq.~(\ref{Delta2def}) while
the  coordinates of fixed points $\left\{\left( x_{1\pm}, x_{2\pm}\right),  \left( x_{1}, x_{2}\right)\right\}$ are respectively defined in
Eq.~(\ref{one_fix_gen}) and Eq.~(\ref{two_fix_flat}). The eigenvalues of the Jacobian at the fixed points
 $\left\{\left( \l_{\pm 1},  \l_{\pm 2} \right), \left( \l_{1},  \l_{2} \right)\right\} $ are  respectively defined in
 Eq.~(\ref{one_fix_lam_gen}) and Eq.~(\ref{gen_two_fix_lam}). }
\label{Tab_hopf}
\end{table} }
\begin{figure}[hbtp]
\includegraphics[width=18cm,height=20cm]{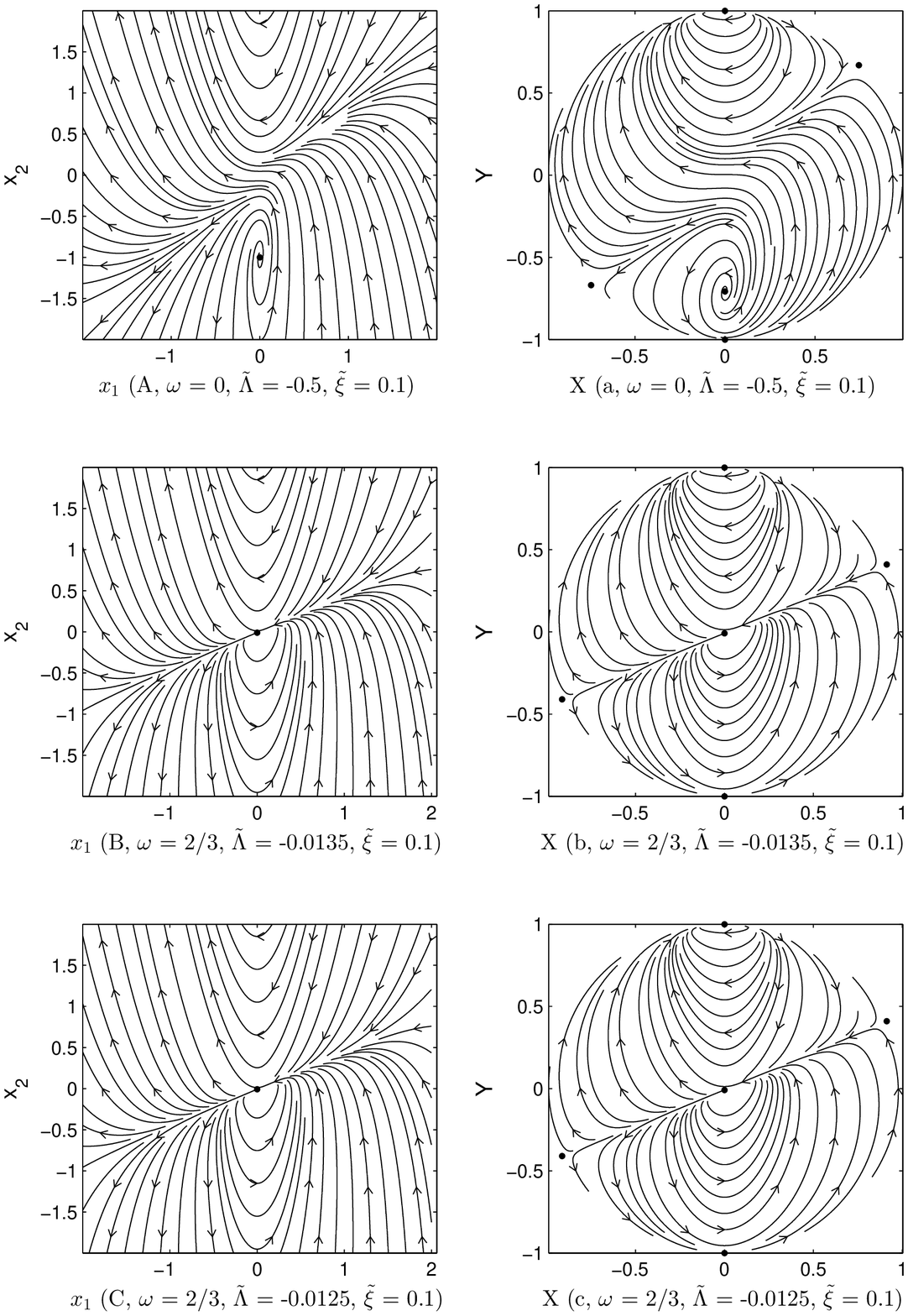}
\caption{{\footnotesize
Uncompact (left panel) and compact (right panel) phase portraits  when viscosity is included, $\xit = 0.1$. Representative cases are
$\left(\o = 0, \Lat = -0.5\right)$, $\left(\o = {2\over 3}, \Lat = -0.0135\right)$ and
$\left(\o = {2\over 3}, \Lat = -0.0125\right)$.
$x_1$ and $x_2$  respectively denote the dimensionless $\Ht$ and $\rt$ as defined in Eq.(\ref{gre3}). $X$ and $Y$ are the coordinates on the Poincar\'{e}
sphere as defined in Eq.(\ref{XYZpoin}). The dotted circles represent fixed points. }}
\label{fig1_cos_omega_lam_xi_ex_bihopf}
\end{figure}
\begin{figure}[hbtp]
\includegraphics[width=18cm,height=20cm]{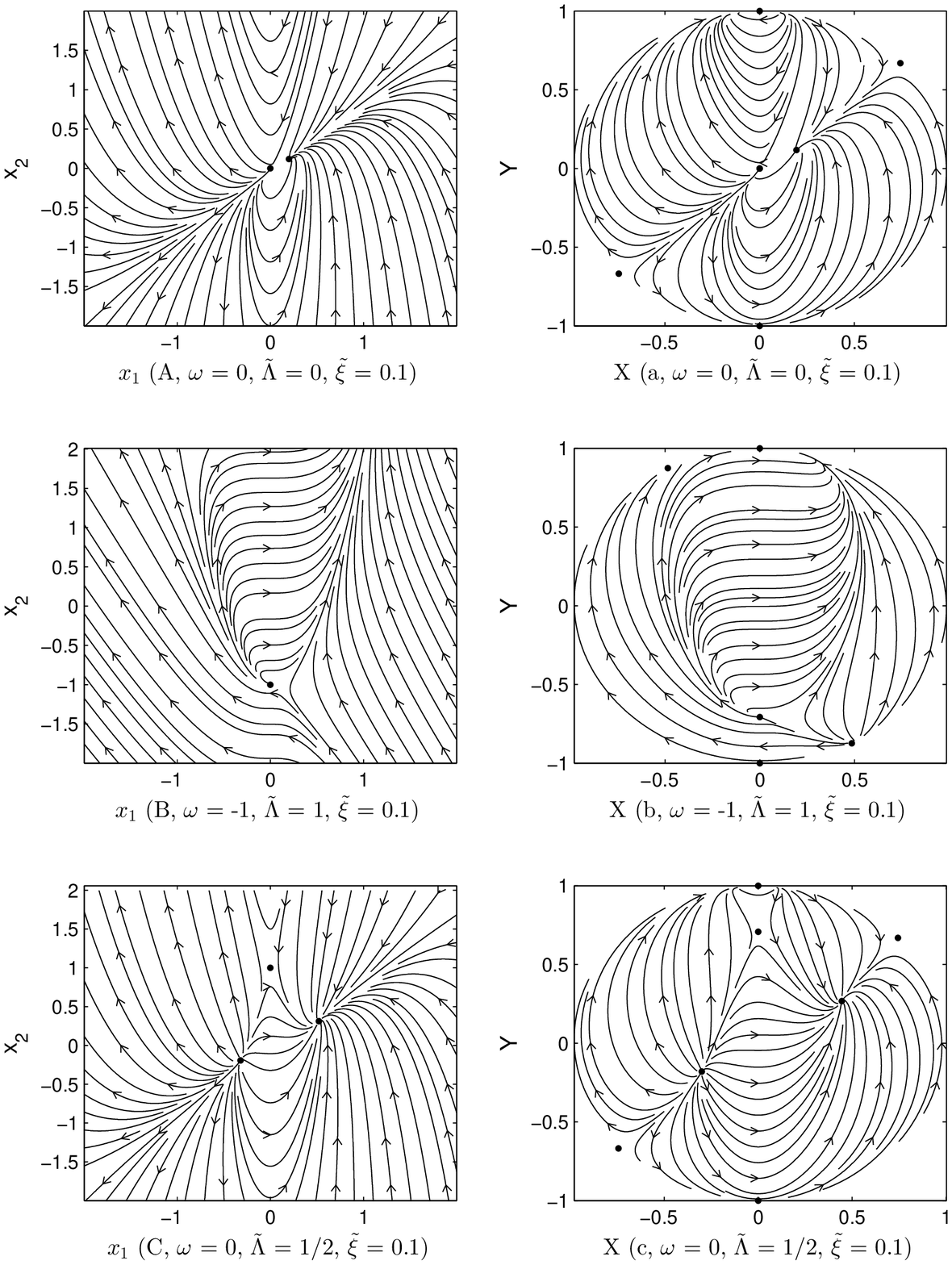}
   \caption{{\footnotesize Uncompact (left panel) and compact (right panel) phase portraits  when viscosity is included, $\xit = 0.1$. Representative cases are
 $\left(\o = 0, \Lat = 0\right)$, $\left(\o = -1, \Lat = 1\right)$ and $\left(\o = 0, \Lat = 0.5\right)$. $x_1$ and $x_2$ respectively
denote the dimensionless $\Ht$ and $\rt$ as defined in Eq.(\ref{gre3}). $X$ and $Y$ are the coordinates on the Poincar\'{e}
sphere as defined in Eq.(\ref{XYZpoin}). The dotted circles represent fixed points.
}}
\label{fig2_cos_omega_lam_xi_ex_bihopf}
\end{figure}
\begin{figure}[hbtp]
\includegraphics[width=18cm,height=20cm]{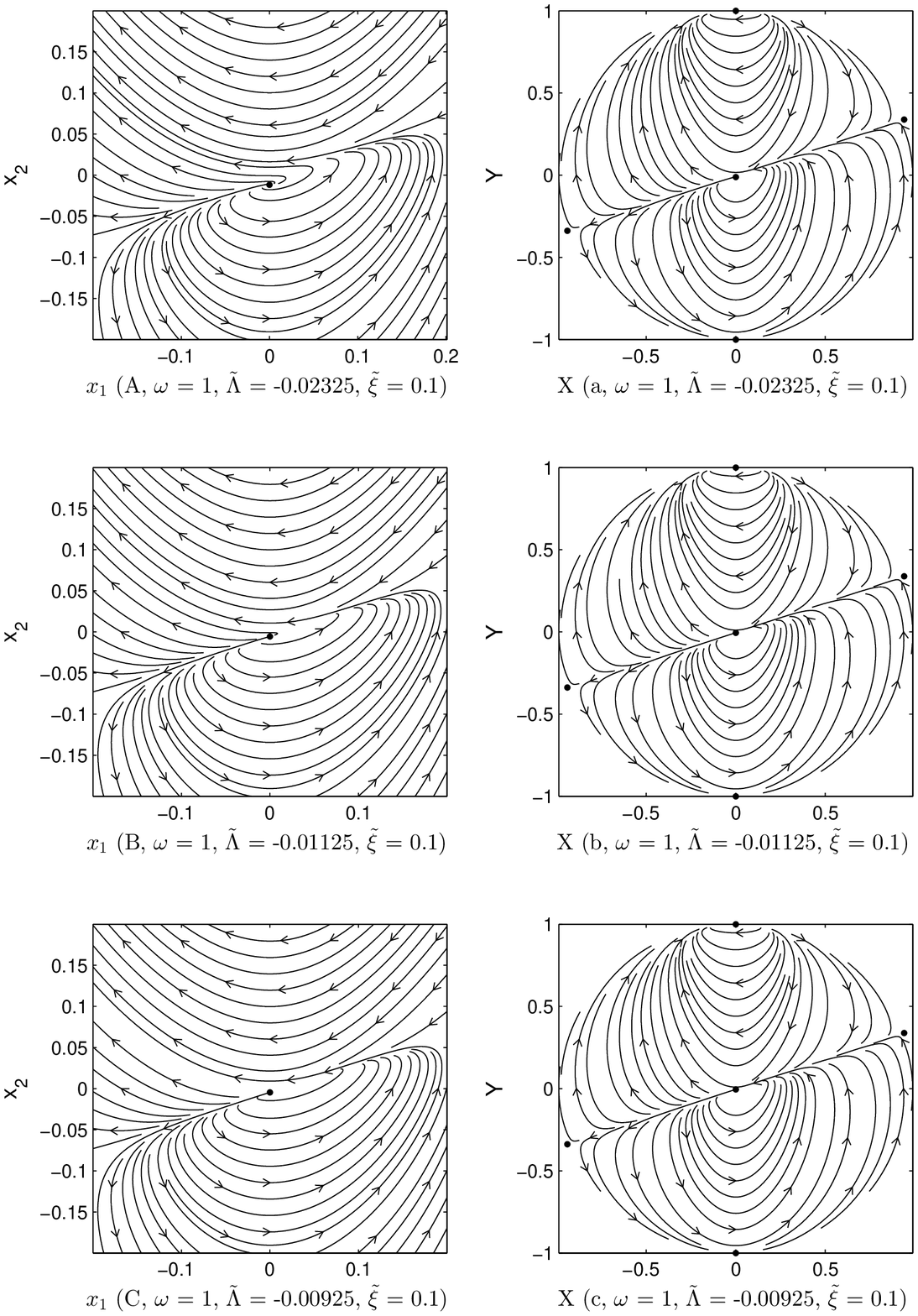}
\caption{{\footnotesize
Uncompact (left panel) and compact (right panel) phase portraits  when viscosity is included, $\xit = 0.1$. Representative cases are
$\left(\o = 1, \Lat = -0.02325\right)$, $\left(\o = 1, \Lat = -0.01125\right)$ and   $\left(\o = 1, \Lat = -0.00925\right)$.
$x_1$ and $x_2$ respectively denote the dimensionless $\Ht$ and $\rt$ as defined in Eq.(\ref{gre3}). $X$ and $Y$ are the coordinates on the Poincar\'{e}
sphere as defined in Eq.(\ref{XYZpoin}). The dotted circles represent fixed points. }}
\label{fig1_tot_cos_omega_lam_xi_bihopf}
\end{figure}
\begin{figure}[hbtp]
\includegraphics[width=18cm,height=20cm]{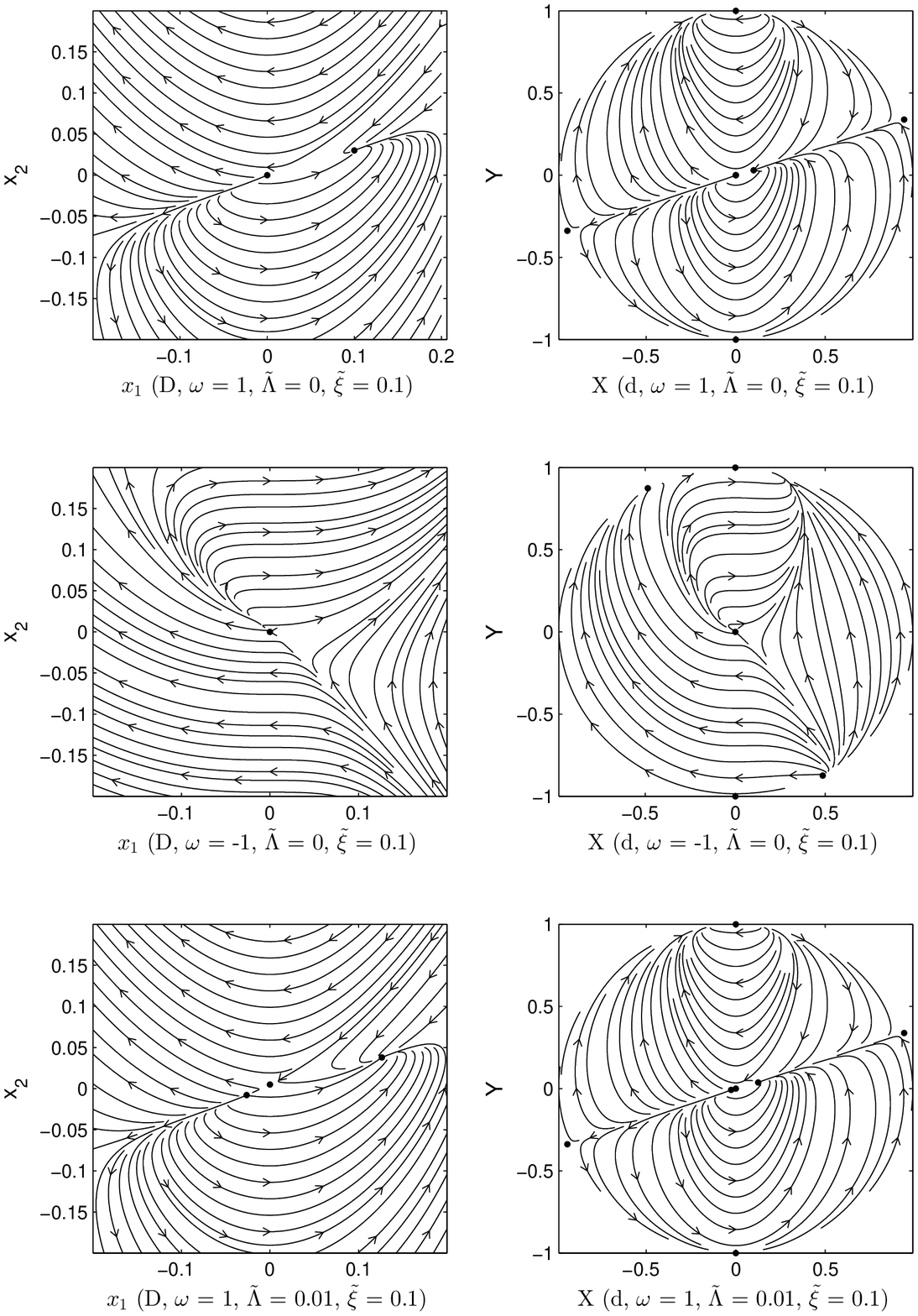}
\caption{{\footnotesize Uncompact (left panel) and compact (right panel) phase portraits  when viscosity is included, $\xit = 0.1$. Representative cases are
 $\left(\o = 1, \Lat = 0\right)$, $\left(\o = -1, \Lat = 0\right)$ and $\left(\o = 1, \Lat = 0.01\right)$. $x_1$ and $x_2$ respectively
denote the dimensionless $\Ht$ and $\rt$ as defined in Eq.(\ref{gre3}). $X$ and $Y$ are the coordinates on the Poincar\'{e}
sphere as defined in Eq.(\ref{XYZpoin}). The dotted circles represent fixed points.
}}
\label{fig2_tot_cos_omega_lam_xi_bihopf}
\end{figure}
\begin{figure}[hbtp]
\includegraphics[width=18cm,height=20cm]{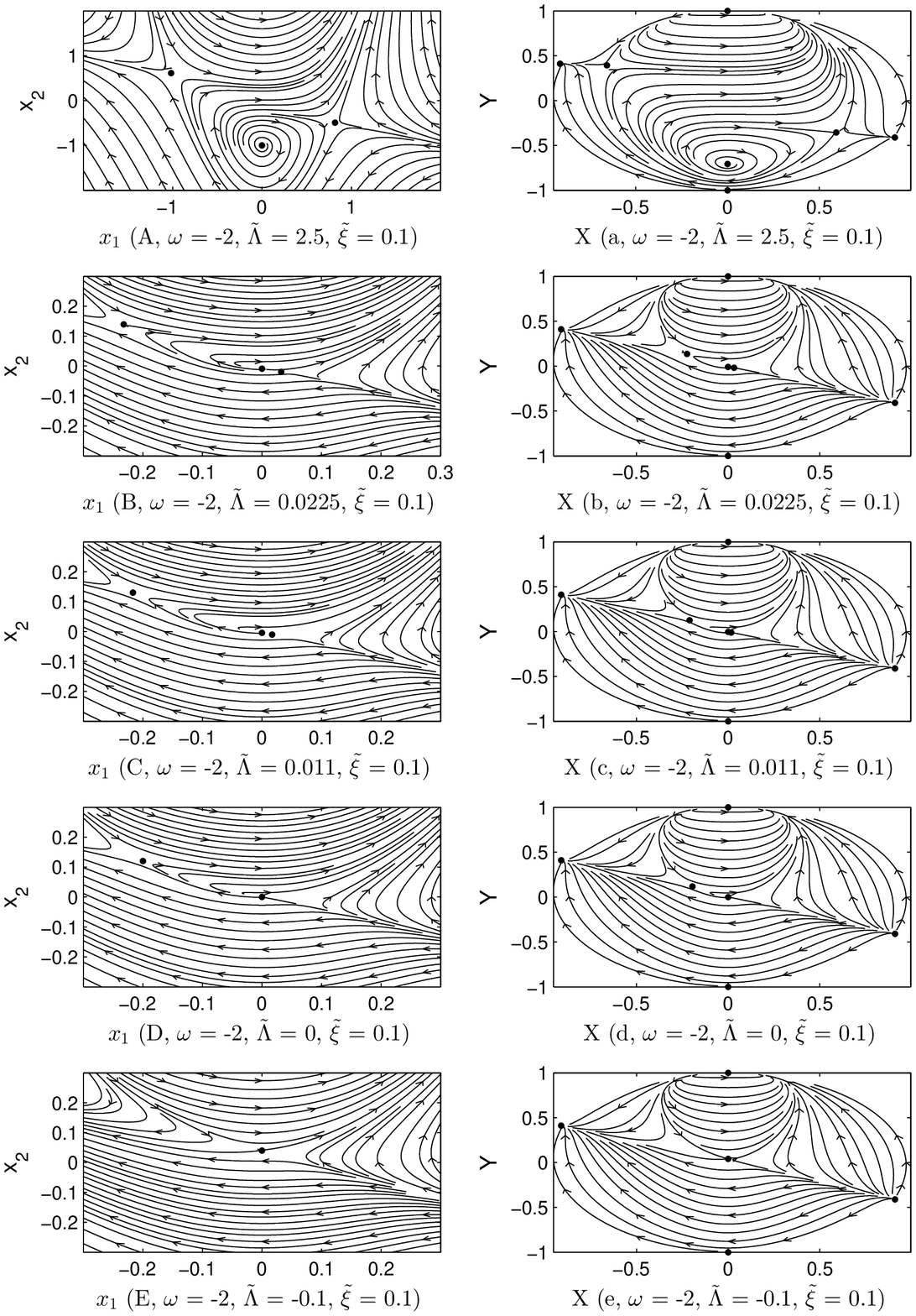}
\caption{{\footnotesize Uncompact (left panel)  and compact (right panel) phase portraits  when viscosity is included, $\xit = 0.1$. Representative cases are
 $\o = -2$ while $\Lat = 2.5, 0.1575, 0.011, 0$ and $-0.1$. $x_1$ and $x_2$ respectively
denote the dimensionless $\Ht$ and $\rt$ as defined in Eq.(\ref{gre3}). $X$ and $Y$ are the coordinates on the Poincar\'{e}
sphere as defined in Eq.(\ref{XYZpoin}). The dotted circles represent fixed points.
}}
\label{fig_omeg_eq_min2_cos_omega_lam_xi_bihopf}
\end{figure}
\clearpage

The second possibility where we have two fixed points that are given as,
\bea
x_{1\pm} = {3\,\xit \pm \sqrt{\D_2}\over 3\,\left(1 + \o\right)},&& x_{2\pm} = {2\,\left(3\,\xit \pm \sqrt{\D_2}\right)\,\xit \over \left(1 + \o\right)^2},\;\;\;
\o \neq -1,
\label{two_fix_flat}
\eea
where
\be
\D_2 = 3\,\left(1 + \o\right)^2\,\Lat + 9\,\xit^2.
\label{Delta2def}
\ee
The reality of the fixed points are ensured when $\D_2 \ge 0$ (or equivalently $ \Lat \ge -{3\xit^2\over \left(1 + \o\right)^2}$).
The real fixed points $\left(x_{1+}, x_{2+}\right)$ and $\left(x_{1-}, x_{2-}\right)$, when realized,  are always located on the parabola describing
flat solution $x_1^2 = {1\over 3}\,\left(x_2 + \Lat\right)$.

The Jacobian at these fixed points, Eq.(\ref{two_fix_flat}), are found to be
\bea
\left[{\partial f_i \over \partial x_j}\right]_{\pm} &=&
\left(
\begin{array}{cc}
3\,\xit - 2\, {\left(3\,\xit \pm \sqrt{\D_2}\right)\over 3\,\left(1 + \o\right)} & -{1\over 6}\,\left(1 + 3\,\o\right)\\\\
6\,\xit {\left(3\,\xit \pm \sqrt{\D_2}\right)\over \left(1 + \o\right)} & -\left(3\,\xit \pm \sqrt{\D_2}\right)
\end{array}
 \right),
 \label{genjac_two_fix}
 \eea
where the sign $(\pm)$ respectively denotes the fixed points $\left(x_{1+}, x_{2+}\right)$ and $\left(x_{1-}, x_{2-}\right)$. The resultant
eigenvalues and their associated eigenvectors are
\bea
\l_{+1} = -\sqrt{\D_2},\; \l_{+2} = -{2\,\left(3\,\xit + \sqrt{\D_2}\right)\over 3\,\left(1 + \o\right)},&&\hspace{-6mm} \mathbf{e}_1=\left(1,\;  {2\,\left(3\,\xit
+ \sqrt{\D_2}\right)\over \left(1 + \o\right)}\right)^T,\;\mathbf{e}_2=\left({1+ 3\,\o \over 18\,\xit},\; 1\right)^T\hspace{-2mm},\mathbf{(+)},\nn\\
\l_{-1} = \sqrt{\D_2},\; \l_{-2} = -{2\,\left(3\,\xit - \sqrt{\D_2}\right)\over 3\,\left(1 + \o\right)}, &&\hspace{-6mm} \mathbf{e}_1=
\left(1,\;  {2\,\left(3\,\xit - \sqrt{\D_2}\right)\over \left(1 + \o\right)}\right)^T,\;
\mathbf{e}_2 = \left({1+ 3\,\o \over 18\,\xit},\; 1\right)^T\hspace{-2mm},\mathbf{(-)}.
\label{gen_two_fix_lam}
\eea

The case with two fixed points, along the flat curve solution,  is more involved than the case of a single point along the $x_2$ axis. In the parameter space where
$\D_2 < 0$ there are no fixed points at  all. When $\D_2 =0$ an emergent single fixed point (non-hyperbolic one) appears  whose coordinates,  associated eigenvectors
and eigenvalues are, after using Eq.(\ref{two_fix_flat}) and Eq.(\ref{gen_two_fix_lam}),
\bea
x_{1} = {\xit \over \left(1 + \o\right)}, &&  x_{2} = {6\, \xit^2 \over \left(1 + \o\right)^2},\;\;\; \o \neq -1,\nn\\
\l_1 = 0,\; \l_2 = -{2\,\xit \over \left(1 + \o\right)}&& \mathbf{e}_1=\left(1,\;  {6\,\xit \over \left(1 + \o\right)}\right)^T,
\;\mathbf{e}_2=\left({1+ 3\,\o \over 18\,\xit},\; 1\right)^T, \;\;\; \o \neq -1.
\label{D2zero_fix}
\eea
The eigenvector $\mathbf{e}_1$ corresponding to the zero eigenvalue is in the same direction as that of the tangent of the flat curve solution at
 $\big(x_{1} = {\xit \over \left(1 + \o\right)},\;\; x_{2} = {6\, \xit^2 \over \left(1 + \o\right)^2}\big)$ whenever $(1 + \o ) > 0$ and opposite
 otherwise. As to the direction given by $\mathbf{e}_2$, it represents a stable direction whenever  $(1 + \o ) > 0$ and an unstable for $(1 + \o ) < 0$.

In the parameter space where $\D_2 > 0$, the single fixed point at $\D_2=0$ is shattered into two fixed points as described by Eq.(\ref{two_fix_flat})
and Eq.(\ref{gen_two_fix_lam}).
The fixed point designated by $(+)$ is a stable (sink) fixed point when  $(1 + \o ) > 0$ and of a saddle type for $(1 + \o ) < 0$.
The other fixed point designated by $(-)$ doesn't behave in a simple manner as the one designated by $(+)$. When $\D_2 = 9\,\xit^2$, that leads to $\Lat= 0$
provided that $\o \neq -1$, the fixed point turns out
to be at the origin $( x_{1-} =0,\; x_{2-}=0)$ and the associated eigenvalues and eigenvectors are,
\bea
\l_1 = 3\,\xit ,\;\; \l_2 = 0, && \mathbf{e}_1=\left(1,\;  {6\,\xit \over \left(1 + \o\right)}\right)^T,
\;\mathbf{e}_2=\left(1,\; {18\,\xit \over (1+ 3\,\o)}\right)^T.
\label{fix(-)9xi^2}
\eea
The direction $\mathbf{e}_1$ corresponds to a stable direction while $\mathbf{e}_2$ has a zero eigenvalue which means that fixed point is a non-hyperbolic one.
Apart from this value of $\D_2$ and as $0 <  \D_2 < 9 \,\xit^2 $ the fixed point, $(-)$, is an unstable (source) for $(1 + \o ) < 0$,  while of saddle type for
 $(1 + \o ) > 0$. The behavior is switched off for $\D_2 > 9\,\xit^2 $,  which means getting unstable (source) fixed point for $(1 + \o ) > 0$, while a saddle type
 for $(1 + \o ) < 0$.

 The corresponding bifurcation diagram can be simplified by considering a fixed value for $\o$ and depicting the condition $\D_2 = 0$ as a parabola
 curve in the plane $( \Lat , \xit )$ given by $\Lat = - {3\,\xit^2 \over (1 + \o )^2}$. This parabola divide the plane $( \Lat , \xit )$ into
 five distinct regions\footnote{Here the boundary is counted as a region if it has a distinct behaviour for the fixed points} and each region has a characteristic behavior for the fixed points. All these behaviors are displayed in the bifurcation
 diagram in Fig.(\ref{figbi_two_pm_xi}) showing a similar  behavior  to that of saddle-node bifurcation.
\begin{figure}[hbtp]
\centerline{\epsfig{file=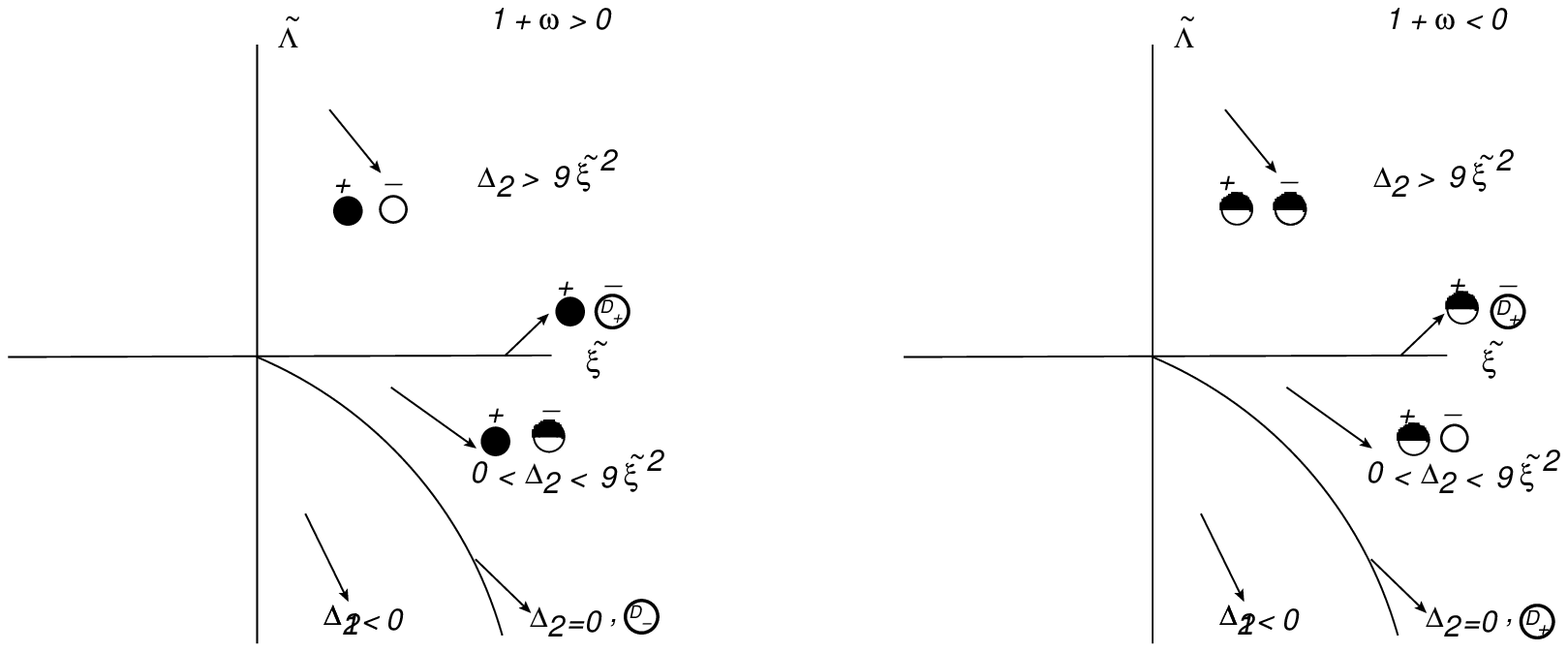,width=15cm,height=6cm}}
\caption{{\footnotesize
The bifurcation diagram  for all the five  possible regions in the $(\Lat ,\;\xit)$  plane as divided by the solid curve, $\Lat = -{3\,\xit^2\over \left( 1 + \o\right)^2}$ and the $\xit$ axis.
The hollow circle, solid circle and half-filled circle  indicate  respectively, an unstable (source) fixed point,  a stable (sink) fixed point and a saddle. The circled $D_+$
denotes degenerate fixed point having one zero eigenvalue and one positive  while the circled $D_-$ denotes degenerate fixed point having one zero eigenvalue and
one negative eigenvalue. The $+$ and $-$ signs over  the circles indicates that fixed point coordinates are given according to Eq.(\ref{two_fix_flat}). }}
\label{figbi_two_pm_xi}
\end{figure}
The phase space diagrams, compact and uncompact ones, are also
 displayed for representative  cases as in  Figures Figs.~(\ref{fig1_cos_omega_lam_xi_xpm_SD}) and (\ref{fig2_cos_omega_lam_xi_xpm_SD}).
 The finding for these representative cases are summarized in Table~(\ref{Tab_SD}) with the same notations used in Table~(\ref{Tab_hopf}).
{\sf
\begin{table}[h]
\centering
\scalebox{0.8}{
\begin{tabular}{lllllll}
\toprule
$\o$ & $\Lat$  & $\D_1$ & $\D_2$ & $\left( x_{1+}, x_{2+}\right)$ & $\left( x_{1-}, x_{2-}\right)$ & $\left( x_{1}, x_{2}\right)$ \\
     &         &         &        &  $\left( \l_{+1},  \l_{+2} \right)$ &  $\left( \l_{-1},  \l_{-2} \right)$ &  $\left( \l_{1},  \l_{2} \right)$ \\
\toprule
0 & -0.03  & -0.03 & 0 & $\left( 0.1, 0.06\right)$ , Degenerate  & $\left( 0.1, 0.06\right)$, Degenerate  & $\left( 0, - 0.06\right)$, Repulsive center \\
 &        &       &       & $\left( 0, -0.2 \right)$     &   $\left( 0, -0.2 \right)$    &  $\left( 0.15 + 0.0866\,i , 0.15 - 0.0866\,i\right)$ \\
\midrule
0 & -0.02  & 0.01 & 0.03 & $\left( 0.1577, 0.0946\right)$ , Stable (Sink) & $\left( 0.0423, 0.0254\right)$, Saddle & $\left( 0, - 0.04\right)$, Unstable (Source) \\
  &        &       &       & $\left( -0.1732, -0.3155 \right)$     &   $\left( 0.1732, -0.0845 \right)$    &  $\left( 0.2 , 0.1 \right)$ \\
\midrule
0 & 0   & 0.09 & 0.09 & $\left( 0.200, 0.1200\right)$ , Stable (Sink) & $\left( 0, 0\right)$, Degenerate & $\left( 0, 0\right)$, Degenerate \\
 &        &       &       & $\left(-0.3, -0.4 \right)$     &   $\left( 0.3, 0 \right)$    &  $\left( 0.3 , 0 \right)$ \\
\midrule
0 & 0.02  & 0.17 & 0.150 & $\left( 0.229, 0.1375\right)$ , Stable (Sink) & $\left( -0.0291, -0.0175\right)$, Unstable (Source) & $\left( 0, 0.04\right)$, Saddle \\
&        &       &       & $\left(-0.3873, -0.4582 \right)$     &   $\left( 0.3873, 0.0582 \right)$    &  $\left( 0.3562 , -0.0562 \right)$ \\
\bottomrule
\bottomrule
-2& -0.03  & 0.2100 & 0 & $\left( -0.1, 0.06\right)$ , Degenerate & $\left(- 0.1, 0.06\right)$, Degenerate  & $\left( 0, 0.0120\right)$, Saddle \\
&        &       &       & $\left( 0, 0.2 \right)$     &   $\left( 0, 0.2 \right)$    &  $\left( 0.3791,  -0.0791\right)$ \\
\midrule
-2 & -0.02  & 0.1700 & 0.03 & $\left( -0.1577, 0.0946\right)$ , Saddle & $\left(- 0.0423, 0.0254\right)$, Unstable (Source) & $\left( 0, 0.008\right)$, Saddle \\
  &        &       &       & $\left( -0.1732, 0.3155 \right)$     &   $\left( 0.1732, 0.0845 \right)$    &  $\left( 0.3562 , -0.0562 \right)$ \\
\midrule
-2& 0  & 0.09 & 0.09 & $\left(- 0.200, 0.1200\right)$ , Saddle & $\left( 0, 0\right)$, Degenerate & $\left( 0, 0\right)$, Degenerate  \\
   &        &       &       & $\left(-0.3, 0.4 \right)$     &   $\left( 0.3, 0 \right)$    &  $\left( 0.3 , 0 \right)$ \\
\midrule
-2 & 0.02  & 0.010 & 0.150 & $\left(- 0.229, 0.1375\right)$ , Saddle & $\left( 0.0291, -0.0175\right)$, Saddle & $\left( 0, -0.008\right)$, Unstable (Source) \\
      &        &       &       & $\left(-0.3873, 0.4582 \right)$     &   $\left( 0.3873, -0.0582 \right)$    &  $\left( 0.2 , 0.1 \right)$ \\
\bottomrule
\end{tabular}
}
\caption{\footnotesize Results for the representative cases of having fixed points along the flat curve solution and also including the possible one along the $x_2$ axis.
The first set are for $\o =0$ while the second one for $\o = -2$ exhibiting all possible scenarios for the fixed points along the flat curve solution.
The quantities $(\D_1$, $\D_2)$ are respectively defined in Eq.~(\ref{Delta1def}) and Eq.~(\ref{Delta2def}),
while  the coordinates of fixed points  $\left\{\left( x_{1\pm}, x_{2\pm}\right),  \left( x_{1}, x_{2}\right)\right\}$ are respectively defined in
 Eq.~(\ref{one_fix_gen}) and Eq.~(\ref{two_fix_flat}).  The eigenvalues of the Jacobian at the fixed points
 $\left\{\left( \l_{\pm 1},  \l_{\pm 2} \right), \left( \l_{1},  \l_{2} \right)\right\} $ are  respectively defined in
 Eq.~(\ref{one_fix_lam_gen}) and Eq.~(\ref{gen_two_fix_lam}). }
\label{Tab_SD}
\end{table} }
\begin{figure}[hbtp]
   \includegraphics[width=18cm,height=20cm]{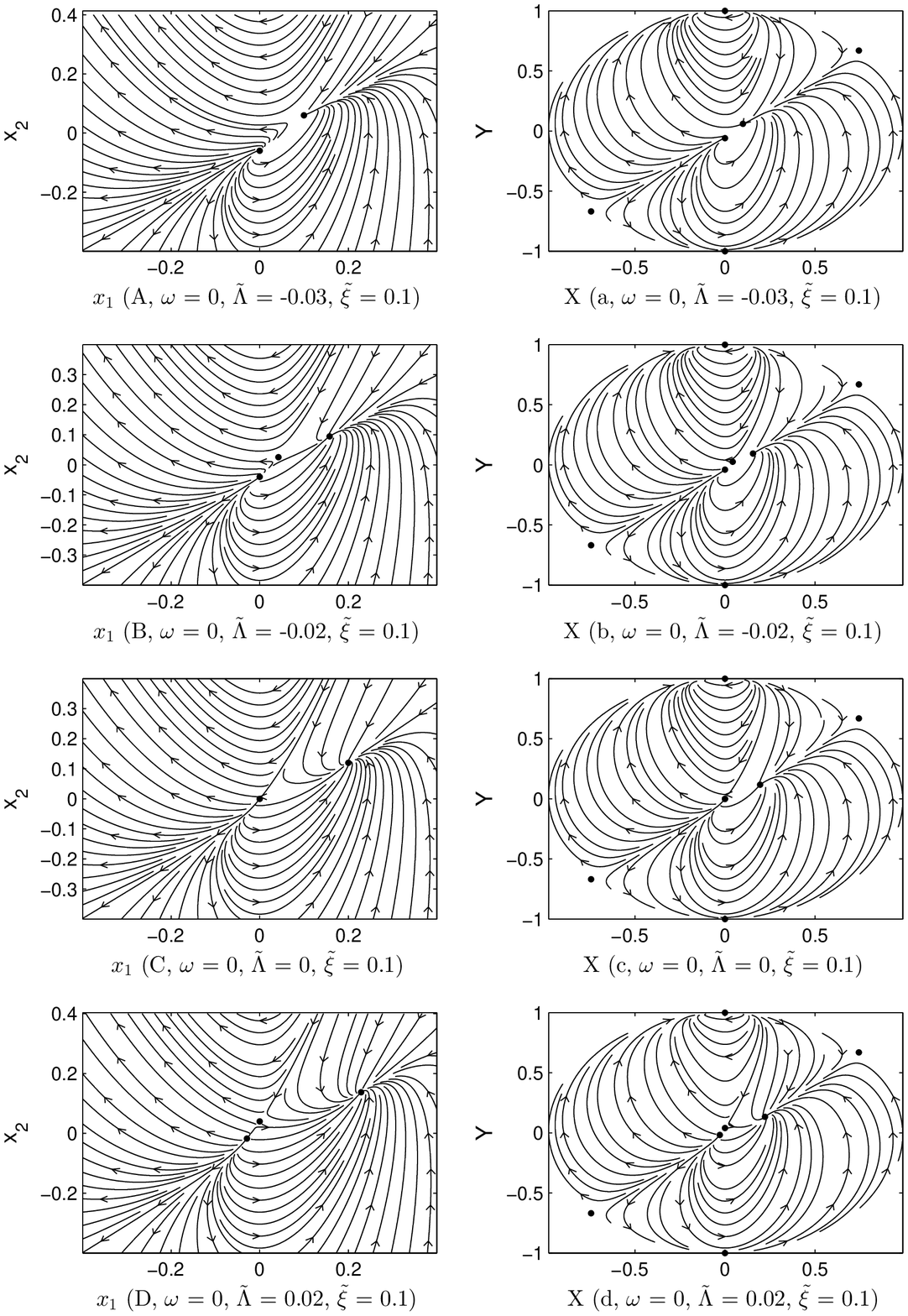}
\caption{{\footnotesize
Uncompact (left panel) and compact (right panel) phase portraits  when viscosity is included, $\xit = 0.1$ for dust case $(\o = 0)$ but with different $\Lat$. Representative cases are
$ \Lat = \left\{-0.03, -0.02, 0, 0.02\right\}$. $x_1$ and $x_2$ respectively
denote the dimensionless $\Ht$ and $\rt$ as defined in Eq.(\ref{gre3}). $X$ and $Y$ are the coordinates on the Poincar\'{e}
sphere as defined in Eq.(\ref{XYZpoin}). The dotted circles represent fixed points. }}
\label{fig1_cos_omega_lam_xi_xpm_SD}
\end{figure}
\begin{figure}[hbtp]
   \centerline{\includegraphics[width=18cm,height=20cm]{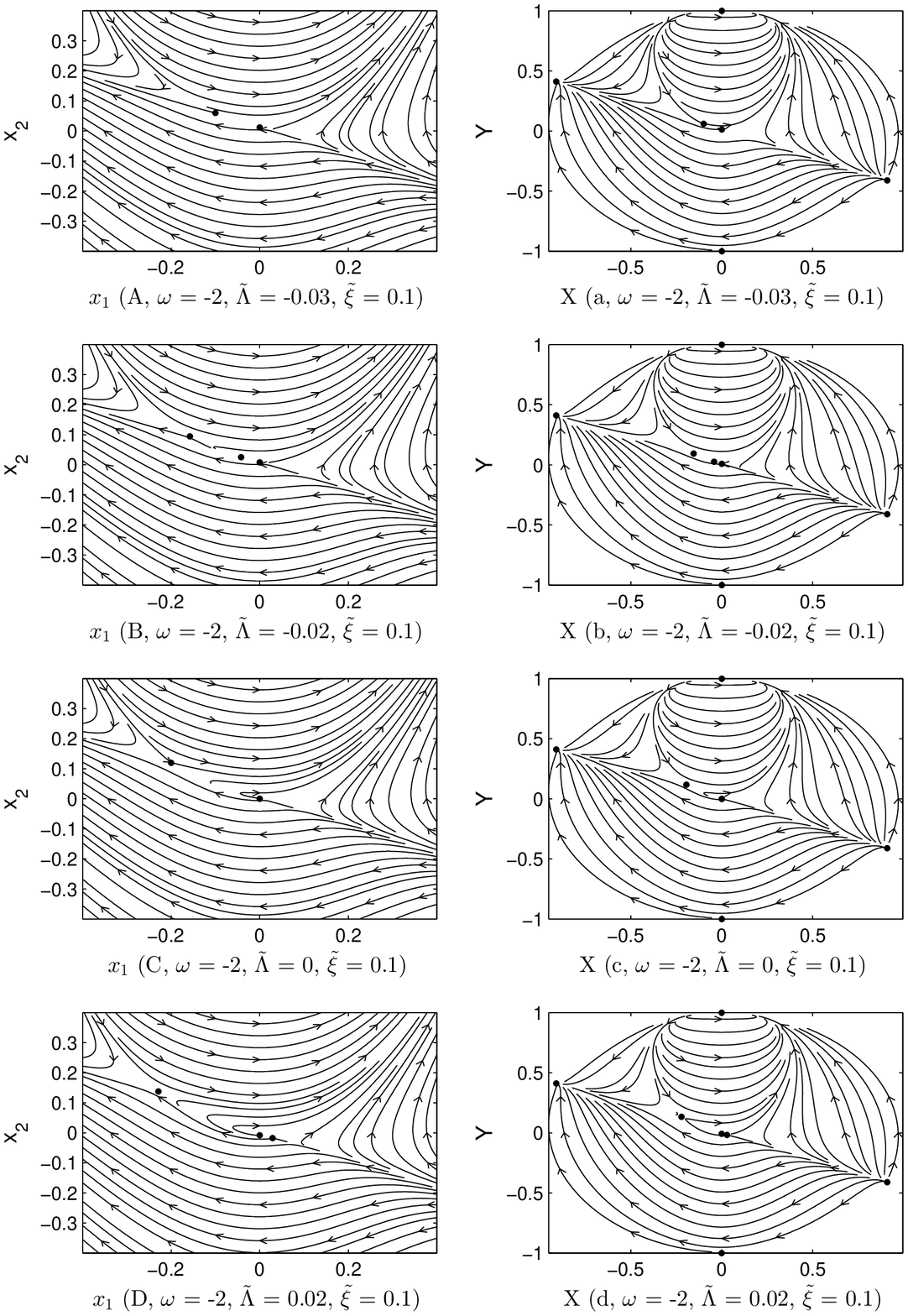}}
\caption{{\footnotesize
Uncompact (left panel) and compact (right panel) phase portraits  when viscosity is included, $\xit = 0.1$ for dust case $(\o = -2)$ but with different $\Lat$. Representative cases are
$ \Lat = \left\{-0.03, -0.02, 0, 0.02\right\}$. $x_1$ and $x_2$ respectively
denote the dimensionless $\Ht$ and $\rt$ as defined in Eq.(\ref{gre3}). $X$ and $Y$ are the coordinates on the Poincar\'{e}
sphere as defined in Eq.(\ref{XYZpoin}). The dotted circles represent fixed points. }}
\label{fig2_cos_omega_lam_xi_xpm_SD}
\end{figure}
\clearpage

The fixed points, at infinity, is determined by the zeros of the  function $G^{\mbox{m}+1}\left(\th\right)$, defined in Eq.(\ref{dynth2}), which for  Eqs.(\ref{eqco}) amounts to
\be
 G^{\mbox{m}+1}\left(\th\right) \stackrel{\mbox{m}=2}{{\scalebox{3}[1]{=}}} G^{3}\left(\th\right) =   - \cos^2{\th}\,\left[ \sin{\th}\, \left(2 + 3\,\omega\right) -18\,\xit\,\cos{\th}\right].
\label{infeqco}
 \ee
 For nonvanishing value of $\xit$, we have four fixed points corresponding to
 $\th =\left\{{\pi\over 2}, {3\,\pi\over 2}, \varphi, \varphi + \pi \right\}$ where $\varphi = \tan^{-1}\left(18\,\xit \over 2 + 3\,\o\right)$. The non vanishing value of $\xit $ prevents the
occurrence of an infinite number of fixed points at infinity when $\o = -{2\over 3}$ and reducing them to just a pair of fixed points at $\th =\left\{{\pi\over 2}, {3\,\pi\over 2}\right\}$.
Considering the flow only along the circle at infinity, the two fixed points at $(\th = {\pi\over 2})$ and $(\th = {3\,\pi\over 2})$ are behaving as  saddles but of nonhyperbolic type,
for any values of the relevant parameters,  since  ${d G^{3}\left(\th\right)\over d \th}$ is vanishing  at $\th = {\pi\over 2}$ or $ {3\,\pi\over 2}$.
Again by considering the flow along the circle at infinity, the other two fixed points corresponding to $\th =\left\{ \varphi, \varphi + \pi \right\}$ can be shown to be of opposite type such that one is stable
and the other is unstable depending on the sign of $\left(2 + 3\,\o\right)$ and which quadrant the angle $\varphi$ belongs to.

As we have seen the presence of viscosity prevents the occurrence of an infinite number fixed points wherever they are;  at the finite domain or the circle at infinity.
 Moreover, the occurrence of periodic orbits are prohibited by the presence of viscosity. In fact, the absence of these two kinds of behaviors is crucial since it is among the basic requirements for the dynamical system to have  structural  stability  according to the criteria presented in \cite{wigg,perko}.  In fact,  Peixoto theorem for a flow defined on a compact  two-dimensional as in \cite{wigg,perko}, which
in our case the  flow induced on the Poincare sphere,  can be used to decide the presence of  structural stability or not in the considered cosmological models.   According to Peixoto theorem \cite{perko,wigg},  the hyperbolcity of the fixed points
is a necessary conditions to attain structural stability which can't be satisfied in our case since we have always non-hyperbolic fixed points at infinity corresponding to  $(\th = {\pi\over 2})$ and $(\th = {3\,\pi\over 2})$.

It is also interesting to notice that when constant  bulk viscosity is included, the curve $\rt + \pt=0$ (phantom divide curve) which turns out to be a straight line given by $\rt  +\pt - 6\,\xit \Ht=0$ is not a solution curve as can be checked explicitly. Thus, there could be a solution curve that might cross the phantom divide curve in a finite time. The crossing of phantom divide can be noticed, as for examples, from the phase portraits presented in Fig.11(A,a) and Fig.12(C,c). Another equally interesting feature is the absence of  Milne type solution $(x_2 = 0)$ which acts as  phantom divide curve when viscosity is not present.

The cosmological model incorporating bulk viscosity in one of its simplest form can still lead to some interesting consequences that
could be relevant to the actual physical universe. We find that our parameters $(\o, \Lat, \xit)$ could be adjusted to have three fixed
points one along the $x_2$ $(\rt)$-axis and the other two along the flat curve solution. The one along the $x_2$ axis is a repulsive
center and it represents a static universe. It is implausible to consider that our physical universe started in the neighborhood of
this repulsive center since it contradicts the standard scenario of initial big-bang and early inflation. Thus we are left
with the two fixed points along the flat curve solution.

The bifurcation diagram in Fig.(\ref{figbi_two_pm_xi}) is of a great help in identifying the parameter space region that could be relevant in describing the real universe. The region characterized by $1+\o >0$ and $\Lat >0$ contains
two fixed points along the flat curve solution designated by $-$(unstable)  and $+$(stable) that represent de-Sitter universe. These solutions curves, which interpolate between two de-Sitter universes are guaranteed to be nonsingular since $x_1$, $x_2$ and their time derivative are finite.
These solution curves are also generic which means that they do not have zero measure. This behavior is evident from the phase portrait in  Fig.\ref{fig2_cos_omega_lam_xi_ex_bihopf}(C, c). One can see easily from the graph that the solution curves, connecting the two fixed point and coasting along and near the  flat curve solution are not of measure zero either. The presence of nonvanishing positive $\Lat$ is crucial for this finding where the properties of fixed points are drastically changed, when $\Lat =0$, as can be inferred from Eqs.( \ref{two_fix_flat}--\ref{gen_two_fix_lam}). This set of generic nonsingular solutions, coasting near the flat curve solution, is missed in the work \cite{belin2} since a cosmological constant was not included.

Before discussing briefly the case of variable viscosity coefficient , $\xit\, = \alpha\,x_2$, we summarize the important features associated with the found
fixed points at a finite domain as follows:\\
\begin{itemize}
\item $\o \neq -1/3$ case: \\
a) We have two fixed points, $ x=\left(\Ht_{\pm},{6\,\xit\,\Ht_{\pm} \over 1+ \o}\right)$, where $\Ht_{\pm}$ are the solutions of the algebraic equation, $\Ht^2-{2\xit \over 1+\o }\,\Ht-{\Lat \over 3}=0$,
these two fixed points are de Sitter universes, which allow for nonsingular solutions interpolating between them.\\
b) We still have the previous case fixed point, $ x=\left(0,{2 \Lat \over 1+3\,\o}\right)$, which is Einstein Static universe if $\Lat {1+\o \over 1+3\,\o}>0$, with ${R} \times S^3$ topology, or a static universe with ${R} \times H^3$ topology, if $\Lat {1+\o \over 1+3\,\o}<0$.\\
\item $\o=-1/3$, and $\o \neq -1$ case: We have two fixed points, $ x=\left(\Ht_{\pm},9\,\xit\,\Ht_{\pm}\right)$, where $\Ht_{\pm}$ are the solutions of the algebraic equation, $\Ht^2-3\,\xit\,\Ht-{\Lat \over 3}=0$, these two fixed points are de Sitter universes, which allow for nonsingular solutions interpolating between them.
\end{itemize}

\subsection*{Comments on the variable visocisty case,  $\mathbf \xit\, = \alpha\,x_2$}
For the case of variable viscosity coefficient , $\xit\, = \alpha\,x_2$, we find that there are four finite fixed points. These fixed points together with corresponding eigenvalues of their associated Jacobians are summarized in Table~(\ref{alah_fixed})
{\sf
\begin{table}[H]
\centering
\scalebox{1}{
\begin{tabular}{cccc}
\toprule
 $\left( x_{1}, x_{2}\right)$ & $\left( \l_{1},  \l_{2} \right)$ &  $\left( x_{1}, x_{2}\right)$ & $\left( \l_{1},  \l_{2} \right)$  \\
\midrule
    $\left( 0,\;\; {2\,\Lat\over 1 + 3\,\o}\right)$  &    $\left({3\,\alpha \,\Lat + \D\over 1 + 3\,\o} , \;\;{3\,\alpha \,\Lat - \D\over 1 + 3\,\o}\right)$& $\left( \sqrt{{\Lat\over 3}},\;\; 0\right)$ & $\left(-2\,\sqrt{{\Lat\over 3}} ,\;\; -\sqrt{3\,\Lat}\,\left(1 + \o\right) + 6\,\alpha\,\Lat \right)$ \\
$\left( {1+\o\over 6 \alpha},\;\; {\left( 1 + \o \right)^2 - 12\,\Lat\,\alpha^2\over 12\,\alpha^2}\right)$  &
$\left( -{1+\o\over 3\,\alpha},\;\; {\left( 1 + \o \right)^2 - 12\,\Lat\,\alpha^2\over 4\,\alpha}\right)$&
$\left( -\sqrt{{\Lat\over 3}},\;\; 0\right)$ &
$\left(+2\,\sqrt{{\Lat\over 3}} ,\;\; +\sqrt{3\,\Lat}\,\left(1 + \o\right) + 6\,\alpha\,\Lat \right),$ \\
\bottomrule
\end{tabular}}
\caption{\footnotesize Fixed points for variable viscosity coefficient, $\xit\, = \alpha\,x_2$. }
\label{alah_fixed}
\end{table} }
where $\D = \sqrt{9\,\alpha^2\,\Lat^2 + \Lat\,\left(1 + \o\right) \,\left(1 + 3\,\o\right)^2}$. All fixed points lies along the flat curve solution except
the point $\left( 0,\;\; {2\,\Lat\over 1 + 3\,\o}\right)$.
Regarding the fixed points, at infinity, is determined by the zeros of the  function $G^{\mbox{m}+1}\left(\th\right)$, defined in Eq.(\ref{dynth2}), which for    variable viscosity coefficient, $(\xit\, = \alpha\,x_2, \alpha \neq 0)$ amounts to
\be
 G^{\mbox{m}+1}\left(\th\right) \stackrel{\mbox{m}=3}{{\scalebox{3}[1]{=}}} G^{4}\left(\th\right) =   -18\,\alpha\,\sin{\th}\,\cos^3{\th},
\label{infeqvar}
 \ee
 leading to  four fixed points corresponding to $\th = \left\{0,\pi,{\pi\over 2}, {3\,\pi\over 2}\right\} $.

As mentioned before the full study for the case of variable viscosity coefficient, $\xit\, = \alpha\,x_2$, would be a subject of a future work.
However we would like to draw the attention of the reader to the possibility of having nonsingular solutions in this case which is a relevant feature for constructing cosmological models, for example if we take $\o =0, \Lat = \left(0, \;\;0.5\right)$ and $\alpha = 0.2$, the noncompact and compact phase portraits for this case are depicted in Fig.(\ref{fig_xi(rho)}).
\begin{figure}[H]
   \includegraphics[width=18cm,height=20cm]{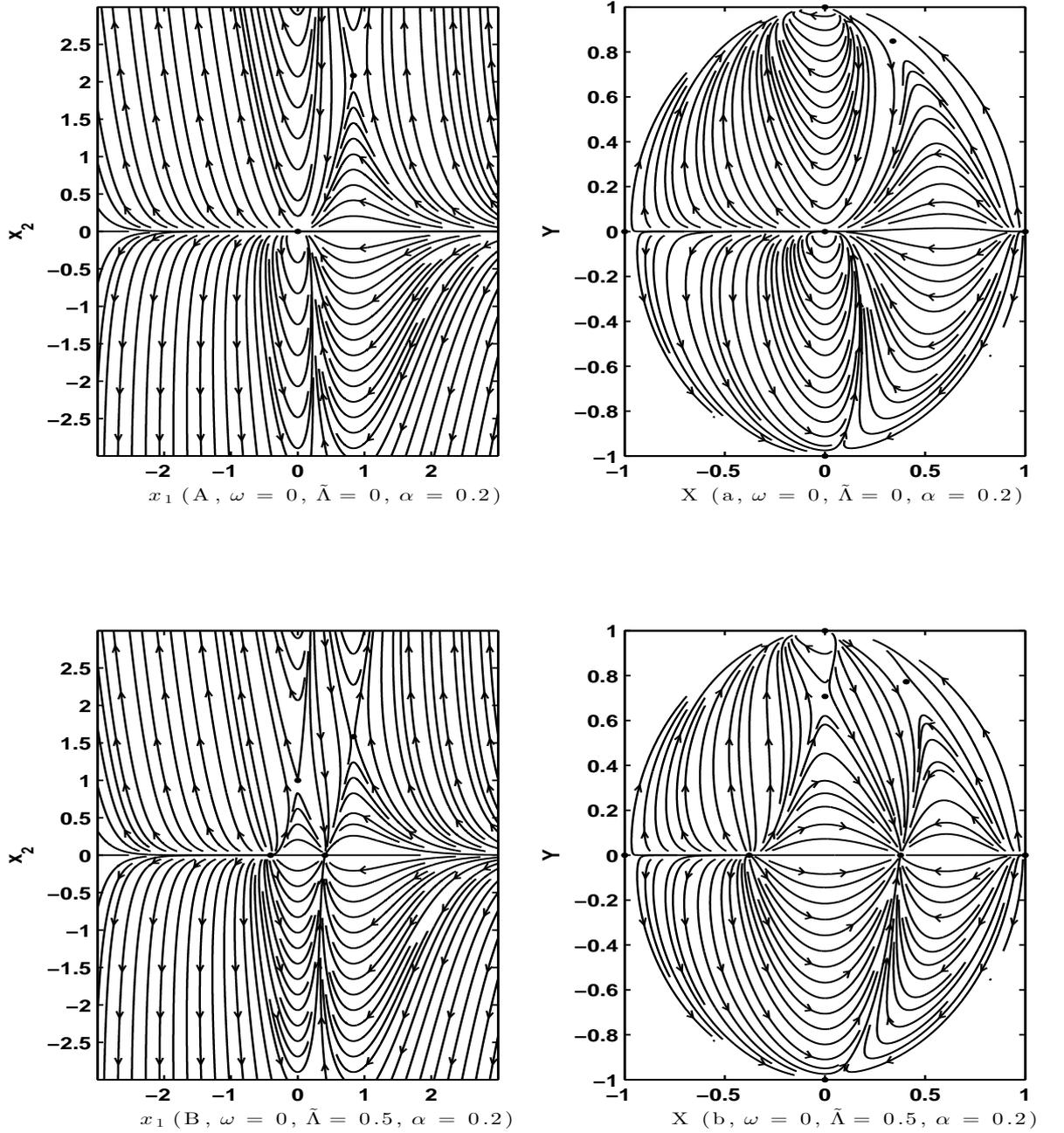}
\caption{{\footnotesize
Uncompact (left panel) and compact (right panel) phase portraits  when variable viscosity coefficients,  is included, $\xit\, = \alpha\,x_2, \,(\alpha = 0.2)$ for dust case $(\o = 0)$ but  $\Lat = \left(0, \;\;0.5\right),$. $x_1$ and $x_2$ respectively
denote the dimensionless $\Ht$ and $\rt$ as defined in Eq.(\ref{gre3}). $X$ and $Y$ are the coordinates on the Poincar\'{e}
sphere as defined in Eq.(\ref{XYZpoin}). The dotted circles represent fixed points. }}
\label{fig_xi(rho)}
\end{figure}
As can be checked, from Fig.(\ref{fig_xi(rho)}), the nonsingular solution along the flat curve connecting the two  fixed points  $\left( \sqrt{{\Lat\over 3}},\;\; 0\right)$ and $\left( {1+\o\over 6 \alpha},\;\; {\left( 1 + \o \right)^2 - 12\,\Lat\,\alpha^2\over 12\,\alpha^2}\right)$  are not generic in the sense of having zero measure in solution space. The fixed point $\left( {1+\o\over 6 \alpha},\;\; {\left( 1 + \o \right)^2 - 12\,\Lat\,\alpha^2\over 12\,\alpha^2}\right)$  is of a saddle type and the nonsingular solution is a separatix connecting the above mentioned two fixed points. Any small deviation from that nonsingular solution would give
other solutions with completely different characters and this was observed long time ago in \cite{belin2}. It was also noticed in \cite{belin2} that there is
a group of solution curves starting at finite time in the past with zero $x_2$ and positive infinite $x_1$ which are gradually building up (in $x_2$) till reaching a maximum positive value, then decaying to the fixed  point at $\left( \sqrt{{\Lat\over 3}},\;\; 0\right)$ at $t=\infty$. The trips downwards, after  reaching a maximum $x_2$, can be tuned to be coasting near the flat curve solution in order to mimic the expansion history of the observed universe. These solution curves can be also proved to be nonsingular which means that all invariant constructed out of Riemann curvature tensor are finite. The proof is simple because all curvature invariant can be written in terms of density $(\rt = x_1) $ and pressure $ (\pt = \o\, x_2 -3\,\alpha \, x_2\, x_1 )$   while $x_2 \propto e^{-18\,\alpha x_1}$ as $(x_1 \rightarrow \infty,\; x_2 \rightarrow 0)$. The only region of potential singularity, for this particular kind of solutions, resides in the region  $(x_1 \rightarrow \infty,\; x_2 \rightarrow 0)$ while for the other remaining region both $x_1$ and $x_2$ are finite. Other class of nonsingular generic solution, for nonvanishing positive $\Lat$, are the ones connecting the two fixed points at  $\left(- \sqrt{{\Lat\over 3}},\;\; 0\right)$ and   $\left(\sqrt{{\Lat\over 3}},\;\; 0\right)$ that are respectively unstable and stable. These nonsingular solutions in the positive $x_2$ region can't coast near the flat curve because they are confined between the two separatrices of the saddle fixed point at  $\left( 0,\;\; {2\,\Lat\over 1 + 3\,\o}\right)$ and thus are of no relevance for the real universe.

So far we are interested in applying the dynamical system tools to explore both the dynamics of cosmological models and the relevant parameter space in case of a single fluid with viscosity in the presence a cosmological constant. The aim is to produce an expansion history that matches the observed universe that started in the past with early-time inflation and ends at the future with late-time acceleration described by a de Sitter fixed point. As evident from Fig.\ref{fig_xi(rho)}(A,a) where the universe starts from a big-bang and ends at a late-time acceleration attributed to viscosity or starting from early-time inflation induced by viscosity and ending as empty expanding universe $(x_1 =0, x_2 =0 )$. When $\Lat$ is nonvanishing, as evident from Fig.\ref{fig_xi(rho)}(B,b), we still have the scenario of initial big-bang that ends at a late-time acceleration attributed to viscosity and in addition there could be early-time inflation caused by the combined effect of viscosity and cosmological constant and finally late-time acceleration due to the presence of $\Lat$.

It is worthy to mention that the expansion history is not the whole story and the model should be tested against several observational data among them are cosmic microwave background (CMB) anisotropies coming from different observations. It was shown in \cite{Gio} that cosmological models including a viscous fluid as the sole source of inflation have a serious drawback, namely, the dominance of the tensor modes against the scalar modes of perturbation which is in contradiction with observations.

\subsection{ Analysis of  bulk viscosity in a spatially flat case}
Since cosmological observations show that the universe is spatially flat with a high degree of accuracy, it is convenient to restrict the dynamical study to the spatially flat case. This kind of restriction could give us a simplified picture concerning the dynamics of the universe as was done in  \cite{Adel}.
\begin{itemize}
\item {\bf Constant bulk viscosity}\\
There is only one equation governing the dynamics since $x_2  = 3\,x_1^2 -\Lat$ in the flat case. Using the first equation in the
set of Eqs.(\ref{eqgen}) one can find,
\be
 \dot{x} _1 = -{3\over 2}\,\left(1 + \omega\right)\,\left(x_1 - {\xit \over 1 + \omega} \right)^2 +  {3\over 2}\,{\xit^2 \over 1 + \omega} +
 {\Lat \over 2}\,\left(1 + \omega \right).
 \label{flazeta_eq}
 \ee
In fact Eq.(\ref{flazeta_eq}), after shifting the variable $x_1$ as $z=x_1 - {\xit \over 1 + \o}$ and rescaling the time variable as $\tau = {3\over 2} \left| 1 + \o\right|\,\tt$,  can be castted into
\be
{dz\over d\tau} = \pm z^2 +  \mu,
\label{saddle-node}
\ee
where $\mu = {\xit^2 \over (1 + \omega) \left|1 + \o\right|} + {\Lat \over 3}\,{\left(1 + \omega \right)\over  \left|1 + \o\right|}$. This matches the normal form of saddle node bifurcation as
listed in Eq.(\ref{bifur_types}). The parameter $\mu$ has the critical value, zero, when $\displaystyle \Lat = -\,{3\,\xit^2 \over \left(1 + \omega\right)^2}.$

The fixed points corresponding to this flow, in Eq.(\ref{flazeta_eq}),  are
 \be
 x_{1\,\pm} = {1\over 3}\, {3\,\xit \pm \sqrt{\Delta_2} \over 1 + \omega},\;\;  x_{2\,\pm} = {2\,\left(3\,\xit \pm \sqrt{\Delta_2}\right)\,\xit \over 3\,\left( 1 + \omega\right)^2}
\label{fix_flatzeta}
 \ee
The flows as depicted in Fig.~\ref{fix_zeta_alpha}(A) single out a trajectory starting from a big-bang singularity and ending at a de-Sitter universe represented by the fixed point $x_{1\,+}$ as the only possible candidate describing our physical universe. This scenario of starting with a big-bang and ending up with late acceleration could be achieved in the presence of viscosity without the need for including a cosmological constant. Also, the big-bang can occur in both closed and open universe and still ending up with a late acceleration (fixed point $x_{1+}$) as evident from the plots in  Fig.\ref{fig2_cos_omega_lam_xi_ex_bihopf}(C, c).
The other two remaining possibilities are not good candidates for describing our physically observed universe as explained as follows.  The first one staring from $x_{1-}$ and ending up
with $x_{1+}$ (starting with small value for Hubble parameter, negative for positive $\Lat$,  and ending with a larger one) can't describe the actual universe since the opposite behavior is required. As to the second one starting with $x_{1-}$ and going to $x_{1} = -\infty$ which means passing through contracting phase and ending with a big crunch and this is clearly doesn't match the behavior of the observed universe which, at present, is expanding with acceleration.
\begin{figure}[H]
\includegraphics[height=4cm,width=16cm]{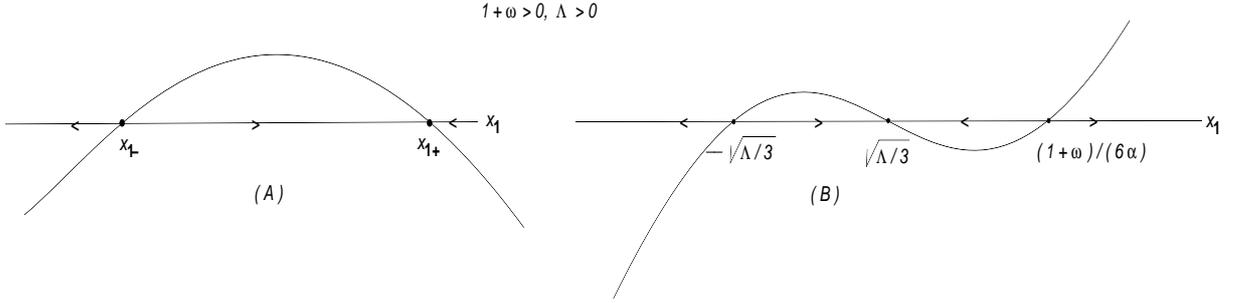}
        \caption{\footnotesize The curve determining the fixed points $\dot{x} _1 =0$ for a spatially flat universe: (A) for constant viscosity coefficient $\xit$ while (B) for varying viscosity coefficient $\xit = \alpha\,x_2$. It is understood that the vertical axis represent $\dot{x} _1$ which is not shown for convenience.}
        \label{fix_zeta_alpha}
     \end{figure}
\item {\bf Variable  bulk viscosity, $\mathbf {\xit\left(x_2\right) = \alpha\, x_2}$ }\\
Generically, the bulk viscosity coefficient is a function of energy density, $x_2$, for simplicity we assume a linear dependence as $\xit\left(x_2\right) = \alpha\, x_2$, which is a physically reasonable assumption (see \cite{murphy}). Inserting this form of varying $\xit$ into the
cosmological equations in Eq.(\ref{eqgen}) would give the following fixed points,
\be
  \left( x_1 =  \pm \sqrt{{\Lat \over 3}} ,\; x_2 =  0 \right), \;\;\;\;
\left( x_1 =  {1 +  \o \over 6\, \alpha } , \; x_2 = {  \left( 1 + \o \right)^2  - 12 \, \Lat \, \alpha^2 \over  12\,\alpha^2} \right) .
\label{fix_alpha}
\ee
Here we are interested in the dynamical equations restricted to flat case ($ k =0$), which amounts to a single equation for $x_1$ as,
  \be
 \dot{x} _1 = 9\,\alpha \,\left(x_1 - { 1 + \omega \over 6 \alpha} \right)\, \left( x_1^2 - {\Lat \over 3}\right).
 \label{flatalpha_eq}
 \ee
In fact Eq.(\ref{flatalpha_eq}), after shifting the variable $x_1$ as $z=x_1 - {1 + \o  \over 6 \alpha}$ and rescaling the time variable as $\tau = 9\,\alpha\,\tt$,  can be castted into
\be
{dz\over d\tau} = f(z)\equiv z^3 + b\,z^2 + a\,z,
\label{trans_3rd}
\ee
where $\displaystyle  a =\left({1 + \omega\over 6\,\alpha}\right)^2 - {\Lat \over 3}$ and $\displaystyle b = {1 + \omega\over 3\,\alpha}$. This matches the normal form of the transcritical  bifurcation, as listed in Eq.(\ref{bifur_types}),  but extended to a third order term. When adopting the form in Eq.(\ref{trans_3rd}), then one can determine the fixed points through  $f(z) =0$ and their degeneracy by evaluating  $f'(z)$ at the fixed points to discover regions where $f'(z)=0$. All these findings are summarized in Table~(\ref{trans_3rd_fix}).
\begin{table}[H]
\centering
\scalebox{1}{
\begin{tabular}{ccc}
\toprule
Fixed point ($f(z)=0$)& $f'(z)\,$ evaluated at the fixed point & Degeneracy (where $f'(z)=0$)\\
\midrule
$z_0 = 0$ & $a$  & $a=0\;\;(\Lat = 3\,\left({1 + \omega\over 6\,\alpha}\right)^2)$ \\
\midrule
$z_1 =\displaystyle{ -{b\over 2} +{ \sqrt{b^2 -4\,a}\over 2}}$ & $ \displaystyle{{b^2 - 4\,a\over 2} - {b\over 2}\,\sqrt{b^2 - 4\,a}}$ &  $a=0,\;b >0$ or $b^2 = 4\,a\, (\Lat =0) $\\
\midrule
$z_2 = \displaystyle{-{b\over 2} - { \sqrt{b^2 -4\,a}\over 2}}$ &  $\displaystyle{{b^2 - 4\,a\over 2} + {b\over 2}\,\sqrt{b^2 - 4\,a}}$ & $ a=0,\;b <0$ or $b^2 = 4\,a\, (\Lat =0)$\\
\bottomrule
\end{tabular}}
\caption{\footnotesize Fixed points for dynamical system ${dz\over d\tau} = f(z)\equiv z^3 + b\,z^2 + a\,z$ corresponding to the flat case with variable viscosity coefficient, $\xit\, = \alpha\,x_2$. }
\label{trans_3rd_fix}
\end{table}
It is important to check the degeneracy of the fixed points where $f'(z)=0$ since the bifurcation arises due to the presence of theses degenerate fixed points.
As evident from Table~(\ref{trans_3rd_fix}), the regions of degeneracy are where $a=0\,(\Lat = 3\,\left({1 + \omega\over 6\,\alpha}\right)^2)$ or $b^2 = 4\,a\, (\Lat =0)$. Crossing these degeneracy regions induces  bifurcation as shown in a bifurcation diagram depicted in Fig.~(\ref{trans_3rd}).
\begin{figure}[H]
\includegraphics[height=5cm,width=7cm]{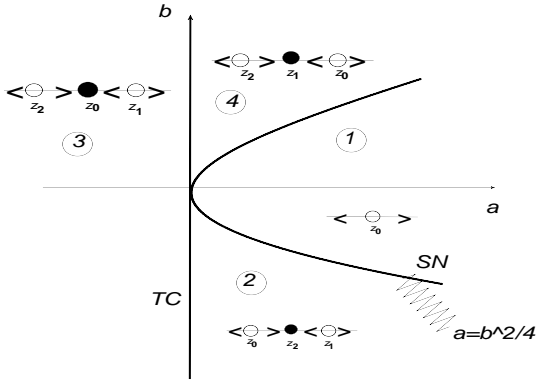}
        \caption{\footnotesize Bifurcation diagrams and schematic portrait where we have four topologically different areas that are separated by
saddle node (SN) and transcritical (TC) bifurcations. The solid dots and the circle dots in each phase portraits represent stable fixed points (sinks) and
unstable fixed points (sources) respectively. The thick lines represents lines of bifurcation such as $b$-axis for TC and the parabola, $a = b^2/4$, for SN.  The thin line just represents jus an axis.}
\label{trans_3rd}
\end{figure}

Going back to the cosmological equation as expressed in Eq.(\ref{flatalpha_eq}), the fixed points are clearly $  x_1 =\pm \sqrt{{\Lat \over 3}},  x_1 = { 1 + \omega \over 6 \alpha} $ .  The flow behavior as depicted in  Fig.~\ref{fix_zeta_alpha}(B) reveals
an interesting trajectory connecting  $x_1 = { 1 + \omega \over 6 \alpha} $   (early inflation) and $x_1 =  \sqrt{{\Lat \over 3}}$  (late acceleration)  provided that $\alpha$ satisfies $ { 1 + \omega \over 6 \alpha}   >>  \sqrt{{\Lat \over 3}}$. Moreover this solution,  connecting $x_1 = { 1 + \omega \over 6 \alpha} $   and $x_1 =  \sqrt{{\Lat \over 3}} $, is nonsingular but non generic as discussed before.

To get more physical insight for the cosmological model solutions and with the help of Eq.(\ref{gre1}), Eq.(\ref{gre3}) an Eq.(\ref{xpar}), one can compute the deceleration parameter $q$ to get
\be
q \equiv -{\ddot{a}\over H^2\,a} = {1\over x_1^2}\, \left[ {1\over 2}\,\left( {x_2\over 3} + \o\, x_2 - 6\,\xit\,x_1\right) - {\Lat \over 3}\right].
\label{decel}
\ee
The deceleration parameter $q$, evaluated at the fixed points given in Eq.(\ref{fix_flatzeta}) and Eq.(\ref{fix_alpha}) and located at the flat curve solution,  turns out to be $-1$  which means acceleration.
In case of constant viscosity coefficient and for the solution starting from a big-bang singularity and reaching a fixed point $x_{1+}$ as shown in  Fig.~\ref{fix_zeta_alpha}(A), the $q$ starts positive (deceleration)  and then the combined effect of cosmological constant and viscosity
tends to decrease $q$ till reaching zero and then becoming negative and equal to $-1$ (acceleration) at $x_1 = x_{1+}$. While in the case of varying coefficient of viscosity and for the solution connecting
 $x_1 = { 1 + \omega \over 6 \alpha} $   (early inflation) and $x_1 =  \sqrt{{\Lat \over 3}}$  (late acceleration), the $q$  evolves in a continuous way that  starts and ends with value $-1$ and having intermediate region where
$q$ is positive (deceleration).
\end{itemize}

\section{Discussion and conclusion}
The purpose of this work is to emphasis the importance of dynamical systems tools, especially, that of degenerate cases with bifurcations, classify them through calculating their normal forms and identify them with the known forms of bifurcations. In order to apply these normal form calculations we better have equation of states beyond the linear ones (this takes us beyond the $\Lambda CDM$ model). It is natural to consider viscous cosmological models for this purpose since any real fluid shows dissipative phenomena, which in the case of FRW models have nonlinear equation of states.

We present a complete dynamical study for a bulk viscous cosmology with a single fluid in the presence of a cosmological constant. For the sake of illustration and clarification we don't study the bulk viscous cosmological model in a single step containing the  three parameters namely $\o, \Lat$ and $\xit$ (constant viscosity coefficient), but our investigation is carried out in three different stages.
The first stage, we consider only $\o$ to be nonvanishing and then $\Lat$ is included while finally $\xit$ is introduced. In each of these stages, the fixed points, whether they are at the  finite domain of the phase space or at infinity, are studied and classified.
Also, the normal forms are obtained for each stage together with phase space portraits for meaningful representative cases. Suitable and convenient bifurcations diagrams are plotted for illustrating the changing behavior of fixed points as the relevant parameters vary.

The case of varying viscosity coefficient in its full generality,  $\xit\left(x_2\right) = \alpha\, x_2$, is  briefly studied and the full study would be a subject of a future work. The flat case is studied in detail for both constant and varying viscosity coefficient and is shown to produce standard bifurcation like saddle node and transcritical bifurcation.

The dynamical system corresponding to bulk viscous cosmological model is shown,  in Section~2,  to be a two dimensional one. The resulting dynamical system can be classified, for constant bulk viscosity, as a degenerate  Bogdanov-Takens system  following the classification carried out in
\cite{kuz}. This  point concerning the classification is a novel result up to the best of our knowledge. Another issue besides the classification which is worthy to be discussed is the structural stability which means that the qualitative behavior of the system is unaffected by small perturbations. In two dimensional dynamical system,  simple criteria can be established for testing structural stability utilizing  Peixoto theorem for a flow defined on a compact  two-dimensional space as in \cite{perko,wigg}. In our study the flow induced on the Poincar\'{e} sphere  can be used to shed some light on the structural stability
of the considered cosmological models.  One of the basic criteria is to have  a finite number of fixed points and  periodic orbits which are hyperbolic. Here the finiteness of the number of fixed points, as shown in section~6,  can be achieved by introducing a  non vanishing
viscosity while the hyperbolicity of fixed points in the finite domain of phase space can be attained by restricting the relevant parameters $(\o, \Lat, \xit)$, as an example,  $\Lat > 0$ and $\o + 1 >0$.  Unfortunately,  as can be inferred from Eq.(\ref{infeqco}),  we have at infinity fixed points, $(\th ={\pi\over 2}, {3\,\pi\over 2})$, that are always  nonhyberbolic for any choice of the parameters. Thus the structural stability for the considered cosmological models  can't be achieved even after introducing viscosity and for any chosen region in the parameter space $(\o, \Lat, \xit)$. The finding concerning structure stability is not changed when including varying viscosity coefficients where $\xit\left(x_2\right) = \alpha\, x_2$.

The late acceleration behaviour is a confirmed feature of the observed universe due to the
observations of distant supernova type Ia \cite{nova1,nova2} and cosmic microwave background anisotropy measurements \cite{anis1,anis2}. The bulk viscous fluid can provide us with a source of this late acceleration, even in the absence of cosmological constant $\Lat$,  as discussed in Section~6  and illustrated in Fig.\ref{fix_zeta_alpha}(A).
In this case,  the cosmological model interpolates between big-bang and late acceleration. This induced late acceleration can be attributed to the effect of negative pressure associated with viscosity as is clear from the expression of $T_{\mu\nu}$ in Eq.~(\ref{em}).

The bulk viscous fluid with viscosity coefficient dependent on density as $\xit(x_2) = \a\,x_2$ and in conjunction with cosmological constant can provide us with a nonsingular cosmological model. The model interpolates between an  inflation point, $x_1 =\left( { 1 + \omega \over 6 \alpha}\right) $ , and a late acceleration point,  $x_1 =  \sqrt{{\Lat \over 3}}$,  as discussed in Section~6  and illustrated in  Fig.\ref{fix_zeta_alpha}(B). The complete study of this model including viscosity coefficient $\xit$ dependent on $x_2$ along the lines presented for the one of constant $\xit$ would be a subject for the future work. To confront the introduced cosmological  models with observational  data like Type Ia supernova, one should include an additional fluid
component that represents matter besides the dark energy component represented by $\Lat$ and a viscous fluid. The introduced parameter $\Lat$, $\xit$ and $\a$ might enhance the agreement with observational data but that needs a detailed study which would be a subject for a future work.

Finally, it is worthy to mention that producing correctly the expansion history of the universe is curial but it is not the whole story. There is another important check which is the test against cosmological perturbations. Unfortunately,  all models that attribute  the early inflation solely due to viscosity effect were shown in \cite{Gio} to be ruled out because of producing dominant tensor modes against the scalar modes of perturbation which  is not consistent with observational evidence. A possible remedy according to \cite{Gio}, is  to  introduce a scalar field component to produce a viable early inflation. The dynamical system
tools,  presented in this work,  can be also  applied in the presence of scalar field component but this might be a subject for future work.





\begin{thebibliography}{99}
\bibitem{Strogatz} S. H. Strogatz, {\it Non Linear Dynamics and Chaos }, Perseus Books, (1994).

\bibitem{stewart} C. B. Collins and J. M.  Stewart, Monthly Notices of the Royal Astronomical Society, 153(4),419 (1971)

\bibitem{collin1} C. B. Collins,  Comm. Math. Phys. 23 (2), 137 (1971).
\bibitem{collin2} C. B. Collins,  Comm. Math. Phys. 27 (1), 37  (1972).

\bibitem{belin1} V. A. Belinski and  I.M. Khalatnikov, Sov. Phys. JETP 42, 205–210 (1976).

\bibitem{belin2} V. A. Belinski and I.M. Khalatnikov, Sov. Phys. JETP 45, 1–9 (1977).

\bibitem{Adel} A. Awad, Phys. Rev. D, 87, 103001 (2013).
\bibitem{Andron1} Supriya Pan,  Jaume de Haro,  Andronikos Paliathanasis and Reinoud Jan Slagter, Mon. Not. Roy. Astron. Soc. 460 (2), 1445  (2016) .

\bibitem{Andron2} G. Papagiannopoulos, Spyros Basilakos and Andronikos Paliathanasis, Eur. Phys. J. C, 80  (2020).

\bibitem{review1} S. Bahamonde, C. G. B$\ddot{o}$hmer, S. Carloni,
E. J. Copeland, W. Fang and N.  Tamanin,  Phys. Rep. 775-777, 1  (2018).
\bibitem{bonnet} E. J. Kim and S. Kawai,  Phys. Rev. D 87, 083517, (2013).

\bibitem{kohli1}  I.  S. Kohli and  M.  C.  Haslam, J. Geom and Phys, 123, 434 (2018).

\bibitem{ekart} C. Eckart,  Phys. Rev. 58, 919 (1940).
\bibitem{is1} W. Israel,  Ann. Phys. 100, 310 (1976).
\bibitem{is2} W. Israel and J. M.  Stewart,  Ann. Phys. 118, 341 (1979).

\bibitem{WZ} Winfried Zimdahl, Phys.Rev. D 53, 5483 (1996).
\bibitem{BdHOS}I. Brevik, O. Gron, J. de Haro, S. D. Odintsov and E. N. Saridakis,, Int. J.  Mod.  Phys. D 26, 1730024 (2017).
	

\bibitem{murphy} G.L. Murphy, Phys. Rev. D 8, 4231 (1973).

\bibitem{colist} R. Colistete, J. C. Fabris, J. Tossa and W. Zimdahl,  Phys. Rev.  D 76, 103516 (2007).



\bibitem{nova1} A.G. Riess et al., Supernova search team, Astron. J. 116, 1009 (1998).

\bibitem{nova2} S. Perlmutter et al., The Supernova Cosmology Project, Astophys. J. 517, 565, (1999).

\bibitem{anis1} D.N. Spergel et al., Astrophys. J. Suppl. 148 (2003).

\bibitem{anis2}  M. Tegmark et al., The SDSS Collaboration, Phys. Rev. D 69,  103501 (2004).

\bibitem{jou} D. Pav'on, J. Bafaluy and D. Jou, Class. Quantum Grav. 8, 347 (1991).
\bibitem{zak}  M. Zakari and D. Jou, Phys. Rev. D 48, 1597 (1993).
\bibitem{maart} R. Maartens, Class. Quantum Grav. 12, 1455 (1995).

\bibitem{orest_anp}  M. Szydlowski and O.  Hrycyna, Ann. Phys. 322 , 2745 (2007).

\bibitem{odin} S. D. Odintsov, V. K. Oikonomou and P. V.  Tretyakov, Phys. Rev.  D 96 (4), 044022 (2017).

\bibitem{mathew_2017} A. Sasidharan and T. K.  Mathew, JHEP, 06 , 138 (2017).

\bibitem{kuz} Yu. A. Kuznetsov, {\it Elements of Applied Bifurcation Theory}, Springer-Verlag, NY, (2004).

\bibitem{coley1} A. A. Coley, {\it  Dynamical systems and cosmology},  Dordrecht Boston London: Kluwer
Academic Publishers, (2003).






\bibitem{perko} L. Perko, {\it Differential Equations and Dynamical Systems}, Springer-Verlag, NY, (2000).

\bibitem{wigg} S. Wiggins, {\it Introduction to applied nonlinear dynamical systems and chaos}, second ed.,
Texts in Applied Mathematics, vol. 2, Springer-Verlag, NY, (2003).

\bibitem{dumo} F. Dumortier, R. Roussarie,  J. Sotomayor, and H. Zaladek,
{\it Bifurcation of Planar Vector Fields}, Lecture Notes in Math., Springer-Verlag, NY, (1991).

\bibitem{Gio} M. Giovannini, Phys. Rev.  D 93, 083521  (2016).

\end{thebibliography}
\end{document}